\documentstyle[preprint,aps,harvard,epsf]{revtex}

\newcommand{\be}{\begin{equation}}
\newcommand{\ee}{\end{equation}}
\newcommand{\bea}{\begin{eqnarray}}
\newcommand{\eea}{\end{eqnarray}}
\newcommand{\vp}{\varphi}

\citationstyle{dcu}
\begin{document}

%\special{papersize=8.5in,11in}
\title{Electroweak baryogenesis}
\author{Mark Trodden\footnote{\tt trodden@theory1.phys.cwru.edu.}}
\address{Particle Astrophysics Theory Group\\
         Department of Physics\\
         Case Western Reserve University\\
         10900 Euclid Avenue\\
         Cleveland, OH 44106-7079.}
\maketitle
\begin{abstract}
   Contrary to naive cosmological expectations, all evidence suggests that 
   the universe contains an abundance of
   matter over antimatter. This article reviews the currently popular
   scenario in which testable physics, present in the standard model of
   electroweak interactions and its modest extensions, is responsible
   for this fundamental cosmological datum. A pedagogical explanation of
   the motivations and physics behind electroweak baryogenesis is
   provided, and analytical approaches, numerical studies, up to date 
   developments and open questions in the field are also discussed.
\end{abstract}

\vfill

\noindent CWRU-P6-98 \hfill Submitted to {\it Reviews of Modern Physics}

\noindent \hfill Typeset in REV\TeX

\newpage
\tableofcontents
\newpage
\markright{Trodden 12/15/98 Revised}
\section{Introduction}
\label{Intro}

The most basic distinction drawn between particles found in nature
is that between particle and antiparticle. Since antiparticles were first 
predicted \cite{Dirac1,Dirac2} and observed \cite{Anderson1,Anderson2}, it has 
been clear that there is a high degree of symmetry between particle and
antiparticle. This means, among other things, that a world
composed of antimatter would behave in a similar manner to
our world. This basic tenet of particle physics, the symmetry between matter
and antimatter, is in stark contradiction to the wealth of everyday and
cosmological evidence that the universe is composed almost entirely of
matter with little or no primordial antimatter. 

The evidence that the universe is devoid of antimatter comes from a
variety of different observations. On the very small scale, the absence of
proton-antiproton annihilations in our everyday
actions, constitutes strong evidence that our world
is composed only of matter and no antimatter.
Moving up in scale, the success of satellite launches,
lunar landings, and planetary probes indicate that our solar system 
is made up of the same type of matter that we are, and that there
is negligible antimatter on that scale. To determine whether antimatter
exists in our galaxy, we turn to cosmic rays. Here we see the first
detection of antimatter outside particle accelerators. Mixed in with the many
protons present in cosmic rays are a few antiprotons, present at a
level of around $10^{-4}$ in comparison with the number of protons
(see for example \citeasnoun{SA 88}). 
However, this number of antiprotons is 
consistent with their secondary production through accelerator-like
processes, $p+p\rightarrow 3p + {\bar p}$, as the cosmic rays stream
towards us. Thus there is no evidence for primordial antimatter in 
our galaxy. Finally, if matter and antimatter galaxies were to coexist
in clusters of galaxies, then we would expect there to be a detectable
background of $\gamma$-radiation from nucleon-antinucleon annihilations 
within
the clusters. This background is not observed and so we conclude that
there is negligible antimatter on the scale of clusters (For a review of the 
evidence for a baryon asymmetry see \citeasnoun{GS 76}).

A natural question to ask is, what is the largest scale on which we can
say that there is no antimatter? This question was addressed by
\citeasnoun{GS 76},and by \citeasnoun{FS 85}, and in particular has been 
the subject of a careful 
recent analysis by \citeasnoun{CRG 97}. If large domains of matter and
antimatter exist, then annihilations would take place at the 
interfaces between them. If the typical size of such a domain was small
enough, then the energy released by these annihilations would
result in a diffuse $\gamma$-ray background and a distortion of the cosmic 
microwave radiation, neither of which is observed.
Quantitatively, the result obtained by the latter
authors is that we may safely conclude that the universe consists
entirely of matter on all scales up to the Hubble
size. It therefore seems that the universe is fundamentally 
matter-antimatter asymmetric.

\begin{figure}
  \centerline{\leavevmode\epsfysize=20cm \epsfbox{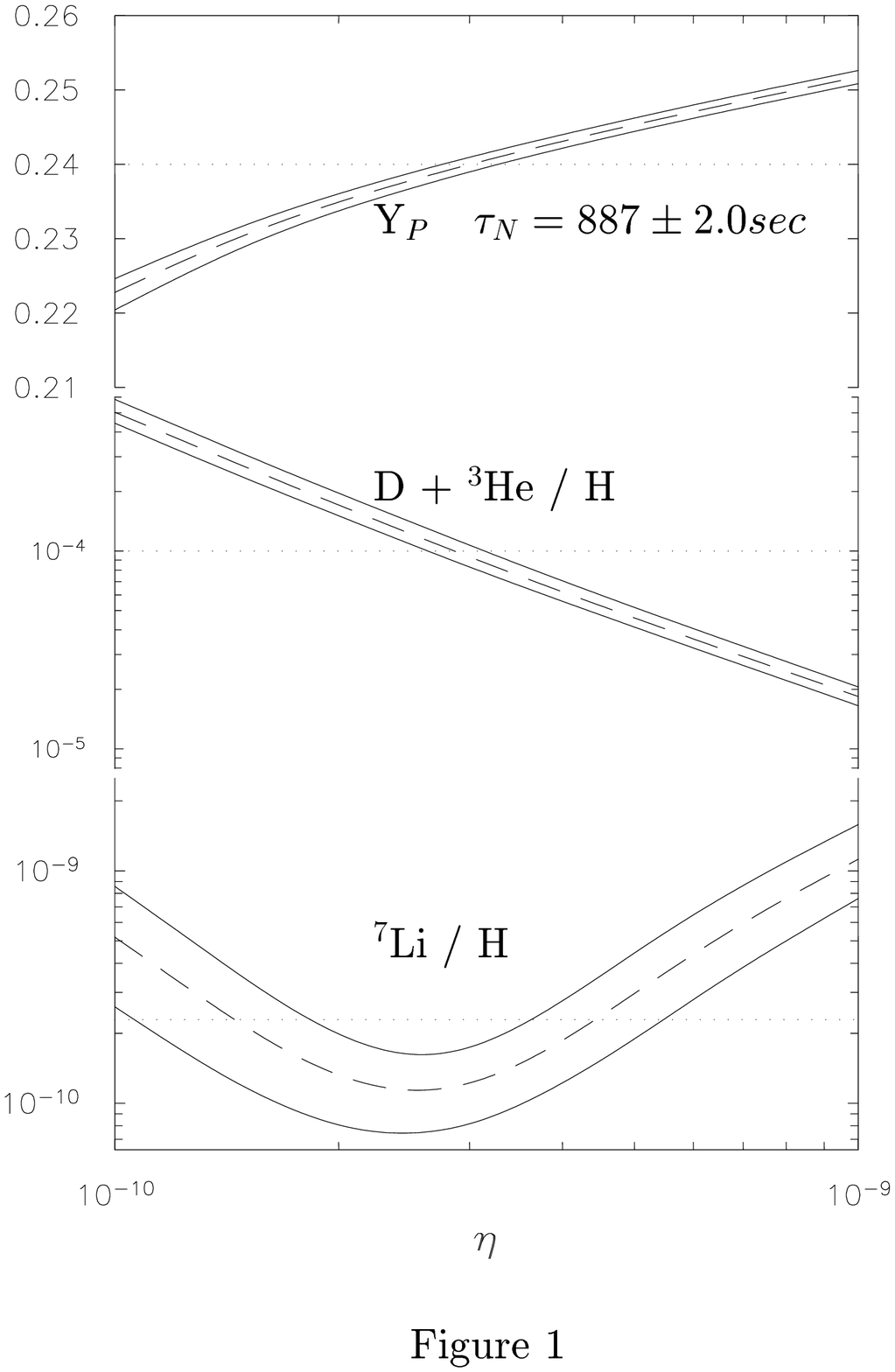}}
  \caption{}
\label{bbnfig}
\end{figure}

The above observations put an experimental upper bound on the amount
of antimatter in the universe. However, strict quantitative estimates of
the relative abundances of baryonic matter and antimatter may also be 
obtained from the standard 
cosmology. Primordial nucleosynthesis (for a review see \citeasnoun{CST 95}) 
is one of the most powerful tools of the standard cosmological 
model. The theory allows accurate predictions of the cosmological
abundances of all the light elements, H, $^3$He, $^4$He, D, B
and $^7$Li, while requiring only a single input parameter.
Define $n_b$
to be the number density of baryons in the universe. Similarly
define $n_{\bar b}$ to be the number density of antibaryons, and
the difference between the two to be $n_B$. Then, if
the entropy density in the universe is given by $s$, the single
parameter required by nucleosynthesis is the baryon to entropy ratio 

\be
\eta \equiv \frac{n_B}{s} = \frac{n_b-n_{\bar b}}{s} \ ,
\ee
and one may conservatively say that calculations of the primordial light 
element abundances are correct if

\be
1.5\times 10^{-10} < \eta < 7\times 10^{-10} \ .
\label{nucleo}
\ee
Although the range of $\eta$ within which all light element abundances 
agree with observations is quite narrow (see figure \ref{bbnfig}), its 
existence at
all is remarkable, and constitutes a strong confirmation of the standard
cosmology. For recent progress in nucleosynthesis see \citeasnoun{TFB 96}
and \citeasnoun{CH 97}.

The standard cosmological model provides a complete and accurate
description of the evolution of the universe from extremely early
times (a few minutes) to the present day ($10-20$ billion years)
given a host of initial conditions, one of which is the value of 
$\eta$. This standard picture is based on classical, fluid sources
for the Einstein equations of General Relativity (GR).
While the success of the standard cosmology is encouraging, there
remains the question of the initial conditions. One approach is just
to consider the initial values of cosmological parameters as given.
However, the values required for many parameters are extremely
unnatural
in the sense that the tiniest deviation from them leads to a universe
entirely different from the one that we observe. One well known example
of this is the initial value of the mass density of the universe, the
naturalness of which is at the root of the {\it flatness problem} of
the standard cosmology.

The philosophy of {\it modern} cosmology, developed over
the last thirty years, is to attempt to explain the required initial
conditions on the basis of quantum field theories
of elementary particles in the early universe. This approach has
allowed us to push our understanding of early universe cosmology
back to much earlier times, conservatively as early as $10^{-10}$
seconds, and perhaps much earlier.

The generation of the observed value of
$\eta$ in this context is referred to as {\it baryogenesis}. A first
step is to outline the necessary properties a particle physics theory 
must possess. These conditions were first identified by \citeasnoun{Sakharov} 
and are now referred to as the three {\it Sakharov Criteria}. They are:

\begin{itemize}

\item Violation of the baryon number ($B$) symmetry.

\item Violation of the discrete symmetries $C$ (charge conjugation)
      and $CP$ (the composition of parity and $C$)
\item A departure from thermal equilibrium.

\end{itemize}

The first of these is rather obvious. If no processes ever occur in which $B$
is violated, then the total number of baryons in the universe must
remain constant, and therefore no asymmetry can be generated from symmetric
initial conditions. The second Sakharov criterion is required 
because, if $C$ and $CP$ are exact symmetries, then one can prove
that the total rate for any process which produces an excess of baryons is
equal to the rate of the complementary process which produces an
excess of antibaryons and so no net baryon number can be created. 
That is to say that the thermal average of $B$, which is odd under
both $C$ and $CP$, is zero unless those discrete symmetries are
violated. Finally, there are many ways to explain the third criterion.
One way is to calculate the equilibrium average of $B$:

\bea
\langle B\rangle_T & = & \hbox{Tr}(e^{-\beta H}B) \nonumber \\
& = & \hbox{Tr}[(CPT)(CPT)^{-1}e^{-\beta H}B)] \nonumber \\
& = & \hbox{Tr}(e^{-\beta H}(CPT)^{-1}B(CPT)] \nonumber \\
& = & -\hbox{Tr}(e^{-\beta H}B) \ ,
\eea
where in the third step I have used the requirement that the Hamiltonian 
$H$ commutes with $CPT$, and in the last step used the properties of $B$ that 
it is odd under $C$ and even under $P$ and $T$ symmetries. Thus
$\langle B\rangle_T = 0$ in equilibrium and there is
no generation of
net baryon number. This may be loosely described in the
following way. In quantum field
theories in thermal equilibrium, the number density of any particle
species, $X$ say, depends only on the energy of that species, through 

\be
n_{eq}(X)=\frac{1}{e^{(E-\mu)/T} \pm 1}\ ,
\ee
where $\mu$ is the chemical potential corresponding to baryon number.
Since the masses of particle and antiparticle are equal by virtue
of the CPT theorem, and $\mu=0$ if baryon number is violated, we have that 

\be
N_{eq}(X)=\int\frac{d^3p}{(2\pi)^3}n_{eq}=N_{eq}({\bar X})\ ,
\ee
and again there is no net asymmetry produced.

The focus of this article is to review one popular scenario for 
generating the baryon asymmetry of the universe (BAU), as quantified in
equation~(\ref{nucleo}), within the context of modern cosmology.
In general, such scenarios involve calculating $n_B$, and then dividing
by the entropy density

\be
s=\frac{2\pi^2}{45}g_* T^3 \ ,
\ee
where $g_*$ is the effective number of massless degrees of freedom at
temperature $T$.
While there exist many attempts in the literature to explain the
BAU (for a review see \citeasnoun{AD 92}), I will concentrate on those
scenarios which involve anomalous electroweak physics, 
when the universe was at a temperature of $10^2\,$GeV (for
earlier reviews see \citeasnoun{NTreview}, \citeasnoun{CKNreview}, and
\citeasnoun{review}). The production of the BAU through these models is 
referred to as {\it electroweak baryogenesis}.

In the next section I will describe baryon number violation in the electroweak 
theory both at zero and at nonzero temperature. In section \ref{CP} I shall
move on to the subject of CP violation, explaining how this arises in
the standard model and how it is achieved in some popular extensions.
Section \ref{EWPT} contains an account of the electroweak phase 
transition, including a discussion of both analytic and numerical 
approaches.
Having set up the framework for electroweak
baryogenesis, I turn in section \ref{local} to the dynamics in the case
where baryon production occurs close to a phase boundary during a phase
transition. In section \ref{nonlocal}, I then extend these ideas to include
the effects of particle transport, or diffusion. Section \ref{MSSM}
contains a description of how baryogenesis is implemented in a popular 
extension of the standard model, the minimal supersymmetric standard 
model (MSSM).
In section \ref{defects}, I explain how, in some extensions of the electroweak
theory, baryogenesis may be mediated by topological defects, alleviating the 
constraints on the order of the phase transition. Finally, in section
\ref{conclusions} I summarize the results and comment on open questions
and future directions in the field.

It is my hope that this article will fulfill its intended role as both
a review of the background and basic material for beginners in the
field, and as a summary of and commentary on the most recent results and
directions in the subject. However, the focus of this article, as in any such
endeavor, is quite idiosyncratic, and I apologize to any of my 
colleagues whose work has been omitted or incorrectly detailed. A different
focus can be found in other accounts of the subject and, in particular, for a
comprehensive modern review of numerical approaches I recommend 
\citeasnoun{review}.

A note about conventions. Throughout I use a metric with signature $+2$
and, unless explicitly stated otherwise, I employ units such that
${\hbar}=c=k=1$ so that Newton's constant is related to the
Planck mass through $G=M_{pl}^{-2}$.

\section{Baryon Number Violation in the Electroweak Theory}
\label{baryon}
The standard model of unified electromagnetic and weak nuclear 
interactions \cite{{SG 61},{SW 67},{AS 68}} is based on the gauge groups
$SU(2)\times U(1)$. The
model is described by the Lagrangian density

\be
{\cal L} = (D_{\mu}\phi)^{\dagger}D^{\mu}\phi - 
\frac{1}{4}F_{\mu\nu}F^{\mu\nu} - \frac{1}{4}W_{\mu\nu}^aW^{a\mu\nu}
+ V(\phi) + {\cal L}_f\ .
\label{ewlag}
\ee
Here, the field strengths are 

\bea
F_{\mu\nu} & = & \partial_{\mu}B_{\nu} - \partial_{\nu}B_{\mu}
\ , \nonumber
\\ 
W_{\mu\nu}^a & = & \partial_{\mu}A_{\nu}^a
-\partial_{\nu}A_{\mu}^a +g\varepsilon^{abc}A_{\mu}^b A_{\nu}^c \ ,   
\nonumber
\eea
where $B_{\mu}$ is the hypercharge gauge field and $A_{\mu}^a$ are
the weak isospin gauge fields.
The covariant derivative is

\bea
D_{\mu} & = & \partial_{\mu} - \frac{i}{2}g{\bf \tau}.{\bf A_{\mu}}
-\frac{i}{2}g'B_{\mu} \ , \nonumber
\eea
and the Higgs potential is

\be
V(\phi) = \frac{\lambda}{4}(\phi^{\dagger}\phi - v^2)^2 \ .
\ee
In the above, $g$ is the $SU(2)$ coupling constant, $g'$ is the $U(1)$
coupling constant, $\lambda$ is the Higgs self-coupling, and $v=246\,$ GeV
is the vacuum expectation value (VEV) of the Higgs. 
The potential $V(\phi)$ is chosen so that the gauge symmetry
is spontaneously broken down to the $U(1)$ of electromagnetism that
is realized by the vacuum today, and ${\cal L}_f$ denotes the fermionic
sector of the theory that I will describe in a moment. Note that it is
conventional to define

\be
\alpha_W \equiv \frac{g^2}{4\pi} \simeq \frac{1}{29} \ ,
\label{alphaweak}
\ee
and to write the ratio of $g'$ to $g$ as

\be
\tan\theta_W \equiv \frac{g'}{g} \ ,
\label{weakangledef}
\ee
where the experimentally measured value of the {\it weak mixing angle} 
$\theta_W$ is given by

\be
\sin^2\theta_W \simeq 0.23 \ .
\label{weakangleval}
\ee

It will be useful to briefly describe a slightly different 
formulation of the $SU(2)+$Higgs theory, equivalent to considering the 
purely bosonic part of the standard model and ignoring the $U(1)$ 
hypercharge gauge field. The Lagrangian may be written in the form

\be
{\cal L}=
-\frac{1}{2}\hbox{Tr}(W_{\mu\nu}W^{\mu\nu})
-\frac{1}{2}\hbox{Tr}(D^{\mu}\Phi)^{\dagger}D_{\mu}\Phi
-\frac{\lambda}{4}\left[\hbox{Tr}(\Phi^{\dagger}\Phi)-v^2\right]^2 \ .
\label{SM action}
\ee
In this form, the standard Higgs doublet 
$\phi=(\phi_1,\phi_2)$ is related to the matrix $\Phi$ by

\be
\Phi({\bf x},t)=\left(\matrix{\phi_2^* & \phi_1 \cr
-\phi_1^* & \phi_2} \right) \ .
\label{higgsmatrix}
\ee
For reference, $g=0.65$, the gauge boson mass is 
$m_W=\frac{1}{2}gv$ and the Higgs
boson mass is $m_H=\sqrt{2\lambda}v$.
Note that

\be
\Phi^{\dagger}\Phi = (\vp_1^*\vp_1 + \vp_2^*\vp_2)\ 
\left(\matrix{1 & 0 \cr 0 & 1} \right) \ ,
\ee
so that one can write

\be
\Phi = \frac{\sigma}{\sqrt{2}}\, U \ ,
\label{sigmaU}
\ee
where $\sigma^2 = 2\left(\vp_1^*\vp_1 + \vp_2^*\vp_2 \right) =
{\rm Tr} \Phi^\dagger \Phi$,  and $U$ 
is an $SU(2)$
valued field which is uniquely defined at any spacetime point where
$\sigma$ does not vanish. Without loss of generality, impose the 
condition that at all times

\be
\lim_{|{\bf x}|\rightarrow \infty} \sigma({\bf x},t) =  v \ ,
\label{sigbc}
\ee

\be
\lim_{|{\bf x}|\rightarrow \infty} U({\bf x},t) =  
\left(\matrix{1 & 0 \cr 0 & 1}\right) \ .
\label{Ubc}
\ee
In $A_0=0$ gauge,
a vacuum configuration is of the form

\bea
\Phi & = & \frac{v}{\sqrt{2}}\,U \nonumber \\
A_j & = & \frac{1}{ig}\partial_jUU^{\dagger} \ .
\label{vacuum}
\eea
At any time $t$ when $\sigma({\bf x},t) \neq 0$ for all ${\bf x}$ we have that
$U({\bf x},t)$ is a map from ${\bf R}^3$ with the points at infinity
identified, that is $S^3$, into $SU(2)$ and therefore $U({\bf x},t)$
can be associated with an integer-valued winding

\be
N_H(t) = w[U]=\frac{1}{24\pi^2} \int d^3x\, \epsilon^{ijk} \hbox{Tr}
[U^{\dagger}\partial_iUU^{\dagger}\partial_jUU^{\dagger}\partial_kU] \ ,
\label{higgswinding}
\ee
the Higgs winding number. 
If $\Phi({\bf x},t)$ evolves continuously
in $t$ then $N_H(t)$ can change only  at times when
there is a zero of $\sigma$ at some point in space. 
At such times, $N_H$ is not defined; at all other times,
it is integer-valued. 
Note that the Higgs winding number of a vacuum configuration (\ref{vacuum})
is equal to its Chern-Simons number

\begin{equation}
N_{CS}(t) \equiv \frac{g^2}{32 \pi^2}\int d^3x\, \epsilon^{ijk}
{\rm Tr}\left( A_i \partial_j A_k + \frac{2}{3}ig A_i A_j A_k \right)\ .
\label{NCSdef}
\end{equation}
However, for a general non-vacuum configuration the Chern-Simons
number is not integer-valued.

Finally, the fermionic sector of the full $SU(2)\times U(1)$
theory is described by

\be
{\cal L}_f = {\cal L}_l +{\cal L}_q \ ,
\label{fermionsector}
\end{equation}
where the lepton and quark sector Lagrangians for a single family are:

\begin{equation}
{\cal L}_l = - i {\bar \Psi} \gamma^\mu D_\mu \Psi 
      - i {\bar e}_R \gamma^\mu D_\mu e_R
      + h({\bar e}_R \phi^{\dag} \Psi + {\bar \Psi} \phi e_R )
\label{leptonsector}
\end{equation}

\begin{eqnarray}
{\cal L}_q =& -i ({\bar u} , {\bar d})_L \gamma^\mu D_\mu 
         \pmatrix{u\cr d\cr}_L 
      -i {\bar u}_R \gamma^\mu D_\mu u_R
      -i {\bar d}_R \gamma^\mu D_\mu d_R 
 \nonumber \\
&
-G_d \biggl [ ({\bar u}, {\bar d})_L \pmatrix{\phi^+\cr \phi\cr} d_R
+{\bar d}_R (\phi^{-} , {\phi^*}) \pmatrix{u\cr d\cr}_L \biggr ]
 \nonumber \\
&
 - G_u \biggl [ ({\bar u}, {\bar d})_L \pmatrix{-{\phi^*}\cr
 \phi^{-}\cr} u_R + {\bar u}_R (-\phi, \phi^+) \pmatrix{u\cr d\cr}_L 
       \biggr ]  \ ,
\label{quarksector}
\end{eqnarray}
where I have written $\phi=(\phi^+,\phi^0)$, with $\phi^- = ( \phi^+ )^*$. 
Here, $e$ is the electron, $u$ and $d$ are 
the up and down quarks respectively and $\Psi$ represents the left-handed 
lepton doublet.
The indices $L$ and $R$ refer to left- and right-handed components, and $G_d$
and $G_u$ are Yukawa coupling constants.

When extended to three families, this contains the six types of quarks 
$U^i=(u,c,t)$, $D^i=(d,s,b)$, the electron, muon and tau lepton, and their 
associated neutrinos. Note that gauge invariant fermion mass terms are 
generated through couplings to the Higgs doublet $\Phi$.

\begin{figure}
\centerline{\epsfbox{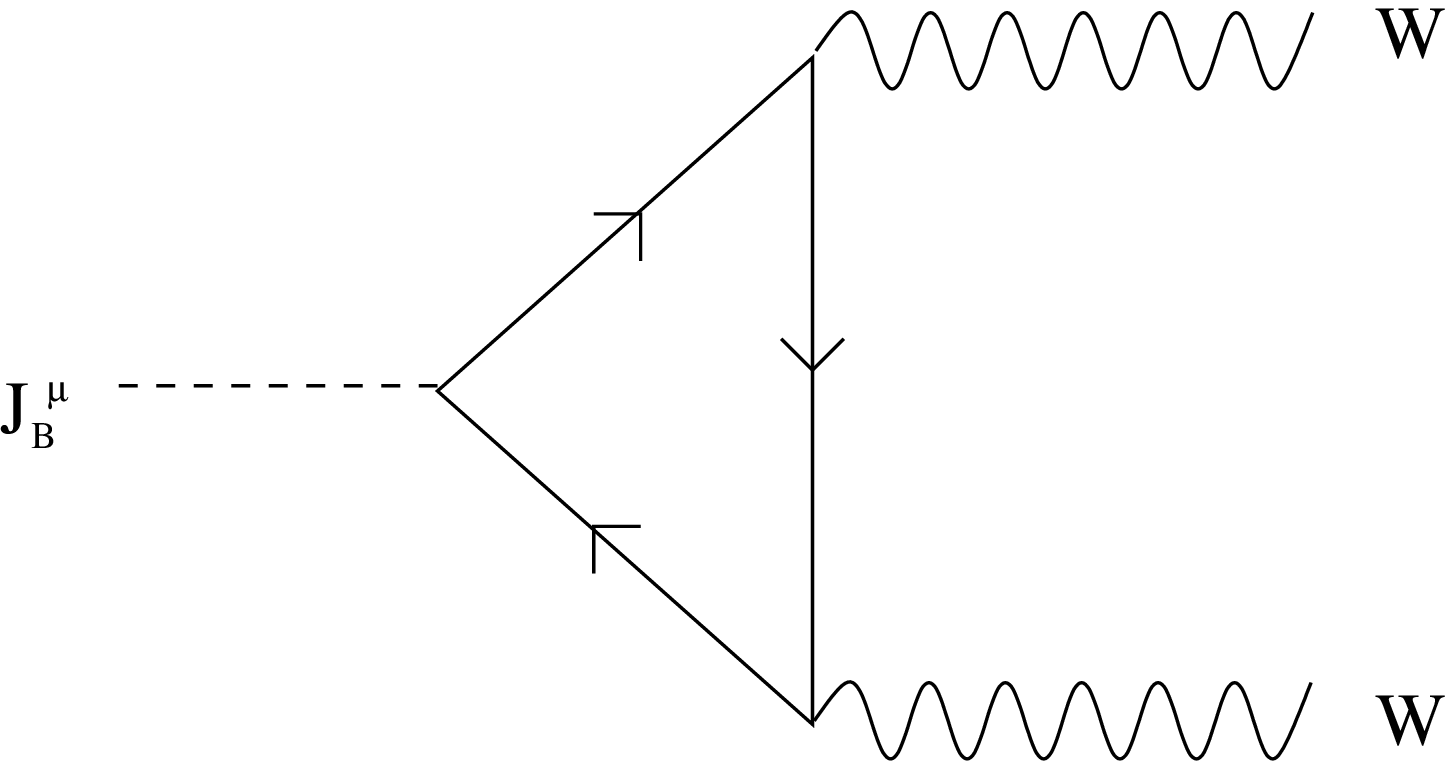}}
  \caption{}
\label{triangle}
\end{figure}

At the classical level, the number of each fermionic species is 
separately conserved in all processes. This is reflected in the fact
that for each species there exists a global $U(1)$  
current, which is exactly classically conserved. In the standard 
model there are a total of 12 different species and so there are 12 separate
conserved global currents. In particular, for
baryons, one may write a vectorlike current

\be 
j_B^{\mu} = \frac{1}{2} {\bar Q}\gamma^{\mu}Q \ ,
\label{baryoncurrent}
\ee
where $Q$ represents quarks, and there is an implied sum over the color and 
flavor indices. Now, due to quantum effects, any axial current 
${\bar \psi}\gamma^{\mu}\gamma^5 \psi$ of a gauge 
coupled Dirac fermion $\psi$, is anomalous \cite{{Adler 69},{B&J 69}}. 
This is relevant to baryon number since the electroweak fermions couple 
chirally to the gauge fields. If we write the baryon current as

\be
j_B^{\mu} = \frac{1}{4}\left[{\bar Q}\gamma^{\mu}(1-\gamma^5)Q
+{\bar Q}\gamma^{\mu}(1+\gamma^5)Q\right] \ ,
\ee
only the axial part of this vector current is important when one calculates 
the divergence. This effect can be 
seen by calculating triangle graphs (see figure \ref{triangle}) and leads
to the following expressions for the divergences of the baryon number and
lepton number currents;

\be
\partial_{\mu}j_B^{\mu} = \partial_{\mu}j_l^{\mu}
=n_f\left(\frac{g^2}{32\pi^2}W_{\mu\nu}^a {\tilde W}^{a\mu\nu}
-\frac{g'^2}{32\pi^2}F_{\mu\nu}{\tilde F}^{\mu\nu}\right) \ ,
\label{anomaly}
\ee
where $n_f$ is the number of families,

\be
\tilde{W}^{\mu\nu} = \frac{1}{2} \epsilon^{\mu\nu\alpha\beta}W_{\alpha\beta}
\ 
\label{Fdual}
\ee
is the dual of the $SU(2)$ field strength tensor, and an analogous
expression holds for ${\tilde F}$

\begin{figure}
\centerline{\epsfbox{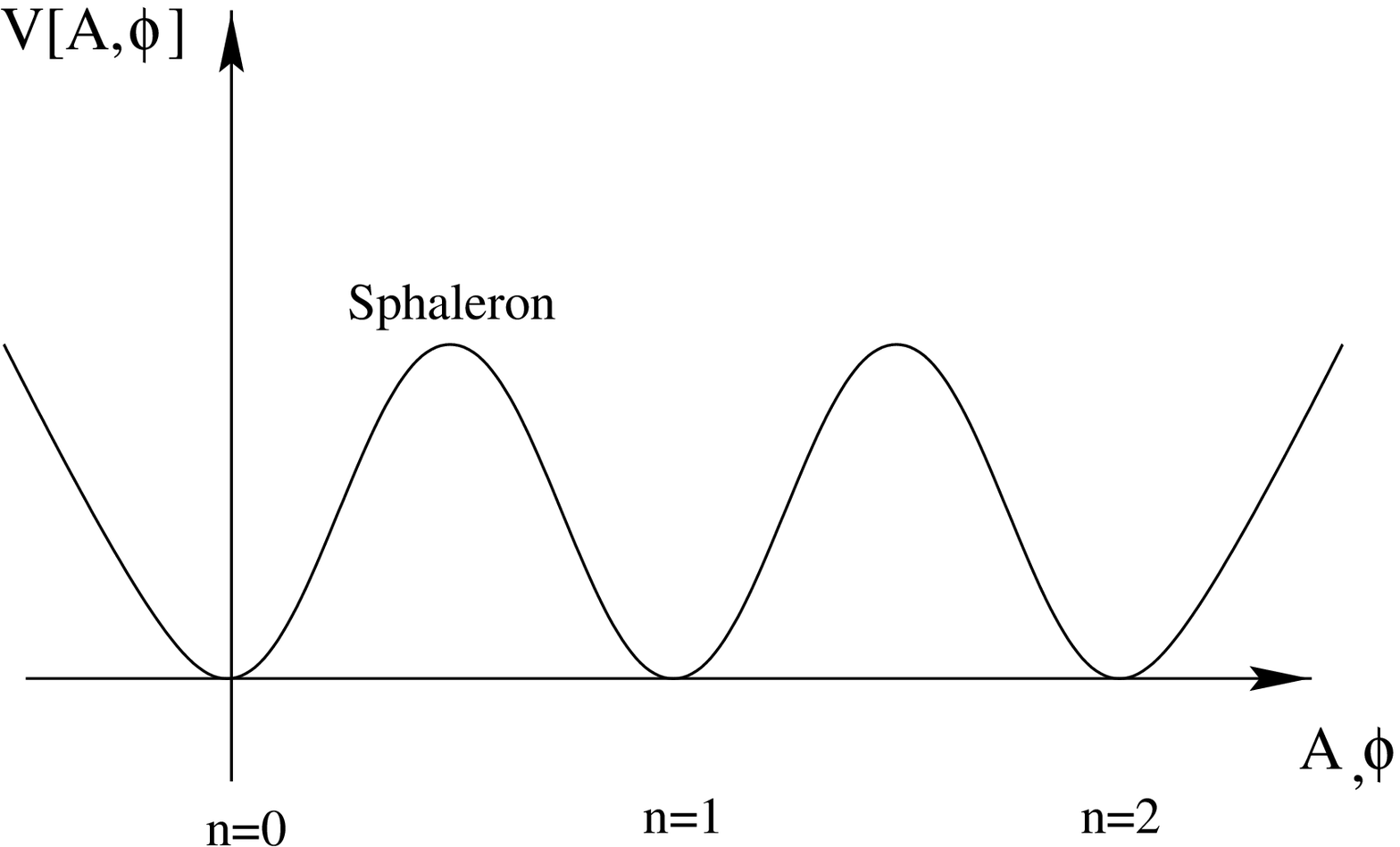}}
  \caption{}
\label{vacua}
\end{figure}

The anomaly is important because of the multiple vacuum structure of the
theory (see figure~\ref{vacua}), as I will describe in the following 
subsections. 
Equation~(\ref{anomaly}) may be written
as

\be
\partial_{\mu}j_B^{\mu} = \partial_{\mu}j_l^{\mu}
=n_f\left(\frac{g^2}{32\pi^2}\partial_{\mu}K^{\mu}
-\frac{g'^2}{32\pi^2}\partial_{\mu}k^{\mu} \right) \ ,
\label{divergences}
\ee
where the {\it gauge non-invariant} currents are defined by

\bea
K^{\mu} & = & \epsilon^{\mu\nu\alpha\beta}
\left(W_{\nu\alpha}^a A_{\beta}^a - \frac{1}{3}g\epsilon_{abc} A_{\nu}^a
A_{\alpha}^b A_{\beta}^c \right) \ , \nonumber \\
k^{\mu} & = & \epsilon^{\mu\nu\alpha\beta}F_{\nu\alpha}B_{\beta} \ .
\eea

For simplicity, consider space to be a 3-sphere and consider the change in
baryon number from time $t=0$ to some arbitrary final time $t=t_f$.
For transitions between vacua, the average values of the field strengths 
are zero at the beginning and the end of the evolution. Since the final
term in (\ref{divergences}) is strictly proportional to the field strength
of the $U(1)$ gauge fields, we may ignore this term. Then, using
(\ref{NCSdef}), the change in baryon number may be written as

\be
\Delta B = \Delta N_{CS} \equiv n_f[N_{CS}(t_f) - N_{CS}(0)]\ .
\label{Bchange}
\ee
Although the Chern-Simons number is not gauge invariant, the change 
$\Delta N_{CS}$ {\it is}. Thus, since $N_{CS}$ is integral
(as we have mentioned), changes in Chern-Simons number result in 
changes in baryon number which are integral multiples of the number of
families $n_f$.

To understand this structure more completely, since the $U(1)$ gauge fields
are unimportant here, I will return to the $SU(2)$ theory.
Consider fermion production in the 
background of the evolving Higgs and gauge fields and ignore
the back-reaction of the fermions on the bosonic fields.  
Consider the dynamics of nonzero energy 
configurations with nonzero Higgs winding. A simple example \cite{ALS 89} is

\bea
\Phi({\bf x}) & = & \frac{v}{\sqrt{2}}U_{[1]}({\bf x}) \nonumber \\
A_\mu({\bf x}) & = & 0 \ ,
\label{winding1}
\eea
where $U_{[1]}({\bf x})$ is a winding number one map, say,

\be
U_{[1]}({\bf x}) = \exp\left(i\eta(r)\mbox{\boldmath $\tau$}
\cdot\hat{\bf x}\right) \ ,
\label{u1}
\ee
with $\eta(0)=-\pi$ and $\eta(\infty)=0$. The 
configuration~(\ref{winding1}) has no
potential energy but does carry gradient energy because the covariant
derivatives $D_i\Phi$ do not vanish.  This configuration
has $N_H=1$ (where $N_H$ is defined in~(\ref{higgswinding})).
If the configuration~(\ref{winding1}) were
released from rest it would radiate away its energy and relax 
towards a vacuum 
configuration. There are two very different ways for this to 
occur \cite{TZi}. If the characteristic size of $U_{[1]}$ is large 
compared to $m_W^{-1}$,
then the gauge field will evolve
until it lines up with the Higgs field making the covariant derivatives zero,
and at late times $N_H$ will still be one.
If the characteristic size is small the configuration 
will shrink,
the Higgs field $\sigma$ will go through a zero, and at late times $N_H$
will be zero. This dynamics is the subject of section \ref{local}.

Note that $N_H$ is not invariant under large gauge transformations. However,
the change $\Delta N_H$ in Higgs winding is gauge invariant, and the two
distinct relaxation processes are distinguished by whether $\Delta N_H$ is
zero or nonzero. To be definite, I will always choose
the gauge such that the
prototypical initial configuration is of the form~(\ref{winding1}) which has
$N_H=1$.

Now, if the fields relax to the vacuum 
by changing the Higgs 
winding then there is no anomalous fermion number production. However, if 
there is no net change in 
Higgs winding during the evolution (for example $\sigma$ never
vanishes) then there is anomalous fermion number production.

To understand these claims consider two sequences of configurations beginning 
with the wound up 
configuration~(\ref{winding1}) and ending at the classical 
vacuum~(\ref{vacuum}). The first sequence ends at the vacuum~(\ref{vacuum})
with $U={\bf 1}$ while the second ends up at $U=U_{[1]}$.
Note that these sequences cannot be
solutions to the classical equations of motion since the initial
configurations carry energy whereas the final ones do not.
Throughout both
sequences the boundary conditions~(\ref{sigbc}) and~(\ref{Ubc}) are 
maintained. For the first sequence, $\sigma$ must vanish at some intermediate 
configuration since the Higgs winding changes. For the second sequence,
the change in Higgs winding is zero and $\sigma$ need not vanish.

Now introduce an $SU(2)_L$ weak fermionic doublet, $\psi$. The fermion is 
given mass through the usual gauge invariant coupling to the Higgs field 
$\Phi$, and for simplicity assume that both the up and down components of 
the doublet have the same mass $m$. The fermion field is quantized in the 
background of the bosonic fields given by the above interpolation. 

The anomaly equation~(\ref{anomaly}) reduces here to

\be
\partial_{\mu}J^{\mu} = \frac{g^2}{32\pi^2}\hbox{Tr}(W{\tilde W}) \ ,
\label{newanomaly}
\ee
which, when integrated, implies that the change in the fermion number from the
beginning to the end of a sequence is given by

\be
\left.\int d^3x\, J^0\right|_{\rm final} - 
\left.\int d^3x\, J^0\right|_{\rm initial} = -w[U] \ ,
\label{intanomaly}
\ee
where $U$ is that of the final configuration~(\ref{vacuum}). For the
first sequence $w$ is one, whereas for the second it is zero. 
Thus fermion number is violated in processes for which the 
configuration~(\ref{winding1}) 
unwinds via gauge unwinding, but is not violated when 
such a configuration unwinds via a Higgs unwinding.

\subsection{Zero temperature results}
We have seen that the vacuum of the electroweak theory is 
degenerate, labeled by $N_H$ and $N_{CS}$. The field theories constructed 
around these vacua are
entirely equivalent and may be obtained from one another through
large gauge transformations. However, transitions between these vacua
result in the anomalous production of fermions via the anomaly
equation~(\ref{anomaly}). Such transitions violate
baryon number and so are classically forbidden since baryon number 
is an exact global symmetry of the theory. In fact, as I shall describe, 
at zero
temperature, baryon number violating processes occur through
quantum tunneling between the classical vacua of the theory.

In the infinite dimensional gauge and Higgs field configuration space, 
adjacent vacua of the electroweak theory are separated by a ridge of
configurations with energies larger than that of the vacuum 
(see figure~\ref{vacua})
The lowest energy point on this ridge is a saddle point solution
\cite{{DHN 74},{NC 80}}
to the equations of motion with a single negative eigenvalue, and
is referred to as the {\it sphaleron} \cite{Manton83,KM84}.
The sphaleron plays a central role in the calculation of the
rate of baryon number violating processes.

The calculation of the tunneling rate between degenerate vacua in 
quantum field theories is well-established (for a review see 
\citeasnoun{colemanreview}). One first constructs the
Euclideanized action $S_E$, obtained from the Minkowski space action by 
performing a Wick rotation 

\be
t \rightarrow -it \equiv \tau \ .
\label{Wick}
\ee
In the case of
pure gauge theories, one then finds the solution to the Euclidean equations
of motion which interpolates between the two vacuum states and 
minimizes the Euclidean action. Such a solution to the full 
four-dimensional Euclidean system is known as an
{\it instanton} and may be seen as a time series of three-dimensional
configurations. The transition rate between the degenerate vacua 
is then proportional to

\be
\exp(-S_E)|_{instanton} \ .
\ee

When the theory in question has a Higgs field, as in the electroweak theory, 
the calculation of the tunneling amplitude is a little more complicated. In
this case, there are no nonzero size configurations that minimize the 
Euclidean action, and a slightly different approach is used. The procedure
is to fix the size of the Euclidean configurations in the functional
integral by explicitly introducing a constraint
\cite{IAi 81}. The functional integral
is then evaluated using these {\it constrained instantons} and, finally,
one performs an integral over the instanton size.

As an estimate of the zero temperature $B$ violating rate in the electroweak 
theory, I shall follow the approach of \citeasnoun{tHooft76}. Consider the
pure gauge $SU(2)$ theory (note no Higgs field). The relevant
configurations are finite action solutions to the Euclidean 
equations of motion. Further, as I described earlier, the
configurations of interest must possess non-zero gauge winding in order
for baryon number violation to take place. If we consider the case
in which the instanton interpolates between adjacent vacua of the
theory, then the topological charge of such a solution is 
$\Delta N_{CS} =1$.
Now, the quantity

\be
\int d^4x \, (W_{\mu\nu}^a - {\tilde W}_{\mu\nu}^a)^2
\label{positive1}
\ee
is positive semi-definite for any gauge field configuration. This
allows us to construct a bound on the Euclidean action of configurations in
the following way. Expanding (\ref{positive1}) gives

\be
\int d^4x \, \left[{\rm Tr}(W_{\mu\nu}W^{\mu\nu}) + 
{\rm Tr}({\tilde W}_{\mu\nu}{\tilde W}^{\mu\nu}) -
2{\rm Tr}(W_{\mu\nu}{\tilde W}^{\mu\nu})\right] \geq 0 \ .
\ee
Now, in terms of the Euclidean action and the Chern-Simons number, this may
then be written as

\be
4S_E -2\left(\frac{16\pi^2}{g^2}\right) N_{CS} \geq 0 \ ,
\ee
which finally yields

\be
S_E \geq \frac{8\pi^2}{g^2}N_{CS} \ .
\ee
The $N_{CS}=1$ instanton saturates this bound,
which means that the rate per unit volume of baryon number violating
processes at zero temperature is approximately

\be
\Gamma(T=0) \sim \exp(-2S_E) \sim 10^{-170} \ .
\ee
This is so small that if the universe were always close to zero 
temperature, not one event would have occurred within the present
Hubble volume ever in the history of the universe.

\subsection{Nonzero temperature results}
We have seen that the rate of baryon number violating processes is
negligible at zero temperature. This is to be expected since such
events occur through quantum tunneling under a high potential barrier 
($\sim 10$TeV). If this were the case
at all temperatures, it is safe to say that electroweak baryon number
violation would have no observable consequences. In this section we
shall see that, when one includes the effects of nonzero temperature, 
{\it classical} transitions between electroweak vacua become possible
due to thermal activation.

\subsubsection{A Mechanical Analogy}
Let us begin with a warmup example. Consider an ideal pendulum of mass
$m$ and length $l$, confined to rotate in the plane. the Lagrangian is

\be
L=\frac{1}{2}ml^2{\dot \theta}^2 -mgl(1-\cos\theta) \ .
\ee
the system possesses a periodic vacuum structure labeled by integer $n$

\be
\theta_n = 2n\pi \ .
\ee
The Schr\"odinger equation describing the quantum mechanics of this system is

\be
\frac{\hbar \omega}{2}\left(-\alpha\frac{d^2}{d\chi^2} +\frac{1}{\alpha}
\sin^2\chi\right)\psi_n(\chi) = E_n\psi_n(\chi) \ ,
\ee
where $\chi\equiv \theta/2$, $\omega\equiv g/l$ and
$\alpha\equiv\hbar\omega/(4mgl) \ll1$.
Since the potential is periodic in the angle $\theta$, the wave-functions
will be periodic and therefore the multiple vacuum structure of the
system is manifest. However, if one performs perturbation theory, Taylor
expanding the potential about a chosen minimum ($\theta=0$) and keeping only 
the first term, then the Schr\"odinger equation becomes that for a simple 
harmonic oscillator and all information about the periodic vacua is lost.
This situation is analogous to most 
familiar calculations in the electroweak theory, in which perturbation
theory is usually a safe tool to use. 
Such an approximation scheme is only valid when the energy of the pendulum is
much less than the height of the barrier preventing transitions between
vacua (c.f. the sphaleron). In that limit, quantum tunneling between vacua
is exponentially suppressed as expected. This may be seen by employing
a different approximation scheme which preserves periodicity -- the
semiclassical (WKB) approximation. 

Now consider raising the temperature of the system. The pendulum is coupled
to a thermal bath and is thermally excited to states of higher and higher
energy as the temperature is raised. As the temperature becomes comparable with
the barrier height, it becomes possible for thermal transitions over the
barrier to occur. Writing the energy of the barrier between vacua as
$E_{bar}$, the rate of these transitions can be shown to be

\be
\Gamma_{pendulum}(T) \propto \exp\left(-\frac{E_{bar}}{T}\right) \ ,
\ee
so that at $T\sim E_{bar}$, the pendulum makes transitions between vacua, 
crossing the point $\theta=\pi$ randomly, at an
unsuppressed rate. The important point here is that the rate of thermally
activated transitions between vacua is governed by the barrier height, or
more accurately, by the maximum of the free energy of the configurations
which interpolate between the vacua.

The general features of this simple mechanical system are very important for
a nonabelian gauge theory such as the electroweak theory.

\subsubsection{The Electroweak Theory}
The calculation of the thermally activated transition
rate for the infinite dimensional field theory is, of course, much more
complicated than that for the one dimensional example above. The field
theory approach to problems of metastability was first
outlined by \citeasnoun{IAii 81} (see also \citeasnoun{AL 81}), 
using techniques developed by 
\citeasnoun{Langer} for use in condensed matter problems. The general 
framework for evaluating the thermal rate of anomalous processes in the
electroweak theory is due to \citeasnoun{KRS}.

For any sequence of configurations interpolating between vacua, there is
one configuration which maximizes the free energy. Over all such
sequences, there is one maximum energy configuration with the 
smallest free energy (i.e. a saddle point) and it is this configuration, in 
analogy with the 
pendulum example, which governs the rate of anomalous transitions due to 
thermal activation.

In the electroweak theory the relevant saddle point configuration is the 
sphaleron. In the full Glashow-Salam-Weinberg theory, the sphaleron solution 
cannot be obtained analytically. In fact, the
original calculation of \citeasnoun{KM84} was performed
in the $SU(2)+$ Higgs theory, and identified a sphaleron with 
energy in the range

\be
8{\rm TeV} <E_{sph} <14 {\rm TeV} \ , 
\ee
depending on the Higgs
mass. The corresponding object in the full electroweak theory can
then be obtained from the $SU(2)$ sphaleron by a perturbative
analysis in which the small parameter is the weak mixing angle 
$\theta_{W}$. 

To calculate the thermal baryon number violation rate, the following steps
are performed.
One first computes the rate to cross
the barrier over the sphaleron beginning from a given state. This process
is essentially a one degree of freedom process since all field directions 
perpendicular to the negative mode corresponding to the sphaleron are ignored.
The second step is to sum over all such paths, weighting each by the
appropriate Boltzmann factor.  The calculation
for the electroweak theory, properly taking into account the translational and 
rotational zero modes of the sphaleron, was first carried out by
\citeasnoun{AM}. 
A final step is to take account
of the infinity of directions transverse to the sphaleron by performing
a calculation of
the small fluctuation determinant around the sphaleron. This final step
is carried out within the Gaussian approximation, and was originally performed 
by \citeasnoun{C&M 90} (see also \citeasnoun{CLMW 90}), with more recent 
analyses by \citeasnoun{B&S 93}, \citeasnoun{B&S2 94} and 
\citeasnoun{B&S1 94}. If $M_W(T)$
is the thermal $W$ boson mass calculated from the finite temperature 
effective potential (see next section), the approximations which go into this 
calculation are valid only in the regime 

\be
M_W(T) \ll T \ll \frac{M_W(T)}{\alpha_W} \ .
\ee
The rate per unit volume of baryon 
number violating events in this range is calculated to be

\be
\Gamma(T) = \mu\left(\frac{M_W}{\alpha_W T}\right)^3M_W^4 
\exp\left(-\frac{E_{sph}(T)}{T}\right) \ ,
\label{brokenrate}
\ee
where $\mu$ is a dimensionless constant. Here, the temperature-dependent
``sphaleron'' energy is defined through the finite temperature effective
potential by

\be
E_{sph}(T) \equiv \frac{M_W(T)}{\alpha_W} {\cal E} \ ,
\ee
with the dimensionless parameter ${\cal E}$ lying in the range

\be
3.1 < {\cal E} < 5.4 \ ,
\ee
depending on the Higgs mass.

Recent approaches to
calculating the rate of baryon number violating events in the broken
phase have been primarily numerical. Several results using real time
techniques \cite{{ambjorniii},{M&Tii 97}} were found to contain lattice
artifacts arising from an inappropriate definition of $N_{CS}$. These
techniques were later shown to be too insensitive to the true rate when
improved definitions of $N_{CS}$ were used \cite{{M&Ti 97},{A&K 97}}. 
The best 
calculation to date of the broken phase sphaleron rate is undoubtably
that by \citeasnoun{GM 98}. This work yields a fully nonperturbative 
evaluation of the broken phase rate by using a combination of multicanonical
weighting and real time techniques.

Although the Boltzmann suppression in~(\ref{brokenrate}) appears large,
it is to 
be expected that, when the electroweak symmetry becomes restored 
\cite{{K 72},{KL 72}} at a 
temperature of around $100\,$GeV, there will no longer be an exponential 
suppression factor. Although calculation of the baryon number violating 
rate in the 
high temperature phase is extremely difficult, a simple estimate is 
possible. In Yang-Mills Higgs theory at nonzero temperature the infrared
modes (those with wavenumber $k \ll T$)
are well described classically in the weak coupling limit, while the
ultraviolet modes ($k \sim T$) are not. 
Now, the only important scale in the symmetric phase is the 
magnetic screening length given by 

\be
\xi=(\alpha_W T)^{-1} \ .
\ee
Assuming that any time scale must also behave in this way, then on 
dimensional grounds, we 
expect the rate per unit volume of baryon number violating events, an
infrared spacetime rate, to be

\be
\Gamma(T)=\kappa(\alpha_W T)^4 \ ,
\label{unbrokenrate}
\ee
with $\kappa$ another dimensionless constant, assuming that the infrared and
ultraviolet modes decouple from each other. 

The rate of baryon 
number violating processes can be related to the sphaleron rate, which is
the diffusion constant for Chern-Simons 
number~(\ref{NCSdef}) and is defined by

\be
\lim_{V\rightarrow\infty} \lim_{t\rightarrow\infty}
\left[\frac{\langle(N_{CS}(t) - N_{CS}(0))^2\rangle}{Vt}\right] \ ,
\ee
(c.f. (\ref{Bchange})) by a fluctuation-dissipation theorem
\cite{KS 88} (for a good description of this see \citeasnoun{review}).
In
almost all numerical calculations of the baryon number violation
rate, this relationship 
is used and what is actually evaluated is the diffusion constant.
The first attempts to numerically 
estimate $\kappa$ in this way \cite{{AAPS 90},{AAPS 91}} yielded 
$\kappa \sim 0.1 - 1$, but the approach suffered from limited statistics and
large volume systematic errors. Nevertheless, more recent numerical attempts 
\cite{{ambjorni},{GM 96},{ambjorniii},{M&Tii 97}} found 
approximately the same result. However, as I mentioned above, these
approaches employ a poor definition of the Chern-Simons number which
compromises their reliability. 

The simple scaling argument leading to (\ref{unbrokenrate}) has been 
recently criticized by \citeasnoun{ASYi} who 
argue that damping effects in the plasma suppress the rate by an extra 
power of $\alpha_W$. The essential reason for the modification is that
the decoupling between infrared and ultraviolet modes does not hold
completely when dynamics are taken into account.
Since the transition rate involves physics at soft energies $g^2T$ that are 
small compared to the typical hard energies $\sim T$ of  the thermal 
excitations in the plasma, the simplest way of analyzing the problem is to 
consider an effective theory for the soft modes. Thus, one integrates out 
the hard  modes 
and keeps the dominant contributions,  the 
so-called hard thermal loops. It is the resulting typical frequency 
$\omega_c$ of a gauge field configuration with spatial extent  $(g^2 T)^{-1}$
immersed in the plasma that determines the change of baryon 
number per unit time and unit volume. This frequency has been 
estimated to be 

\be
\omega_c\sim g^4T \ ,
\ee
when the damping effects  of the 
hard modes \cite{ASYi,PA 97} are taken into account.  

In recent months the dust has settled around these issues. The analysis
of \citeasnoun{M&Ti 97} demonstrated that, when a reliable definition of 
$N_{CS}$ is used, there is a lattice spacing dependence in the symmetric
phase rate that is consistent with the claims of \citeasnoun{ASYi} and
similar
results were also obtained by \citeasnoun{A&K 97}.
\citeasnoun{ASYii} have constructed the nonlocal infra-red effective
theory which includes the effects of hard thermal loops that they expect to
be responsible for the extra power of $\alpha_W$ in the rate of
baryon number violation. Further, a rigorous field-theoretic derivation
of this theory has been derived \cite{DS 97}. Using ideas developed
by \citeasnoun{H&M 97}, the effects of hard thermal 
loops have also been considered in work by \citeasnoun{MHM 97} (see also
\citeasnoun{EI 97}). In that 
work the authors find

\be
\Gamma(T\gg T_c) = \kappa' \alpha_W (\alpha_W T)^4 \ ,
\label{newunbroken}
\ee 
with

\be
\kappa' = 29\pm 6
\ee
for the minimal standard model value of the Debye mass. Note that, although 
this estimate takes into account physics that did not enter the original 
estimate, this expression is numerically close to (\ref{unbrokenrate}).

Finally, the effective dynamics of soft nonabelian 
gauge fields at finite temperature has been recently addressed 
by \citeasnoun{DB 98}, who finds a further logarithmic correction

\be
\Gamma_{sp}\sim \alpha_W^5 T^4\:{\rm ln}(1/\alpha_W) \ . 
\ee
The physics leading to these corrections has been discussed at length 
by Moore \cite{Moorenew}, who describes in detail both the intuitive
arguments for such a term and the lattice Langevin calculations required to 
provide an accurate numerical determination of its magnitude. 

Now that I have discussed baryon number violation, I shall turn to the
second Sakharov criterion and its realizations in the standard model.

\section{C and CP Violation}
\label{CP}
As I mentioned earlier, it is necessary that both the C and CP symmetries
be violated for baryogenesis scenarios to succeed. One cause of the initial
excitement over electroweak baryogenesis was the observation that the
Glashow-Salam-Weinberg model naturally satisfies these requirements.

Recall that fermions in the theory are chirally coupled to the gauge fields 
as in equation~(\ref{fermionsector}). This means that, for example, only the 
left-handed 
electron is $SU(2)$ gauge coupled. In terms of the discrete symmetries of 
the theory,
these chiral couplings result in the electroweak theory being maximally
C-violating. This is a general feature that remains true in  
extensions of the theory and makes the model ideal for baryogenesis.
However, the issue of CP-violation is more complex.

\subsection{Standard Model CP Violation: the KM Matrix}
CP is known not to be an exact symmetry
of the weak interactions. This is seen experimentally in the neutral 
Kaon system through $K_0$, ${\bar K}_0$ mixing \cite{CCFT 64}. At
present there is no accepted theoretical explanation of this.
However, it is true that CP violation is a natural feature of the
standard electroweak model. When the expression~(\ref{fermionsector})
is expanded to include $N$ generations of quarks and leptons, there exists a
charged current which, in the weak interaction basis, may be written
as

\be
{\cal L}_W = \frac{g}{\sqrt{2}} {\bar U}_L \gamma^{\mu} D_L W_{\mu}
+ (h.c.) \ ,
\label{chargedcurrent}
\ee
where $U_L=(u,c,t,\ldots)_L$ and $D_L=(d,s,b,\ldots)_L$. Now, the
quark mass matrices may be diagonalized by unitary matrices
$V^U_L$, $V^U_R$, $V^D_L$, $V^D_R$ via

\bea
{\rm diag}(m_u,m_c,m_t,\ldots) & = & V^U_L M^U V^U_R \ ,  \\
{\rm diag}(m_d,m_s,m_b,\ldots) & = & V^D_L M^D V^D_R \ .
\eea
Thus, in the basis of quark mass eigenstates, (\ref{chargedcurrent})
may be rewritten as

\be
{\cal L}_W = \frac{g}{\sqrt{2}}{\bar U}_L'K\gamma^{\mu}D_L' W_{\mu}
+(h.c.) \ ,
\ee
where $U_L'\equiv V^U_L U_L$ and $D_L'\equiv V^D_L D_L$. The matrix
$K$, defined by 

\be
K \equiv V_L^U (V_L^D)^{\dagger} \ ,
\label{KM}
\ee
is referred to as the Kobayashi-Maskawa (KM) quark mass mixing
matrix \cite{K&M 73}. For $N$ generations, $K$ contains $(N-1)(N-2)/2$ 
independent phases, and a nonzero value for any of these phases signals CP
violation.
Therefore,  CP violation exists for $N\geq 3$ and when $N=3$, as in the
standard model, there is precisely one such phase $\delta$.
While this is
encouraging for baryogenesis, it turns out that this particular source of
CP violation is not strong enough. The relevant effects are parameterized by
a dimensionless constant which is no larger than $10^{-20}$. This appears
to be much too small to account for the observed BAU and, thus far, 
attempts to utilize this source of CP violation for electroweak
baryogenesis have been unsuccessful.

When one includes the strong interactions
described by quantum chromodynamics (QCD), a second potential source of 
CP violation arises due to instanton effects. However, precision 
measurements of the dipole
moment of the neutron constrain the associated dimensionless parameter
$\theta_{QCD}$
to be less than $10^{-9}$, which again is too small for baryogenesis. 
In light 
of these facts, it is usual to extend the standard model in some minimal 
fashion that increases the amount of CP violation in the theory while not
leading to results that conflict with current experimental data.

\subsection{The Two-Higgs Doublet Model}
One particular way of achieving this \cite{LM 89} is to expand the Higgs 
sector of the
theory to include a second Higgs doublet. In a two-Higgs model, the 
structure described in section II is doubled so that we have scalars 
$\Phi_1$ and $\Phi_2$, and the scalar potential is 
replaced by the most general renormalizable two-Higgs potential \cite{HHG}

\begin{eqnarray}
V(\Phi_1,\Phi_2) & = & \lambda_1(\Phi_1^{\dagger}\Phi_1 - 
v_1^2)^2 +
                   \lambda_2(\Phi_2^{\dagger}\Phi_2 - 
v_2^2)^2  \nonumber \\
                 & + & \lambda_3[(\Phi_1^{\dagger}\Phi_1 - 
v_1^2) + (\Phi_2^{\dagger}\Phi_2 - 
v_2^2)]^2 \nonumber \\
                 & + & \lambda_4[(\Phi_1^{\dagger}\Phi_1)(\Phi_2^{
\dagger}\Phi_2) - (\Phi_1^{\dagger}\Phi_2)(\Phi_2^{
\dagger}\Phi_1)]  \nonumber \\
                 & + & \lambda_5[{\rm Re}(\Phi_1^{\dagger}\Phi_2) -
v_1 v_2\cos\xi]^2 \nonumber \\
                 & + & \lambda_6[{\rm Im}(\Phi_1^{\dagger}\Phi_2)
-v_1 v_2\sin\xi]^2 \ .
\end{eqnarray}
Here $v_1$ and $v_2$ are the respective vacuum expectation values
of the two doublets, the $\lambda_i$ are coupling constants, and $\xi$ is a
phase. To make the CP-violation explicit, let us write the Higgs fields in
unitary gauge as

\begin{equation}
\Phi_1=(0, \vp_1)^T \ \ \ \ , \ \ \ \ \Phi_2=(0, \vp_2 e^{i\theta})^T \ ,
\end{equation}
where $\vp_1$, $\vp_2$, $\theta$ are real, and $\theta$ is the CP-odd phase.
The important terms for CP violation in the potential are the final two, 
with coefficients $\lambda_5$ and $\lambda_6$, since it is these terms which
determine the dynamics of the CP-odd field $\theta$.

There are several mechanisms by which $\theta$ may contribute
to the free energy density of the theory. To be specific, let us concentrate
on the one-loop contribution \cite{TZi} and for simplicity assume that 
$\theta$ is spatially homogeneous. The relevant term is

\be
{\cal F}_B =-\frac{14}{3\pi^2 n_f}\zeta(3)\left(\frac{m}{T}\right)^2
{\dot \theta} n_B \ ,
\label{1loop}
\ee
where $m$ is the finite temperature mass of the particle species dominating
the contribution to the anomaly and $\zeta$ is the Riemann function.

The coefficient of $n_B$ in the above equation can be viewed as a sort of
chemical potential, $\mu_B$, for baryon number

\be
\mu_B =\frac{14}{3\pi^2 n_f}\zeta(3)\left(\frac{m}{T}\right)^2
{\dot \theta} \ .
\ee

The changes in $\theta$ are
dependent on changes in the magnitude of the Higgs fields. In particular, if
a point in space makes a transition from false electroweak vacuum to true
then $\Delta \theta >0$, and sphaleron processes result in the preferential
production of baryons over
antibaryons. For the opposite situation $\Delta \theta <0$, and sphaleron
processes generate an excess of antibaryons. 
The total change in the phase $\theta$ 
(from before the phase transition to $T=0$) may be estimated to be

\be
\Delta \theta \sim \xi-\arctan\left(\frac{\lambda_6}{\lambda_5} \tan\xi
\right) \ ,
\ee
and, as we will see, it is this quantity that enters into estimates of 
the BAU.
This then is how the dynamics of the two-Higgs 
model bias the production of baryons through sphaleron processes in 
electroweak baryogenesis models.
The use of the two-Higgs model is motivated in part by supersymmetry (SUSY), 
which demands at least two Higgs scalars. However, in the minimal 
supersymmetric standard model (MSSM), supersymmetry demands that
$\lambda_5 \equiv \lambda_6 = 0$ and
so the two-Higgs potential is CP invariant. In such
models CP violation arises through soft-SUSY breaking which generates
nonzero values for $\lambda_5$ and $\lambda_6$, as we shall see later.

\subsection{CP Violation from Higher Dimension Operators}
The second popular method of extending the standard model is to view the
model as an effective field theory, valid at energies below some mass
scale $M$. In addition to the standard terms in the Lagrangian, one then 
expects extra, nonrenormalizable operators, some of which will be CP odd
\cite{{misha 88},{DHSSi},{DHSSii},{zhang2},{LRT 97}}.
A particular dimension six example is 

\be
{\cal O}=\frac{b}{M^2}\,\hbox{Tr}(\Phi^{\dagger}\Phi)
\hbox{Tr}(F_{\mu\nu} \tilde{F}^{\mu\nu}) \ ,
\label{newop}
\ee
with $b$ a dimensionless constant.
${\cal O}$ is the lowest dimension CP odd operator 
which can be constructed from
minimal standard model Higgs and gauge fields.
Such a term, with $M=v$, can be induced in the effective action by 
CP violation in the CKM matrix, but in that case the coefficient
$b$ is thought to be tiny.

The operator ${\cal O}$ induces electric dipole moments
for the electron and the neutron \cite{zhang}, and the strongest 
experimental constraint on the size of such an operator comes from the
fact that such dipole moments have not been observed.
Working to lowest order (one-loop) \citeasnoun{LRT 97} find

\begin{equation}
\frac{d_e}{e} = \frac{m_e \sin^2(\theta_W)}{8\pi^2}\frac{b}{M^2}
\ln\left(\frac{M^2 + m_H^2}{m_H^2}\right)\ .
\label{dipole}
\end{equation}
Note that $M^2$ arises in the logarithm without
$b$ because $M$, the scale above which the effective theory
is not valid, is the ultraviolet cutoff for the divergent loop integral.
Using the experimental limit \cite{commins} 

\be
\frac{d_e}{e} < 4 \cdot 10^{-27} {\rm cm}
\ee
 yields the bound

\begin{equation}
\frac{b}{M^2}\ln\left(\frac{M^2 + m_H^2}{m_H^2}\right) <
\frac{1}{(3 {\rm ~TeV})^2}\ .
\label{bound}
\end{equation}
The corresponding experimental limit \cite{{ramseyii},{ramseyi}} on
the neutron electric dipole moment $d_n$ is weaker than that
on $d_e$, but because $d_n$ is proportional to the quark mass rather
than to the electron mass, the constraint obtained using
$d_n$ is comparable to (\ref{bound}).
Any baryogenesis scenario which relies on CP violation introduced
via the operator ${\cal O}$ must respect the bound (\ref{bound}).

\subsection{General Treatment}
Common to both the above extensions of the standard model is the appearance 
of extra, CP violating
interactions, parameterized by a new quantity $\delta_{CP}$ 
(e.g. $\delta=\Delta\theta$ or
$\delta=b/M^2$). Although there exist a number of other ways in which
CP violation may appear in low energy electroweak models (see, for example
\cite{{JM1},{JM2}}), in many parts of this review, for definiteness, 
I shall 
focus on $\Delta\theta$.
However, in the discussion of specific SUSY models I shall
explain how CP violating quantities arise. 
Whatever its origin, the effect of CP violation on anomalous baryon number 
violating processes is to provide a fixed direction for the net change in 
baryon number. When the bias is small, the equation governing this can be 
derived from detailed balance arguments \cite{{KS 88},{DLSFP 90}} and 
may be written as

\be
\frac{dn_B}{dt} = -3\frac{\Gamma(T)}{T} \Delta F \ ,
\label{balance}
\ee
where $\Delta F$ is the free energy difference, induced by the CP
violation, between producing baryons and antibaryons in a given process, and
$\Gamma$ is the rate per unit volume of baryon number violating events. 
I shall use this equation further when considering nonlocal baryogenesis.

\section{The Electroweak Phase Transition}
\label{EWPT}
To begin with I shall lay down some definitions from thermodynamics. If a
thermodynamic quantity changes discontinuously (for example as a function
of temperature) then we say that a {\it first
order phase transition} has occurred. This happens because, at the point at 
which the transition occurs, there exist two separate thermodynamic states
that are in equilibrium. Any thermodynamic quantity that undergoes such a 
discontinuous change at the phase transition is referred to as an 
{\it order parameter}, denoted by $\vp$. 
Whether or not a first order phase transition occurs
often depends on other parameters that enter the theory. It is
possible that, as another parameter is varied, the change in the order
parameter at the phase transition decreases until it, and all other
thermodynamic quantities are continuous at the transition point. In this
case we refer to a {\it second order phase transition} at the point at which
the transition becomes continuous, and a {\it continuous crossover} at
the other points for which all physical quantities undergo no changes. 
In general, we are interested in systems for which the high temperature
ground state of the theory is achieved for $\vp=0$ and the low temperature
ground state is achieved for $\vp\neq 0$.

The question of the order of the electroweak phase transition is central to
electroweak baryogenesis. Phase transitions are the most important phenomena 
in particle
cosmology, since without them, the history of the universe is simply one of
gradual cooling. In the absence of phase transitions, the only departure
from thermal equilibrium is provided by the expansion of the universe. At
temperatures around the electroweak scale, the expansion rate of the universe
in thermal units is small compared to the rate of baryon number violating 
processes. This
means that the equilibrium description of particle phenomena is extremely
accurate at electroweak temperatures. Thus, baryogenesis cannot occur at
such low scales without the aid of phase transitions (for a treatment of 
this argument in the context of non-standard cosmologies, in which the universe
is not radiation dominated at the electroweak scale, see 
\citeasnoun{J&P 97}).

For a continuous transition, the extremum at $\vp=0$ becomes a local 
maximum at $T_c$ and thereafter there is only a single minimum at 
$\vp \neq 0$ (see figure \ref{secondorder}).
At each point in space thermal
fluctuations perturb the field which then rolls classically to the new
global minimum of the FTEP. Such a process is referred to as 
{\it spinodal decomposition}. If the 
EWPT is second order or a continuous crossover, the associated departure from
equilibrium is insufficient to lead to relevant baryon number production
\cite{KRS}.
This means that for EWBG to succeed, we either need the EWPT to be strongly
first order or other methods of destroying thermal equilibrium,
for example topological defects (see section \ref{defects}), to be 
present at the phase transition.

\begin{figure}
  \centerline{\epsfbox{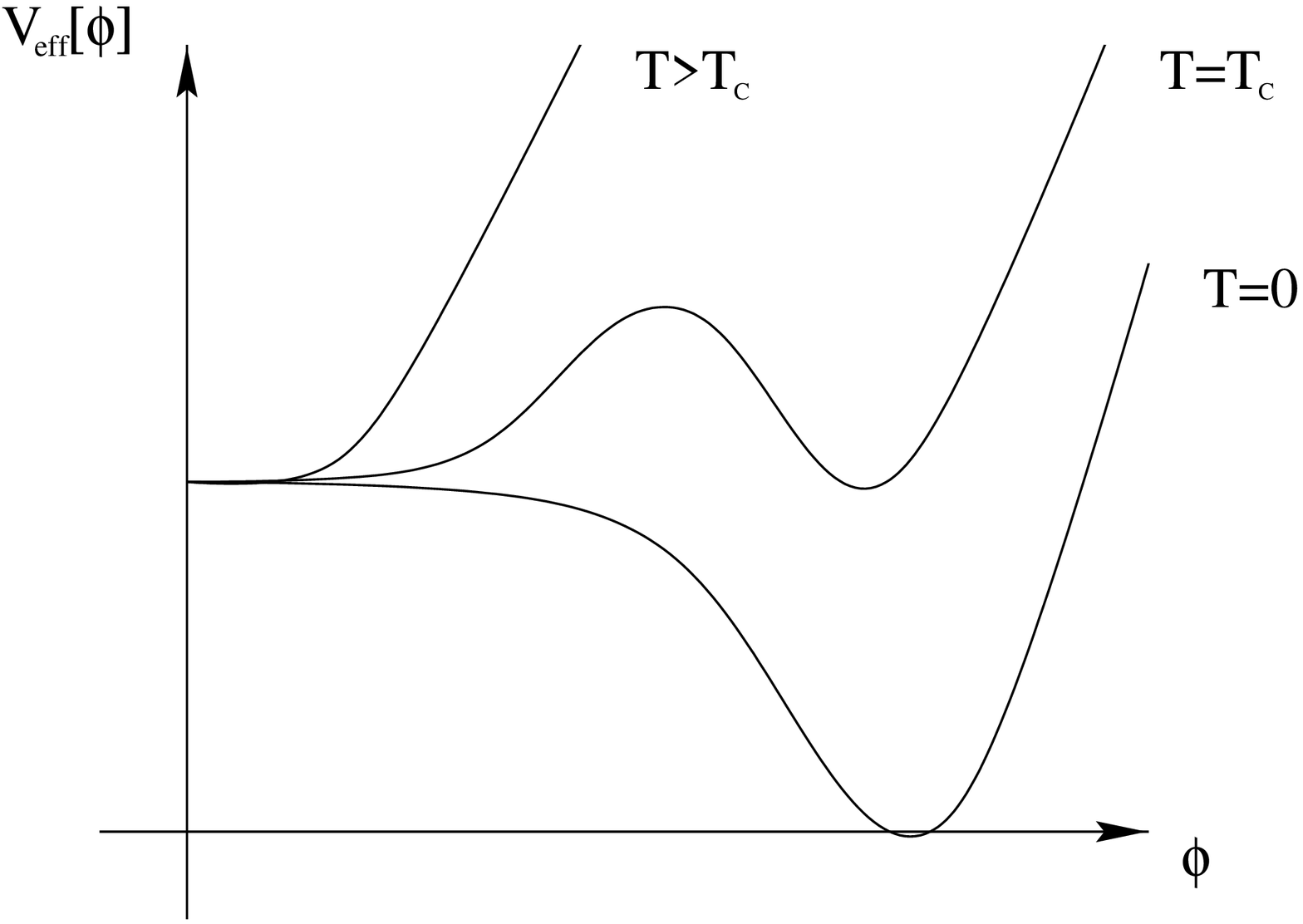}}
  \caption{}
\label{firstorder}
\end{figure}

\begin{figure}
  \centerline{\epsfbox{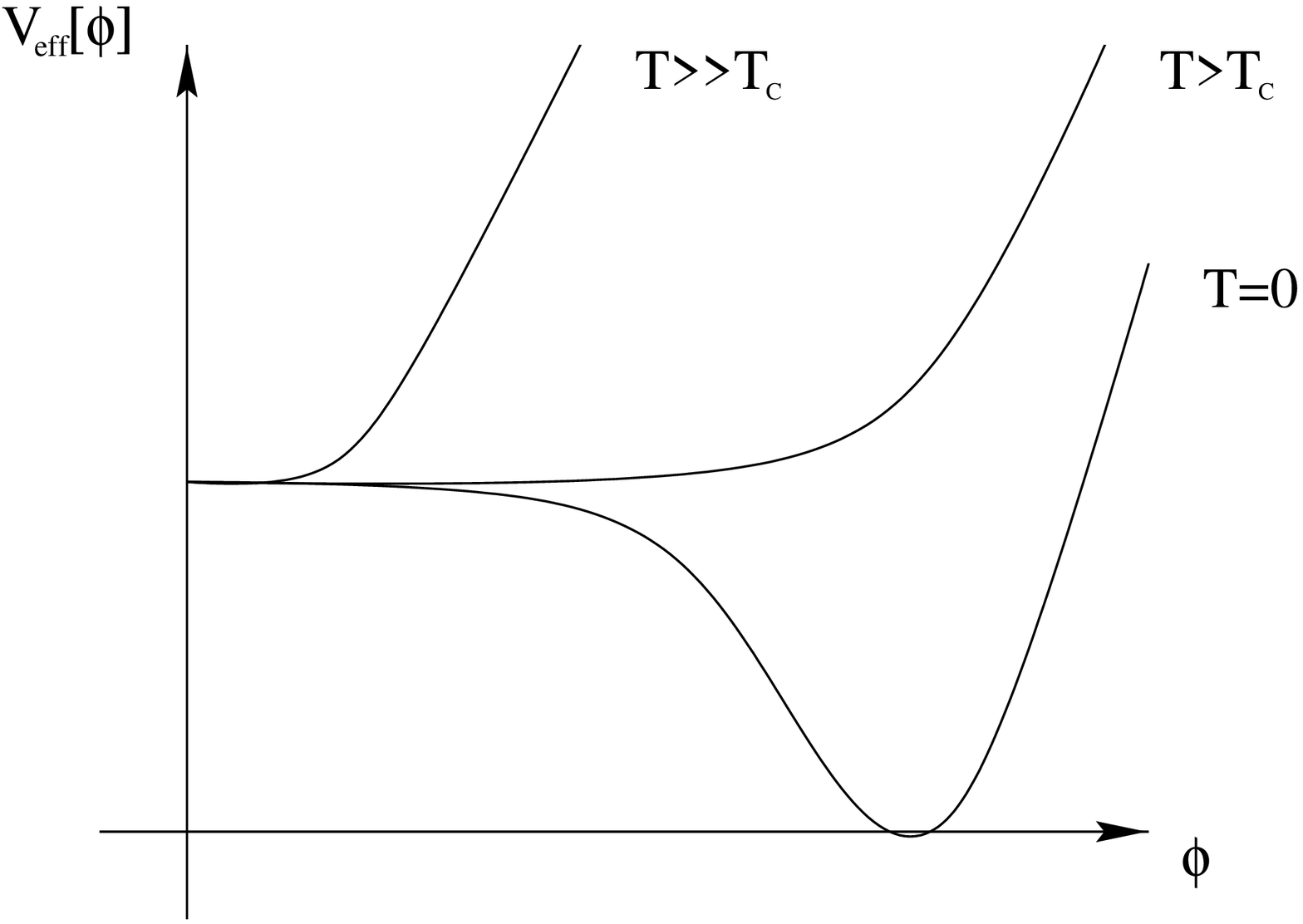}}
  \caption{}
\label{secondorder}
\end{figure}

For a first order transition the extremum at $\vp=0$ becomes separated
from a second local minimum by an energy barrier (see figure \ref{firstorder}).
At the critical temperature $T=T_c$ both phases are equally 
favored energetically and at later times the minimum at $\vp \neq 0$ becomes
the global minimum of the theory. 
The dynamics of the phase transition in this situation
is crucial to most scenarios of electroweak baryogenesis. The essential picture
is that at temperatures around $T_c$ quantum tunneling
occurs and nucleation of bubbles of the true vacuum 
in the sea of false begins. Initially these bubbles are not large 
enough for their volume energy to overcome the competing surface tension and 
they shrink and disappear. However, at a particular temperature below $T_c$ 
bubbles
just large enough to grow nucleate. These are termed {\it critical} bubbles,
and they expand, eventually filling all of space and completing the transition.
As the bubble walls
pass each point in space, the order
parameter changes rapidly, as do the other fields, and this leads to a
significant departure from thermal equilibrium. Thus, if the phase 
transition is strongly enough first order it is possible to satisfy
the third Sakharov criterion in this way.

There exists a simple equilibrium analysis of bubble nucleation
in a first order phase transition (see, for example, \citeasnoun{review}). 
Write the bubble nucleation rate per unit 
volume at temperature $T$ as $R(T)$. Further, note that if we assume that 
bubbles expand at constant speed $v$, then the volume occupied at time $t$ by a
bubble that nucleated at time $t_0$ is

\be
V(t,t_0) = \frac{4\pi}{3}(t-t_0)^3 v^3 \ .
\ee
Then, the fraction of the total volume occupied by the broken phase at time $t$
can be written as

\be
P(t) = 1-\exp[-\Sigma (t)] \ ,
\ee
where the quantity $\Sigma (t)$ is given by

\be
\Sigma (t) = \int_{t_c}^t dt_0\, V(t,t_0) R(T_0) 
\label{Sigma}
\ee
and $t_c$ and $T_0$ are defined through $T(t_c)=T_c$ and $T(t_0)=T_0$ 
respectively. In terms of this quantity, we say that the phase transition is 
complete when $\Sigma \simeq 1$. In order to estimate $\Sigma (t)$ note 
that when bubble nucleation begins, the quantity

\be
x \equiv \frac{T_c - T}{T_c}
\ee
is small. Making this change of variables in (\ref{Sigma}), and using
the time-temperature relationship $t={\tilde M}/T^2$, with ${\tilde M}\sim
10^{18}$GeV we obtain, for small $x$

\be
\Sigma (t) = \frac{64\pi v^3}{3}\left(\frac{{\tilde M}}{T_c}\right)^4
\int_0^x dx\, x^3 \frac{1}{T_c^4} R(T) \ .
\ee
Note that, from this expression, it is clear that when $\Sigma$ is of order 
one, when the phase transition completes or {\it percolates}, the rate per 
unit volume is negligibly small.

In general, the calculation of the bubble nucleation rate is extremely 
complicated. The relevant time scale $\tau_{form}$ for the formation of a critical 
bubble is of the form

\be
\tau_{form} \propto \exp(F_c/T) \ ,
\ee
where $F_c$ is the free energy of the
bubble configuration. We may calculate this by writing the bubble as a
spherically symmetric configuration $\vp(r)$ which extremizes the effective 
action (as defined in the next subsection) and satisfies 

\bea 
\lim_{r\rightarrow \infty} & \vp(r) & = 0 \ , \nonumber \\
& \vp(0) & = v(T) \, \nonumber
\eea
where $v(T)$ is the minimum of the finite temperature effective potential 
$V(\vp,T)$ defined properly in the next subsection. Then the free energy is given 
by

\be
F_c[\vp(r)] \simeq \int dr\, r^2\left[\frac{1}{2}(\vp')^2 + V(\vp,T)\right]
\ee
with the bubble configuration obtained by solving

\be
\frac{1}{r^2}\frac{d}{dr}(r^2\vp') - \frac{\partial V}{\partial\vp} = 0
\ee
subject to the above boundary conditions. However, in the general situation
these equations must be solved numerically, although some progress can be made 
in the {\it thin wall} limit in which the typical size of the critical 
bubbles is much larger than the correlation length of the system.

The precise evolution of critical bubbles in the electroweak phase transition 
is a crucial factor in determining which regimes of electroweak
baryogenesis are both possible and efficient enough to produce the BAU. In 
essence, the
bubble wall dynamics is governed by the interplay between the surface 
tension and the volume pressure. 

The physics of a propagating phase boundary, or bubble wall, have been 
extensively investigated by many authors 
\cite{{EIKR 92},{DLHLL 92},{NT 92},{LMT 92},{YK 92},{Arnold 93},{HKLLM 93},{ML 94},{IKKL 94},{K&L 95},{AH 95},{FKOTT 95},{MP1 95},{MP2 95},{K&Li 96},{K&Lii 96},{DY 96},{FKOT 97}}. The
crucial quantities that one wishes to estimate are the wall velocity,
$v$, and the wall width, $\delta$. As will become clear later, it is 
important to know whether the velocity is less than or greater than the
speed of sound in the plasma, because different mechanisms
of baryogenesis dominate in these regimes. As it turns out, the
dynamics of the bubble wall is qualitatively different in these two
regimes also \cite{{PS 82},{GKKM 83}}.

To be definite, consider a single bubble of broken electroweak symmetry
expanding in a sea of symmetric phase plasma, and for simplicity assume 
that the bubble is
large enough that we may neglect its curvature and idealize the wall as
a planar interface. There are two relevant regimes:

\begin{enumerate}
\item The wall velocity is less than
the speed of sound ($v<c_s\simeq 0.58$): In this case it is said that
{\it deflagration} occurs. The plasma near the bubble wall in the
unbroken phase accelerates away from the wall and a shock wave develops
ahead of the wall moving at speed $v_{sw}>v$. This results in the
heating of the plasma behind the shock front (For a detailed treatment
of deflagration see \citeasnoun{HK 85}). 
\item The wall velocity is greater than the speed of sound ($v>c_s$): 
In this case we refer to {\it detonation} occurring.
In contrast to deflagration, the plasma ahead of the wall is now at
rest, whereas that behind the wall in the broken phase accelerates away. 
\end{enumerate}

Which of these regimes is relevant for a given phase transition depends on
a host of microphysical inputs, making an analytic approach 
extremely difficult. However, recent investigations of the bubble wall 
behavior in the standard model
have been performed \cite{MP1 95,MP2 95}. These authors
find that if $m_H<90\,$GeV (encompassing the whole region of physically
allowed Higgs masses for which a strongly first order phase transition is 
expected to occur), the phase transition proceeds through
deflagration. They find a robust upper bound on bubble velocities

\be
v < 0.45 < v_{sw} \simeq c_s \simeq 0.58, \ \ \ \ \ \ m_H<90\,GeV
\ 
\ee
Other recent work \cite{M&Ti 97} has demonstrated the importance
of friction from infrared W bosons on the wall velocity and suggests that
velocities of the order 

\be
v \sim 0.1 - 0.2
\ee
may be realistic.
Of course, the methods used in this analysis cannot be extended past
the point at which the shock waves originating from different bubbles
would collide, since at that stage the approximation of an isolated
wall is no longer valid. In fact, the dynamics of the phase transition
from that point to completion is quite different from the simple
picture I have described above. What seems clear is that significant
heating of the plasma occurs, perhaps up to $T_c$ \cite{K&Lii 96}, 
as the latent heat of the transition
is released. This should result in an appreciable deceleration of the
bubble walls until finally, the transition 
is completed by relatively slow moving bubbles. In fact, some analytical 
progress may be made in the limit in 
which one assumes that the latent heat of the transition is released
instantaneously into the plasma \cite{AH 95}.
While all these approaches are 
useful in understanding the nature of bubble walls, it is important to 
remember that they are performed in the minimal standard model. When
a particular extension of the electroweak theory is used, the analysis should
be repeated in that context.

In practice it can be a difficult task to 
determine the order of a given phase transition and thus whether, and how, bubble
walls may arise. In the remainder of
this section I shall review
some of the analytical and numerical approaches which are used, and discuss 
their application to the electroweak phase transition.

\subsection{The finite temperature effective potential}
A widely used tool in studying thermal phase transitions
is the {\it finite temperature effective potential} (FTEP) for the order
parameter. I have already mentioned this in passing when discussing bubble
nucleation. Here I will define this object precisely, and explain how it 
is calculated in
a simple model. I will then give the form of the potential for the 
electroweak theory and show how it is used.

Let us begin at zero temperature and consider a single scalar field $\vp$
with external source $J$. The generating functional for this theory is

\be
Z[J] = \int {\cal D}\vp \exp\left[i\int d^4x\, ({\cal L}[\vp] 
+J\vp)\right] \ .
\ee
From this quantity the field theory analog $E[J]$ of the Helmholtz
free energy is defined by

\be
e^{-iE[J]} \equiv Z[J] \ .
\ee
As is well known, functional differentiation of $E[J]$ with respect to the 
source $J$ defines the classical field through

\be
\frac{\delta}{\delta J(x)} E[J] = -\vp_{cl} \ .
\ee
Now, in thermodynamics it is usual to construct the Gibbs free energy of the 
theory by a Legendre transform of the Helmholtz free energy. In field
theory we perform an analogous transformation to define the 
{\it effective action} by

\be
\Gamma[\vp_{cl}] \equiv -E[J] -\int d^4y \,J(y)\vp_{cl}(y) \ .
\label{effaction}
\ee
This functional has the useful property that

\be
J(x) = - \frac{\delta}{\delta \vp_{cl}(x)} \Gamma[\vp_{cl}] \ ,
\ee
which means that, in the absence of an external source, the effective action 
defined by (\ref{effaction}) satisfies

\be
\frac{\delta}{\delta \vp_{cl}(x)} \Gamma[\vp_{cl}] = 0 \ .
\label{effmin}
\ee
Thus, the values of the classical fields in the vacua of the theory are
obtained by solving this equation, i.e., by extremizing the effective action. 

One further simplification is possible. If the vacua of the theory are 
translation and Lorentz invariant, then $\vp_{cl}$ is constant. In that
case, $\Gamma[\vp_{cl}]$ contains no derivatives of $\vp_{cl}$ and
(\ref{effmin}) is an ordinary equation. It is therefore convenient to 
define the {\it effective potential} by

\be
V_{eff}(\vp_{cl}) \equiv -\frac{1}{{\cal V}_4} \Gamma[\vp_{cl}] \ ,
\ee
where ${\cal V}_4$ is the spacetime 4-volume, so that the equation leading 
to the vacua of the theory reduces to

\be
\frac{\partial}{\partial\vp_{cl}} V_{eff}(\vp_{cl}) = 0 \ .
\ee
Note that this equation allows one, in principle, to compute the vacua
of the theory exactly, taking into account all corrections to the
bare potential from quantum fluctuations in the field $\vp$.

Exact analytic calculations of the effective potential are difficult. 
Therefore,
it is usual to use perturbation theory. As an example, in the scalar field 
model above 
choose the bare potential to be

\be
V(\vp) = -\frac{\mu^2}{2}\vp^2 + \frac{\lambda}{4}\vp^4 \ ,
\ee
with $\mu$ an arbitrary mass scale and $\lambda$ a parameter. We write $\vp$
as

\be
\vp(x) = \vp_{cl} + \chi(x) \ ,
\ee
and the aim is to take account of the small quantum fluctuations in $\chi$
around the classical vacuum state $\vp_{cl}$. These fluctuations can contribute
to both the energy of the vacuum state and the potential felt by $\vp_{cl}$,
since the mass of the field $\chi$ is due to $\vp_{cl}$. The perturbation
theory approach to calculating the effects of quantum fluctuations is 
usually phrased in the language of Feynman diagrams. At the one-loop
level, there is only one diagram, which is a single $\chi$ loop. This
loop yields a contribution to the effective potential of

\be
V^{(1)}(\vp_{cl}) = \frac{1}{2(2\pi)^4} \int d^4k\, \ln
[k^2 + 3\lambda\vp_{cl}^2 -\mu^2] \ .
\ee
If we perform the $k_0$ integration and remove an infinite constant, 
this becomes

\be
V^{(1)}(\vp_{cl}) = \frac{1}{(2\pi)^3} \int d^3k\,k[|{\bf k}|^2
+3\lambda\vp_{cl}^2 -\mu^2]^{1/2} \ .
\label{intk3}
\ee
This expression can now be seen to correspond to an integral over the
energy of a $\chi$ particle in all momentum modes. Note that, since this
energy depends on $\vp_{cl}$, normal ordering cannot remove this 
contribution to the energy. 

Since (\ref{intk3}) is divergent, we must choose renormalization conditions
to perform the integration. If we choose to renormalize at the point
$\vp=\mu\lambda^{-1/2}$, and there set $V'(\vp)=V''(\vp)=0$, then the one
loop zero temperature effective potential becomes

\bea
V_{eff}^{(1)}(\vp)= -\frac{\mu^2}{2}\vp^2 + \frac{\lambda}{4}\vp^4
& + & \frac{1}{64\pi^2}(3\lambda\vp^2 -\mu^2)^2 \ln
\left(\frac{3\lambda\vp^2 -\mu^2}{2\mu^2}\right) \nonumber \\
&  & +\frac{21\lambda\mu^2}{64\pi^2}\vp^2
-\frac{27\lambda^2}{128\pi^2}\vp^4 \ .
\eea

We now wish to incorporate the effects of thermal fluctuations into this
picture. Just as quantum fluctuations of fields lead to a modification
of the potential, thermal fluctuations have an analogous effect. The
potential resulting from taking account of thermal fluctuations is referred
to as the {\it finite temperature effective potential} (FTEP).
The formalism we set up for the zero temperature case is applicable
here also because of the well-known connection between zero-temperature 
field theory and thermal field theory. The path integral formulation of field 
theory at nonzero temperature T, describing the equilibrium structure of the
theory, is formally equivalent to the zero temperature formalism performed
over a Euclidean time interval of length $\beta\equiv 1/T$. In addition,
one must impose appropriate boundary conditions on the fields; periodic
for bosons and anti-periodic for fermions. Thus, any field $\chi$ may be
Fourier expanded over this Euclidean time interval, yielding the 
expression

\be
\chi(x,\tau) = \sum_{n=-\infty}^{\infty}\chi_n(x)e^{i\omega_n\tau} \ ,
\ee
with

\be
\omega_n  = \left\{ \begin{array}{ll}
         2n\pi T    & \ \ \ \mbox{Bosons} \\
         (2n+1)\pi T & \ \ \ \mbox{Fermions}
         \end{array} \right. \ .
\label{matsubara}
\ee

Since the Euclidean time coordinate is now finite, the zeroth component of a
particle's four momentum is now discrete. When we perform an integral over
$k_0$, as in the quantum case above, this becomes a sum. This means that
the one loop temperature dependent contribution to the FTEP is

\be
V^{(1)}(\vp,T) = \frac{T}{2(2\pi)^3} \sum_{n=-\infty}^{+\infty}
\int d^3k\, \ln[(2\pi nT)^2 + |{\bf k}|^2 + 3\lambda\vp^2 -\mu^2] \ .
\ee
If we define $m^2(\vp)\equiv 3\lambda\vp^2 -\mu^2$, then for
$m\ll T$ we may perform the sum and integral to first order in lambda. After
renormalization we add this to the bare potential to give the one loop FTEP
as

\be
V_{eff}^{(1)}(\vp,T) = -\frac{\mu^2}{2}\vp^2 + \frac{\lambda}{4}\vp^4
+\frac{m^2(\vp)}{24}T^2 - \frac{\pi^2}{90}T^4 \ .
\ee
This can be thought of as the contribution to the $\vp$-potential from the 
energy of, and interaction with, a thermal bath of particles at temperature 
$T$.

In general, in a theory with spontaneous symmetry breaking, the usual
high temperature behavior of the theory is that of the zero temperature
theory at high energies. That is to say that the full symmetry of the
Lagrangian is restored at high temperatures. At high temperatures the 
global minimum of the FTEP is at $\vp=0$ and at zero temperature the global 
minimum is at $\vp = v$, where $v$ is the usual Higgs vacuum expectation 
value (VEV).

In the standard electroweak theory (ignoring the $U(1)$ terms) the high 
temperature form of the one loop FTEP is

\be
V_{eff}^{(1)}(\vp;T)  = \left(\frac{3}{32}g^2+\frac{\lambda}{4}
+\frac{m_t^2}{4v^2}\right)(T^2-T_*^2)\vp^2
-\frac{3g^2}{32\pi}T\vp^3 + \frac{\lambda}{4}\vp^4 \ ,
\label{FTEP}
\ee
with $\vp\equiv \sqrt{\phi^{\dagger}\phi}$, and $m_t$ the top quark mass. 
Note that, in calculating this quantity for a gauge theory, one also takes 
into account loop diagrams corresponding to fluctuations in the gauge fields.
The lowest order thermal correction to the zero temperature effective 
potential is a temperature dependent mass. It is this contribution that 
causes the
extremum at $\vp=0$ to be the global minimum at high temperatures. Note also 
the presence of a cubic term in the effective potential. This term
arises from the gauge field fluctuations, and 
is responsible for the existence of a barrier separating two degenerate
vacua at the critical temperature, and hence for the prediction that the 
electroweak phase transition be first order. 
Using this one loop potential, we may estimate the critical temperature at
which the vacua are degenerate to be

\be
T_c=m_H\left(\frac{3}{8}g^2 +\lambda -\frac{9}{256\pi^2}g^6 
+\frac{m_t^2}{v^2}\right)^{-1/2} \ .
\ee

Calculations of quantities such as the critical temperature can be
refined if the effective potential is calculated to higher orders in
perturbation theory. In particular, nowadays the two-loop calculation has been
performed \cite{{resum1},{resum2},{3dmasses1},{FKRS1 94},{BFH 95}}. 

It is worthwhile including a cautionary note concerning the FTEP. 
If the interactions are weak, i.e. if

\be
\frac{g^2 T}{\pi m_W(\vp)} \ll 1 \ ,
\label{converge}
\ee
then the effective potential approach
is equivalent to solving the full equations of motion for the gauge
and Higgs fields with all fluctuations taken into account. However, in a
perturbative approximation it is very difficult to go beyond two loops, which
limits the accuracy of the method. Further, when the phase transition is
weak, the infrared degrees of freedom become strongly coupled 
\cite{{AL 80},{GPY 81}}, the
inequality (\ref{converge}) is no longer satisfied,  and we
must turn to numerical approaches.

Note that the higher the Higgs mass is, the weaker the phase transition
gets. The current lower bound on the mass of the Higgs
in the minimal standard model (MSM) comes from combining the results of the
DELPHI, L3 and OPAL experiments at LEP and is 

\be
m_H > 89.3 \, {\rm GeV}
\ee 
\cite{SdJ 98} and so the perturbation expansion might not be expected to 
converge.

\subsection{Dimensional Reduction}
As I have already explained, equilibrium field theory at nonzero 
temperature can be
formulated as zero temperature field theory performed over a finite 
interval of Euclidean time with period $1/T$.

If we expand the fields as in (\ref{matsubara}), one may think of the 
theory as a three dimensional system of an
infinite number of fields $\chi_n$. These fields comprise a tower of
states of increasing masses proportional to the Matsubara frequencies 
$\omega_n$. If the theory is
weakly coupled (as we expect), then we may make a simplifying approximation.
For momenta much less that the temperature, we may perturbatively 
account for the effects of all fields with nonzero thermal masses. 
Thus, since all the fermions are massive, what remains is an effective
theory of only the bosonic fields with $n=0$, the zero modes. 

To see how the remaining
fields contribute to the dynamics of such a theory consider the following 
equivalent construction. Write down the most general, renormalizable 
Lagrangian for the zero modes in three dimensions. For this to be the
effective theory we seek, it is necessary that the free parameters be
determined by matching the one-particle irreducible (1PI) Green's functions
with those of the full theory in four dimensions. The massive degrees of 
freedom are 
important for this matching condition and thus contribute to the
masses and couplings in the three-dimensional effective theory.

The three-dimensional effective theory is much simpler than the full theory.
For the MSM, the particle content is the Higgs doublet, the 3d 
$SU(2)\times U(1)$ gauge fields, and extra bosonic fields corresponding to the
temporal components of the gauge fields in the full theory. For this 
effective theory, perturbative calculations of the coupling constants to
one loop \cite{3dmasses1} and masses to second order \cite{3dmasses2} 
have been performed. 

In the region of the phase transition itself, a further simplification is 
possible. In this regime, the masses of the extra bosons in the theory are
proportional to $gT$ or to $g'T$ and are heavy compared to the effective
Higgs mass of the theory. Thus, we may integrate out these heavy degrees of
freedom and obtain a simple, effective 3d theory which describes the
system near the phase transition \cite{{PG 80},{A&P 81},{SN 83},{NL 89},{3dmasses1},{FKRS1 94},{ML 95}}. The appropriate Lagrangian is

\be
{\cal L} = \frac{1}{4}W^a_{ij}W^a_{ij} + \frac{1}{4}F_{ij}F_{ij}
+(D_i\phi)^{\dagger}D_i\phi + m_3^2 \phi^{\dagger}\phi 
+\lambda_3(\phi^{\dagger}\phi)^2 \ ,
\label{3dLag}
\ee
where $m_3$ and $\lambda_3$ are the effective 3d Higgs mass and self-coupling
in the theory near the phase transition. 

This theory is {\it superrenormalizable} since $m_3^2$ has only linear and 
log divergences and, in the $\overline{MS}$ scheme, $\lambda_3$ and
$g_3$, the 3-dimensional gauge coupling, do not run at all.
Thus, the theory described by (\ref{3dLag}) is very powerful, since the
perturbative calculations of the parameters involved do not suffer from
any infrared divergences. As a result, the theory is valid over a large
range

\be
30\,GeV < m_H < 240\,GeV 
\ee
of Higgs masses.
In addition, removal of the ultraviolet divergences of the theory is
simple. Further, the vastly decreased number of degrees of 
freedom makes the analysis of this theory much more manageable. One might 
wonder if the results from this approach can be checked against those from 
the full 4d theory in any regimes. In fact, at high temperatures, 
calculations of the 4d effective potential in which hard thermal loops are
``resummed'' \cite{{resum1},{resum2},{3dmasses1}} agree well with the 3d
results (for a nonperturbative approach to checking this agreement see
\citeasnoun{MLii 96}. For an alternative treatment of the effective 3d 
theory, in which a cutoff is introduced rather than explicitly using the
superrenormalizability of the theory, see \cite{KNPR 96}

While we may construct the 3d effective theory without worrying about 
divergences in its parameters, perturbative calculations using the theory are 
infrared divergent in the symmetric phase of the system. This means that
perturbation theory is not a particularly effective tool and nonperturbative 
approaches are necessary. It is usual to treat the simpler $SU(2)$ system, in
which the $U(1)_Y$ interactions are ignored, to
gain insight into the nonperturbative dynamics of the electroweak phase
transition. In the context of the 3d effective theory I have described, 
a number of nonperturbative methods have been used 
by different groups.

\subsection{Numerical Simulations -- Lattice Approaches}
The second approach which has been very successful for analyzing the
nature of phase transitions at nonzero temperatures is to simulate the
systems numerically on a lattice. These approaches are technically difficult 
to perform but have the advantage of covering the entire range
of parameter space which might be interesting for EWBG.
Here I shall provide
an outline of how these approaches are performed and give the essential 
results. 
Numerical simulations of the electroweak phase transition have been
performed in both the 3d theory described above and in the full
4d theory \cite{CFHJM 96}. However, in the 4d simulations the situation
is complicated in the physically allowed range of Higgs masses.
Nevertheless, considerable progress has been made.

Reliable quantitative results for the order of the EWPT have been 
obtained from lattice Monte Carlo simulations of the effective 3d
theory \cite{{KRS 93},{FKRS1 94},{FKRS2 94},{3dmasses2},{KLRS1 96},{KLRS2 96},{KLRS3 96},{DKLS 96},{DKLS 97},{GIKPS 97},{GISii 97},{GISiii 97},{GIS 98}}. 
This approach is technically complicated and one must take great care
to ensure that the observed features are not artifacts of the lattice
implementation of the theory. For a discussion of these matters and 
an excellent comprehensive treatment of the method, I
refer the reader to \citeasnoun{KLRS1 96}.
The Monte Carlo calculations are performed over a range of system 
volumes $V$ and lattice spacings $a$. The important results are 
obtained in the limit of extrapolation to infinite volume 
($V\rightarrow\infty$) and zero lattice spacing ($a\rightarrow 0$).
Here, I shall just sketch how these lattice Monte Carlo calculations are 
performed.

The most widely studied lattice model is the pure $SU(2)$ theory. The
parameters of the 3d theory are matched with those of the 4d continuum
theory by renormalization group methods. In particular, the top
quark mass is matched to the known value $m_t\simeq 175\,$GeV.
More recently, the full bosonic 3d $SU(2)\times U(1)$ theory has been
simulated \cite{KLRS2 96}. 

\begin{figure}
  \centerline{\leavevmode\epsfysize=20cm \epsfbox{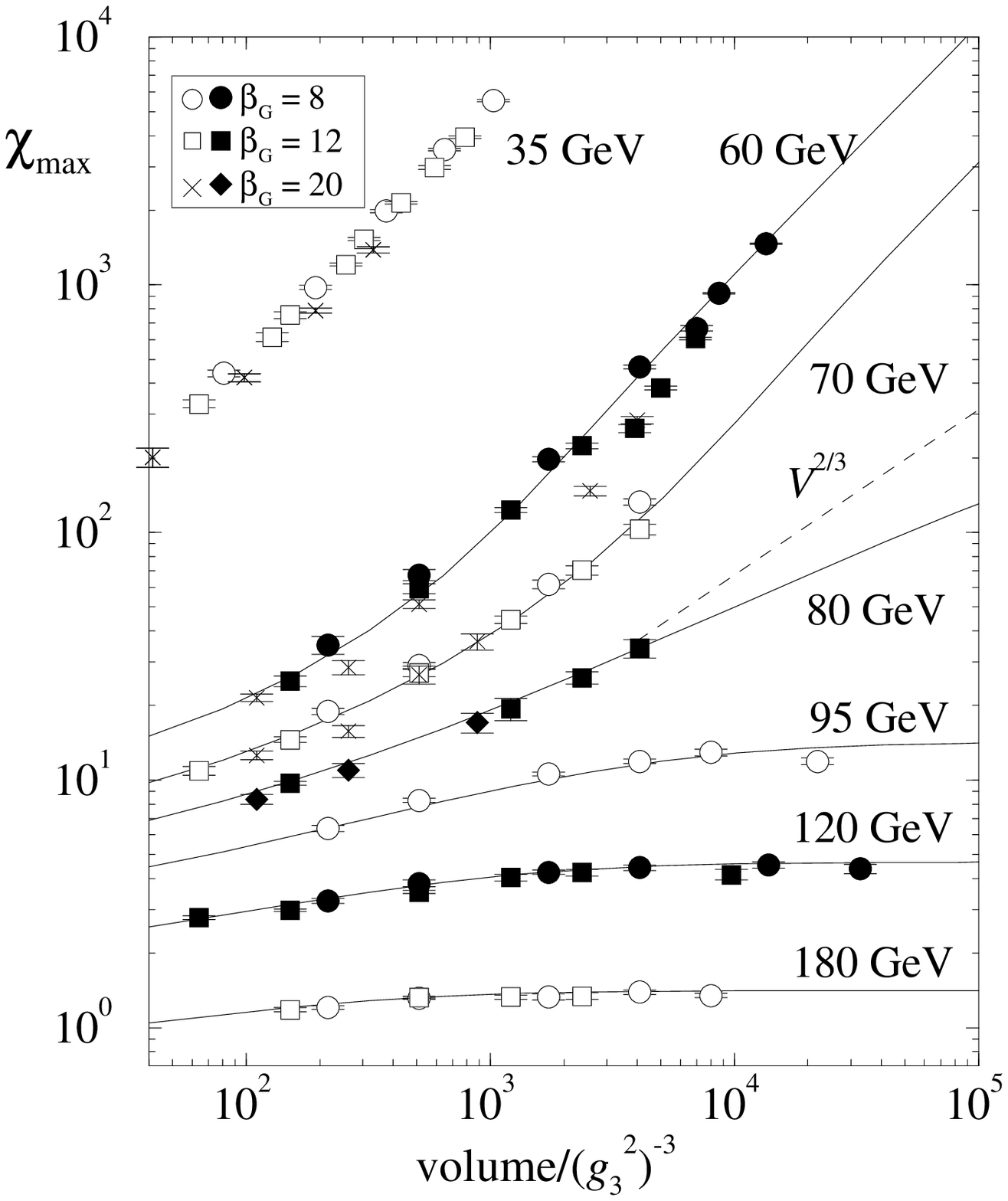}}
  \caption{}
\label{lattice}
\end{figure}

The approach of Kajantie {\it et al.} is to measure the volume 
(finite size) behavior of the susceptibility $\chi$ of the operator
$\phi^{\dagger}\phi$. The susceptibility is defined in terms of quantities
in the 3d effective Lagrangian~(\ref{3dLag}) by

\be
\chi \equiv g_3^2 V \langle(\phi^{\dagger}\phi -\langle\phi^{\dagger}\phi
\rangle)^2\rangle \ .
\label{susceptibility}
\ee
For a given Higgs mass, the procedure is as follows. The first step is, at 
fixed volume $V$, to determine the maximum value $\chi_{max}$ of $\chi$ as a 
function of temperature. This procedure is then performed over a range of 
$V$ to give $\chi_{max}(V)$. From finite size scaling arguments, the following 
behaviors are expected

\begin{equation}
\chi_{max}(V) \propto \left\{ \begin{array}{ll}
         V  & \ \ \ \mbox{First Order Transition} \\
         V^{\gamma/3} & \ \ \ \mbox{Second Order Transition} \\
	 \mbox{const.} & \ \ \ \mbox{No Transition}
         \end{array} \right. \ ,
\end{equation}
with $\gamma\neq 0,3$ a critical exponent.
Using this criterion, it is possible to infer the nature of the phase
transition from the numerical simulations. A summary of the results
\cite{KLRS3 96} is shown in figure~(\ref{lattice}). The general result
of the Monte Carlo approach is that the phase transition is first order for
Higgs masses $m_H \leq 80\,$GeV and becomes a smooth crossover for
$m_H \geq 80\,$GeV. The existence of the endpoint of first order phase 
transitions was shown by \citeasnoun{KLRS3 96}, and the position of the
endpoint was identified by \citeasnoun{GISii 97}. This endpoint, at 
$m_H\sim 80\,$GeV falls into the universality class of the 3d Ising Model.
The endpoint of the transition has also been examined in the 4d simulations
by Aoki \cite{Aoki 97}, with results in agreement with the 3d simulations.

Although the results I quote here are obtained in the $SU(2)$ theory 
(equivalently $\sin^2\theta_W=0$ in $SU(2)\times U(1)$), the
effects of the additional $U(1)$ of the true electroweak theory have been
considered by \citeasnoun{KLRS2 96}. However, while the effects of 
hypercharge strengthen the phase transition in general, the position of the 
endpoint is relatively insensitive to the value od $\sin^2\theta_W$.
I shall mention briefly in the next section how these 
results compare with those obtained from other approaches.

As the Higgs mass is increased to $80\,$GeV from
below, the strength of the first order phase transition decreases.
This behavior is quantified by the tension of the phase interface at $T_c$
and by the latent heat $L$ of the transition. Both these quantities are
measured in the simulations and the expected weakening of the transition
is observed. The Clausius-Clapeyron equation

\be 
\frac{d\Delta p}{dT} = \frac{L}{T} \ ,
\label{Clausius}
\ee
where $\Delta p$ is the difference in pressures between the symmetric and 
broken phases at temperature $T$, is particularly useful here. The 
equation may be written as

\be
\Delta\langle\phi^{\dagger}\phi\rangle \left(\frac{m_H^2}{T_c^3}\right) =
\frac{L}{T_c^4} \ ,
\label{CCequation}
\ee
where $\Delta\langle\phi^{\dagger}\phi\rangle$ is the jump in the squared 
order parameter.
As we shall use later, the measurement of the latent heat from the simulations
yields a value for $\Delta\langle\phi^{\dagger}\phi\rangle$ which we shall
find very useful when discussing local electroweak baryogenesis.

\subsection{Other Approaches}
The application of 
$\varepsilon$-expansion techniques to the electroweak phase transition was
suggested by \citeasnoun{G&K 93} and has been investigated in detail by 
\citeasnoun{A&Y 94}.
Since it is difficult to solve the 3d theory systematically using analytic
methods, the approach
here is to replace the three spatial dimensions by $4-\varepsilon$
dimensions. For example, in massive $\phi^4$ theory, the 4d action is 
transformed to

\be
S = \int d^{4-\varepsilon}x\,\left[(\partial \phi)^2 +m^2\phi^2 +
\mu^{\varepsilon}\lambda\phi^4\right] \ ,
\ee
where a mass scale $\mu$ has been introduced to keep couplings dimensionless.
This class of theories can be solved perturbatively in 
$\varepsilon$ by making use of renormalization group improved perturbation
theory. The procedure is to perform the calculation of a given quantity
in this perturbative scheme to a given order in $\varepsilon$ and then to 
take the 3d limit by taking $\varepsilon\rightarrow 1$. 
This approach
provides an interesting way to study phase transitions. However, its
application to the electroweak theory is complicated and the authors
estimate that the results may be accurate at the 30\% level for Higgs masses
below around $150\,$GeV, and less reliable for higher values. Nevertheless,
the prediction of $\varepsilon$-expansion techniques is that the phase 
transition become weaker but remain first order as $m_H$ increases, even to
values higher than $80\,$GeV. 

In another approach \cite{B&P 95} the one-loop Schwinger-Dyson equations
are studied. This method predicts that the first order transitions become
a smooth cross-over for Higgs masses above $100\,$GeV. Although this result 
is in reasonable agreement with that obtained from numerical approaches, 
it does rely on perturbation theory which, as I have emphasized, 
should not really be trusted for $m_H \geq 80\,$GeV.

Finally, exact renormalization group approaches have been applied to the 
electroweak theory by Wetterich and coworkers \cite{{wetterich1},{wetterich2}}
and by \citeasnoun{B&F 94}. Again, these analyses support the results of
the numerical approaches I described earlier.

With the exception of the $\varepsilon$-expansion, the qualitative conclusion 
of the above approaches is that there exists a critical value of the Higgs 
mass, below which the EWPT is first order, and above which the transition is 
continuous. Since, as I have mentioned, results from LEP \cite{SdJ 98} now give
$m_H > 89.3$ GeV, it seems that the standard electroweak theory undergoes a
continuous transition at high temperatures.

\subsection{Subcritical fluctuations}
\label{mixing}
A central assumption of the analysis of the phase transitions that 
I have followed above is that the initial state be homogeneous. If this
is the case then the existence of a cubic term in the finite
temperature effective potential clearly implies that the phase 
transition proceeds by the nucleation and propagation of critical 
bubbles of true vacuum in the homogeneous sea of false. This result is
obtained within the vacuum decay formalism which relies on the
semiclassical expansion of the effective action. This approach is
valid if the system begins homogeneously in the false vacuum state
so that one may consistently sum over small amplitude fluctuations
about this state, as I described earlier. However, this picture
may not be valid if the initial state of the system is highly inhomogeneous.

The effects of these initial state inhomogeneities have been investigated 
using a 
variety of different approaches by Gleiser and collaborators 
\cite{{G 93},{G 94},{G&R 94},{G 95},{B&G 95},{G&H 96},{B&G 97},{GHK 97}}.
At very high temperatures, the order
parameter is certainly well-localized about the symmetric vacuum state.
However, as we approach the critical temperature we must be sure that thermal
fluctuations do not lead to significant inhomogeneities such that the
saddle point approximation breaks down. In particular, if there is a
sufficient probability for the order parameter to reach the true 
vacuum by purely classical thermal processes (subcritical fluctuations) 
near the critical point,
then we may not apply the nucleation calculation and might expect that
the transition proceeds similarly to spinodal decomposition despite
the first order nature of the effective potential. 

Although quantifying this statement is highly nontrivial, recent
progress has been made numerically by \citeasnoun{B&G 95}.
These authors use a Langevin approach to track the dynamics of the
order parameter in a spontaneously broken theory of a real
scalar field in $3+1$ dimensions. They use a potential which is 
identical in form to the electroweak potential and their initial conditions
are that the scalar field is localized around the false vacuum. The
potential is fixed at the critical temperature, the field evolves
in time and and then the final spatial distribution of the scalar is 
determined for a range of values of the mass. This distribution is
measured by defining $f_0(t)$ to be the fraction of the total volume in the 
false vacuum at time $t$ and evaluating the ensemble average of this quantity
for different values of the scalar field mass.

With this method the authors are making all the assumptions usually 
made in studies of the transition (validity of the effective potential,
homogeneity of the initial state). However, since the dynamics are 
governed by a Langevin equation (with rescaled order parameter $\vp$),

\be
{\ddot \vp}-\nabla^2\vp +\eta{\dot \vp} + 
\frac{\partial V}{\partial \vp}=\xi({\bf x},t)\ ,
\ee
the effects of noise, $\xi$, and viscosity, $\eta$, are taken into account and 
are related by the fluctuation-dissipation equation

\be
\langle\xi({\bf x},t)\xi({\bf x}',t')\rangle =
2\eta T \delta (t-t')\delta^3 ({\bf x}-{\bf x}')\ .
\ee
 
\begin{figure}
  \centerline{\epsfbox{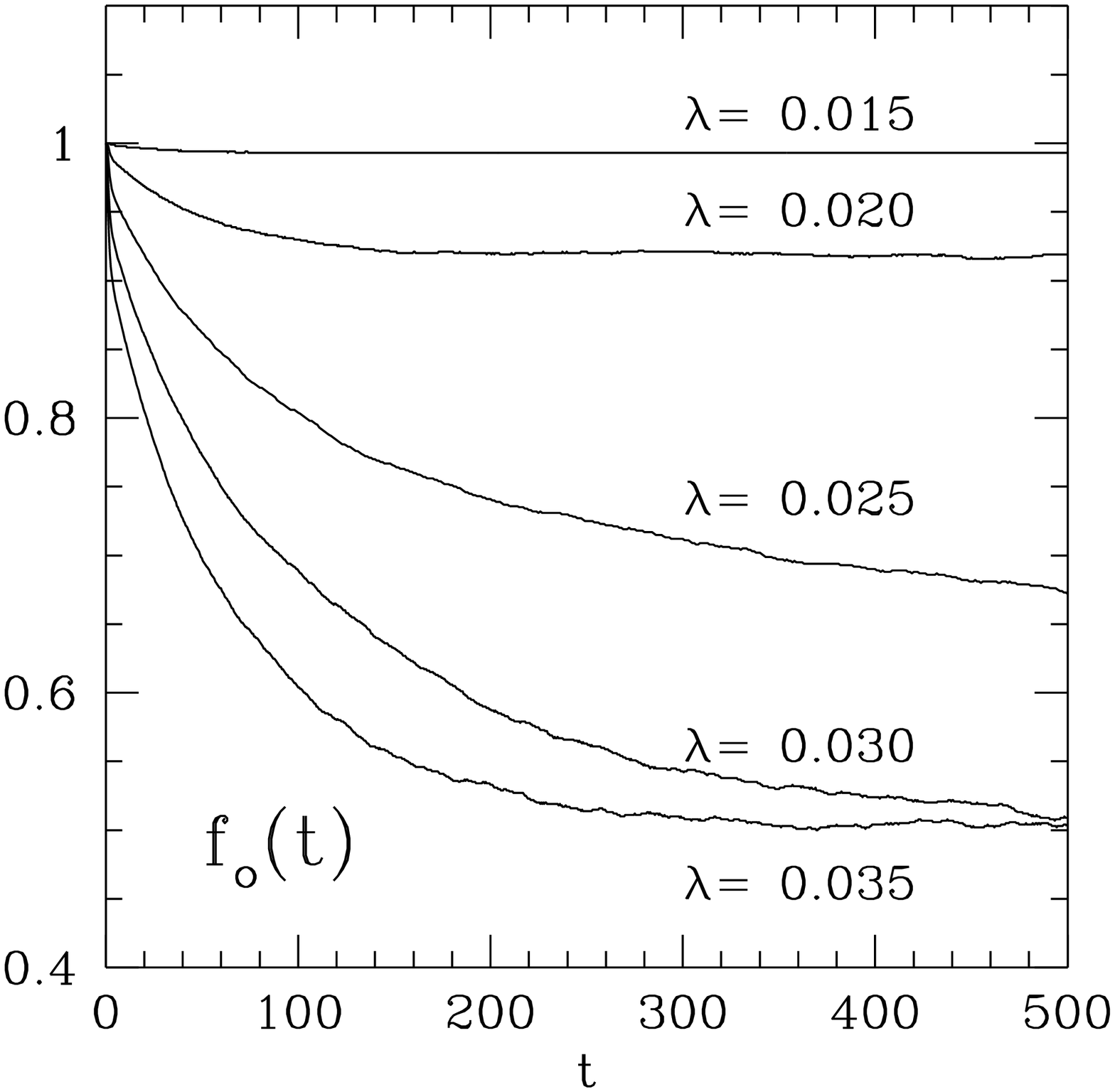}}
  \caption{}
\label{marcelo}
\end{figure}

The results of this simulation are that for all scalar masses in the 
experimentally allowed range of Higgs mass, the authors expect phase 
mixing to occur at the critical temperature (see figure~\ref{marcelo}). 
In other words, by the time of the
transition, half the universe is classically in the false vacuum and
half is classically in the true. 

If this is indeed the case in the electroweak theory then we might
expect the standard bubble nucleation picture not to hold, irrespective of
the validity of the finite temperature effective potential. In fact, the
dynamics of the phase transition may closely resemble that of
spinodal decomposition. It is important to stress, however, that the 
electroweak theory itself is not being simulated in these approaches and when 
the transition from a real scalar field to the modulus of a complex field is 
made, and gauge fields are included, the dynamics may be quite different.

\subsection{Erasure of the Baryon Asymmetry: Washout}
There are essentially two criteria related to the strength of the phase 
transition that are important for electroweak baryogenesis. First, as I 
have commented, if
the phase transition is second order or a crossover, then the only 
departure from
equilibrium arises from the expansion of the universe. At the electroweak 
scale, this is a negligible effect and so no baryon production results. In
such a situation, electroweak baryogenesis may only proceed if other physics
is responsible for displacing the system from equilibrium. Such a situation
is realized if TeV scale topological defects are present at the phase 
transition, as I shall discuss in detail in section \ref{defects}.
If the phase transition is weakly first order, then the dynamics can be
very complicated as I have indicated above.

However, if the phase transition is sufficiently strongly first order, then 
we may be confident that widely
separated critical bubbles nucleate and propagate as I have described. In
this case, electroweak baryogenesis may proceed, but there is a further 
criterion to be
satisfied. Consider a point in space as a bubble wall passes by. Initially,
the point is in the false vacuum and sphaleron processes are copious with
rate per unit volume given by~(\ref{unbrokenrate}). As the wall passes the
point, the Higgs fields evolve rapidly and the Higgs VEV changes from
$\langle\phi\rangle=0$ in the unbroken phase to

\be
\langle\phi\rangle=v(T_c)
\label{vatTc}
\ee
in the broken phase. Here, $v(T)$ is the value
of the order parameter at the symmetry breaking global minimum of the finite 
temperature
effective potential. That is, $v(T_c)$ minimizes the free energy functional
at $T=T_c$. Now, CP violation and the departure from equilibrium, and hence 
biased baryon
production, occur while the Higgs field is changing. Afterwards, the point is
in the true vacuum, baryogenesis has ended, and the rate per unit volume of
baryon number violating events is given by~(\ref{brokenrate}) with
$\langle\phi\rangle$ given by~(\ref{vatTc}). Since baryogenesis is now over, 
it is
imperative that baryon number violation be negligible at this temperature in
the broken phase, otherwise any baryonic excess generated will be
equilibrated to zero. Such an effect is known as {\it washout} of the 
asymmetry. 
The condition that the Boltzmann suppression factor in~(\ref{brokenrate})
be sufficiently large to avoid washout may be roughly stated as \cite{MS 86}

\be
\frac{v(T_c)}{T_c} \geq 1 \ .
\label{washout}
\ee
This is the traditionally used criterion that the baryon asymmetry survive after
the wall has passed, and in this article, this is the criterion I will use.
However, it is worth pointing out that there are a number of nontrivial steps
that lead to this simple criterion. The actual bound is one on the
energy of the sphaleron configuration at the bubble nucleation temperature.
In order to arrive at the criterion (\ref{washout}), one must translate this
into a bound at $T_c$ and then write the sphaleron energy in terms of the VEV
$v(T_c)$. Finally, as I commented earlier, if the evolution of the scale 
factor 
of the universe is non-standard, then this bound can change. For an analysis
of these issues see \citeasnoun{J&P 97}.
It is necessary that this criterion, or one similar to it, 
be satisfied for any electroweak 
baryogenesis scenario to be successful.

\subsection{Summary}
I have given arguments that the minimal standard model has neither enough
CP-violation nor a sufficiently strong phase transition to allow electroweak
baryogenesis to 
take place. These issues have been investigated in depth 
\cite{{F&Si},{F&Sii}} and a detailed analysis was presented by
\citeasnoun{F&Siii}, where the authors found a potential enhancement of 
the Kobayashi-Maskawa
CP-violation by performing a detailed calculation of the effects of particle 
reflection and transport. However, it was later 
argued \cite{{H&S 95},{GHOPQi 95},{GHOPQii 95}} that this potential
effect would be 
erased by quantum decoherence. Separate attempts to enhance the 
CP violation present in the standard model through dynamical effects
\cite{N&T 94} have thus far also been unsuccessful.
It is therefore necessary to turn to extensions of the standard model.
I will do this in some detail for the MSSM in section~\ref{MSSM} but in
the next
two sections I shall focus on the dynamics of electroweak baryogenesis.

\section{Local Electroweak Baryogenesis}
\label{local}
In the previous few sections, I have laid out the necessary criteria for a 
particle physics model to produce a baryonic excess and have demonstrated that
these criteria are satisfied by the standard model of particle physics and
its modest extensions. In the present section I shall describe how these 
separate ingredients can come together dynamically to produce the BAU as
the universe evolves through the electroweak phase transition.

Historically, the ways in which baryons may be produced as a bubble wall, or
phase boundary, sweeps through space, have been separated into two
categories. 

\begin{enumerate}
\item {\it local baryogenesis}: baryons are produced when the baryon number 
violating processes and CP violating processes occur together near 
the bubble walls. 
\item {\it nonlocal baryogenesis}: particles undergo CP violating
interactions with the bubble wall and carry an asymmetry in a quantum number 
other than baryon number into the unbroken phase region away from the wall.
Baryons are then produced as baryon number violating processes convert the
existing asymmetry into one in baryon number.
\end{enumerate}
In general, both local and nonlocal 
baryogenesis will occur and the BAU will be the sum of that generated by the 
two processes. In this section I shall discuss local baryogenesis, and
continue with nonlocal baryogenesis in the next section.

Models of local baryogenesis are some of the earliest viable models in
the field. In this section I want to describe in detail the two major
semi-analytical approaches to calculating the BAU produced by this mechanism. 
It turns out that both approaches fall short of providing a reliable 
quantitative measure of the BAU. However, these attempts provide 
interesting ways of viewing the microphysics behind local baryogenesis.
This is especially important since at present it appears that future 
numerical simulations may be our best hope of obtaining a reasonable 
estimate of this quantity. The most recent analysis of methods of treating
local baryogenesis is due to \citeasnoun{LRT 97} and it is that analysis 
that I shall describe here.

To be specific, I will assume CP violation via the
operator ${\cal O}$, and assume a strongly first order phase transition
and thin bubble walls.
I shall first look at an approach \cite{{ALS 89},{TZi},{TZii},{TZiii}} which
attempts to estimate
the baryon asymmetry
by considering the relaxation of topologically nontrivial field 
configurations produced during the phase transition. Second,
I shall turn to a method introduced by \citeasnoun{DHSSi} 
(see also \citeasnoun{DHSSii}).  
In this treatment, one considers configurations which happen to be near the 
crest of the 
ridge between vacua as the wall arrives, and estimates the extent 
to which their velocity in configuration space is modified
by the operator ${\cal O}$, defined in~(\ref{newop}), during the 
passage of the wall. 

The mechanisms I consider in this section involve the dynamical evolution of
configurations such as~(\ref{winding1}) 
which are released from rest and end up as
outgoing radiation. This complicates the simple situation described in 
section \ref{baryon} somewhat, since in this case it is 
dangerous to use the anomaly
equation~(\ref{newanomaly}) because $\int d^4x\,\hbox{Tr}(W{\tilde W})$ is not
well-defined as an integral and any answer can be obtained for the change in 
fermion number \cite{fiveofus}. Nonetheless, careful calculation shows that
the results of section \ref{baryon} still apply. 
If the configuration~(\ref{winding1}) 
is released and falls apart without ever going through a zero of the
Higgs field, then the analysis of \citeasnoun{FGLR} is directly
applicable and one net antifermion is produced just as in the second 
interpolation. If the Higgs field unwinds by going through a zero, then
one can use arguments presented in \citeasnoun{fiveofus}
to demonstrate that the presence of outgoing radiation in the final 
configuration does not affect the result above, namely that there is no
fermion number violation.

\subsection{Local Baryogenesis Through Unwinding}
\label{T&Z}
Consider using the method of Turok and Zadrozny
to estimate the baryon asymmetry produced by local baryogenesis
in a scenario in which the electroweak phase transition
is strongly first order and the bubble walls are thin~\cite{LRT 97}.
In their original work, Turok and Zadrozny studied the
classical dynamics of topologically nontrivial gauge and Higgs field 
configurations in the presence of CP violation. 
The first step is to consider spherically symmetric non-vacuum
configurations of the form (\ref{winding1}) with Higgs winding $N_H=\pm 1$
and discuss their dynamics when they are released from rest and 
evolve according to the equations of motion.  Solutions to the equations
of motion typically approach a vacuum configuration uniformly throughout
space at late times, and these solutions are no exception.
There are, however, two qualitatively different
possible outcomes of the evolution as I mentioned in section \ref{baryon}.
Without CP violation, for every $N_H=+1$
configuration which relaxes in a baryon producing fashion
there is an $N_H=-1$ configuration which produces anti-baryons.
With the inclusion of the CP violating
operator ${\cal O}$, the hope of this approach is that there will 
be some configurations which produce baryons whose CP
conjugate configurations relax to the $N_H=0$ vacuum without
violating baryon number.

The scenario described above is set in the following dynamical 
context. Imagine that the (thin) bubble wall has just passed,
leaving in its wake a particular configuration, but that this configuration
has not yet had time to relax to equilibrium.  The goal is a 
qualitative understanding of the dynamics of this relaxation.
The first order electroweak phase 
transition can be characterized by the change in the gauge
invariant quantity $\langle \sigma^2 \rangle$, defined in~(\ref{sigmaU}).
Renormalizing $\langle \sigma^2 \rangle$ such that it
is equal to $v^2$ at zero temperature, then for a strongly
first order phase transition it is close to
$v^2$ just below $T_c$ in the low temperature
phase, and is much smaller 
just above $T_c$ in the high temperature phase.
This is a slight motivation for considering initial configurations
in which $\sigma=v$ throughout space, 
even though this is in reality not 
a good description of the non-equilibrium configurations left in the
wake of the wall and is in fact not maintained during the subsequent 
evolution. Similarly, there is no justification for choosing either a 
spherically symmetric configuration, or one with $A_\mu=0$, or one which
is initially at rest.

Note that the analysis of \citeasnoun{LRT 97} that I follow here
treats the non-equilibrium conditions after the passage of a thin wall.
The Lagrangian is simply (\ref{SM action}) plus (\ref{newop}).  
This means that energy 
is conserved (to better than half a percent in numerical
simulations) during the evolution of the gauge and 
Higgs fields.

The strategy now is to solve the equations of motion obtained from
the action (\ref{SM action}) augmented by the addition
of the operator (\ref{newop}).  This is performed in the spherical
ansatz \cite{{ansatzi},{ansatzii}} in which all gauge invariant quantities
are functions only of $r$ and $t$, and the equations
are solved numerically.
First consider initial conditions 
of the form (\ref{winding1}) with $\eta(r)$ of (\ref{u1})
given by 
\begin{equation}
\eta(r) = -\pi \left[ 1- {\rm tanh}\left(\frac{r}{R}\right) \right]
\label{eta}
\end{equation}
where $R$ is a constant parameterizing the size of the configuration.
This configuration satisfies the boundary condition (\ref{Ubc})
and has Higgs winding number $N_H=+1$, and is used 
as the initial condition for the equations of motion, setting all time 
derivatives to zero at $t=0$. Initially, $b$ is set to zero
in (\ref{newop}) and thus there is no CP violation.
In agreement with Turok and Zadrozny, this analysis finds that 
there is a critical value $R_c^+$ of $R$, defined
as follows.  For all $R<R_c^+$ the configuration 
evolves toward a vacuum configuration with $N_H=0$.  No fermions
are produced in this background.  For all $R>R_c^+$, the configuration
evolves toward a vacuum configuration with $N_H=+1$, and
fermions are produced.  The values of $R_c^+$ obtained
in simulations with several different values of $m_H/m_W$ 
are in quantitative agreement with those obtained by \citeasnoun{TZii}.
Repeating this exercise beginning
with $\eta(r)$ given by $-1$ times that in (\ref{eta}), that
is beginning with the CP conjugate configuration having
$N_H=-1$, results in an analogously defined $R_c^-$.
Since $b=0$, as expected $R_c^-=R_c^+$.
The entire procedure is now repeated with $b\neq 0$, that is with CP
violation present and yields $R_c^-\neq R_c^+$. 
Unfortunately, this is not the end of the story.

Consider slightly more general initial conditions, namely exactly as above
except that the time derivative of $\sigma$ is nonzero and
given by
\begin{equation}
\dot\sigma(r) = \gamma v^2 \left[ 1 - {\rm tanh}\left(r/R\right)\right]\ ,
\label{sigmadot}
\end{equation}
with $\gamma$ some constant.  Performing the simulations again reveals
that for some values of $\gamma$, 
$R_c^- < R_c^+$ whereas for other values of $\gamma$, $R_c^- > R_c^+$.
This dooms an analysis in terms of the single parameter $R$.
Clearly, a more general framework is needed.

Consider a family of initial configurations with $N_H=+1$, much more
general than considered to this point, parameterized
by a set of parameters ${\beta_i}$. It would be more general still
to go beyond the spherical ansatz and eventually to
work towards an analysis involving an infinite set of $\beta$'s,
but it seems reasonable to start with some finite set ${\beta_i}$.
For any fixed $b$, define a function $F^+(\beta_1,\beta_2,\ldots)$ 
which has the following properties.

\begin{enumerate}
\item $F^+(\beta_1,\beta_2,\ldots)>0$ for all points in $\beta$-space
which describe configurations which evolve toward the $N_H=+1$
vacuum thereby producing fermions,
\item $F^+(\beta_1,\beta_2,\ldots)<0$ for all 
points in $\beta$-space describing
configurations which evolve towards the $N_H=0$ vacuum.
\end{enumerate}
Considering only  initial configurations described by (\ref{winding1}),
(\ref{u1}) and (\ref{eta}) which are parameterized by the
single parameter $R$, implies $F^+(R) = R-R_c^+$. 
Completely analogously, define a function $F^-(\beta_1,\beta_2,\ldots)$
such that the hypersurface $F^-(\beta_1,\beta_2,\ldots)=0$ 
divides the $\beta$-space
of $N_H=-1$ configurations into those which evolve towards the 
$N_H=-1$ and $N_H=0$ vacua.

\begin{figure}
\centerline{\epsfbox{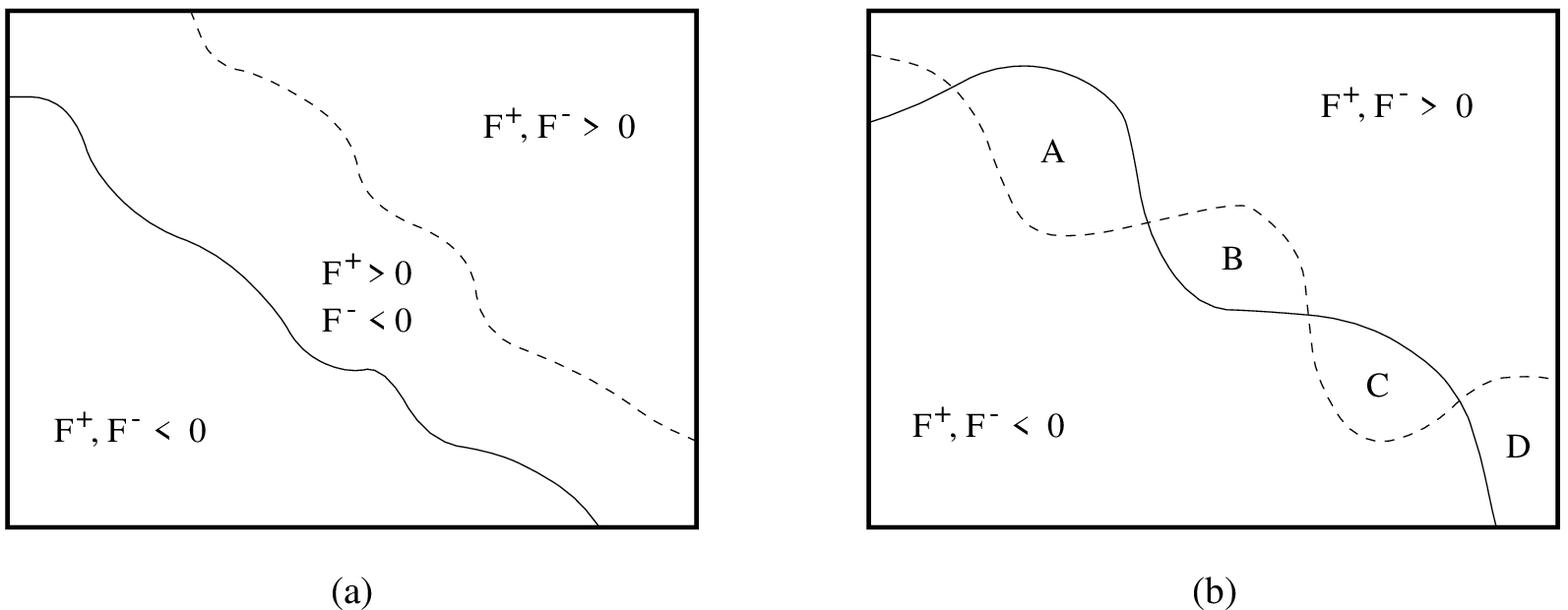}}
\vspace{0.3in}
\caption{}
\label{surf}
\end{figure}

In the absence of CP violation, 
$F^+(\beta_1,\beta_2,\ldots)=0$ and $F^-(\beta_1,\beta_2,\ldots)=0$ define
the same hypersurface. In this case, imagine allowing 
a CP symmetric ensemble 
of configurations with $N_H=+1$ and $N_H=-1$ to evolve.
(in this context, CP symmetric means that the probability for finding
a particular $N_H=+1$ configuration in the ensemble
is equal to that for finding its CP conjugate $N_H=-1$ configuration.)
Since the configurations which anomalously produce fermions
are exactly balanced by those which anomalously produce antifermions, 
the net fermion number produced after relaxation would be zero. 
The aim is to investigate the behavior in the 
presence of the CP violating term ${\cal O}$ of
(\ref{newop}). The hope is that ${\cal O}$ will 
affect the dynamics of $N_H=+1$ configurations 
and $N_H=-1$ configurations
in qualitatively different ways
and that after relaxation to vacuum a net
fermion number will result, even though the initial 
ensemble of configurations
was CP symmetric.  
As described above, when $b\neq 0$ the two hypersurfaces
$F^+=0$ and $F^-=0$ are indeed distinct. 
The configurations represented by points in $\beta$-space 
between the two hypersurfaces yield a net
baryon asymmetry. There are two qualitatively
different possibilities, however, illustrated schematically
in Figure \ref{surf}.  

In Figure \ref{surf}(a), the hypersurfaces $F^+=0$ and $F^-=0$
do not cross. The sign of $b$ has been chosen such that
net baryon number is produced in the region between the
two surfaces.
In Figure \ref{surf}(b), the hypersurfaces cross and
a net baryon number is produced in regions $B$ and $D$,
and net antibaryons in regions $A$ and $C$.
If the hypersurfaces do not cross,
as in Figure \ref{surf}(a), then a simple
estimate of the fraction of configurations which yield a 
net baryon asymmetry is possible.  This fraction
would be proportional to the separation between the 
two hypersurfaces measured in any direction in $\beta$-space with
a component perpendicular to the hypersurfaces --- for example, it
would be proportional to
($R_c^+ - R_c^-$) --- and it would be proportional
to $b/M^2$, the coefficient of ${\cal O}$.
Unfortunately, as described earlier, in the two parameter space of 
$(R,\gamma)$ the hypersurfaces $F^+=0$ and $F^-=0$ do in fact cross.
Thus, in the more general space $(\beta_1,\beta_2,\ldots)$
the picture cannot look like that sketched in Figure \ref{surf}(a)
and must look like that sketched in Figure \ref{surf}(b).

Hence, although the dynamics of the unwinding of topological configurations 
after the phase transition in the presence of the operator ${\cal O}$ may 
lead to a baryon asymmetry, there is at present no way to make a simple 
analytical estimate of this asymmetry.
This may well be a valid way of looking at the microphysics
of electroweak baryogenesis, but it seems that large scale
$3+1$ dimensional numerical simulations of the kind recently pioneered by 
\citeasnoun{M&Tii 97} (but including CP violation via
(\ref{newop}) and 
working in a setting in which the bubble walls are thin and rapidly moving)
are required in order to estimate the contribution to the BAU.
I shall return shortly to a brief discussion of the
large scale numerical simulations which seem necessary.
Before that, however, we shall examine a different
attempt at obtaining a semi-analytical estimate of the magnitude
of the effect.

\subsection{Kicking Configurations Across the Barrier}
The ideas presented in this subsection were originally considered by
\citeasnoun{misha 88} (see also \citeasnoun{DHSSi}). The basic idea is to
consider the dynamics of configurations which are near the crest of the ridge 
between vacua as the bubble wall of the first order electroweak phase transition
arrives. The objective is to understand how the dynamics might be biased by the 
presence of CP violation in the theory and hence how a baryon asymmetry might 
result.

In a sense,
this discussion is more general than that of the previous
section, because it attempts to treat baryon number violating
processes of a type more general than the unwinding of winding number
one configurations.  On the other hand, the treatment
of these more general processes is, of necessity, greatly
over-simplified. Although this discussion follows
that of \citeasnoun{DHSSi} to some extent, they are not always parallel 
and the one here follows \citeasnoun{LRT 97}.

In the high temperature phase, baryon number violating 
processes are not exponentially suppressed.
The barrier crossing configurations typically 
\cite{{AM},{misha 88},{ASYi},{ASYii}}
have sizes given by the magnetic correlation length
\begin{equation}
\xi \sim (\alpha_W T)^{-1}\ .
\label{magcor}
\end{equation}
Imagine dividing space up into cells of this size,
and looking at configurations cell by cell.
The energy in gauge field oscillations with wavelength $\xi$
is of order $T$, but the total energy in a cell is much larger,
as it is presumably of order $T^4\xi^3$. Indeed, this energy is
much larger than the sphaleron energy, which is 
$E_{\rm sph}\sim v/g$.  Most of
the energy is in oscillations of the gauge and Higgs fields on
length scales shorter than $\xi$. These configurations
are crossing the barrier between vacua via regions of the barrier
far above the lowest point on the barrier, that is far above the
sphaleron, and they look nothing like the sphaleron. 
Recall that it is now believed that in each cell of volume $\xi^3$,
the sphaleron barrier is crossed once per time $\xi/\alpha_W$, 
leading to $\kappa\sim \alpha_W$. Now consider what happens when the bubble 
wall hits the configurations just described.

Focus on the configuration in one cell.  It traverses a path
through configuration space, which may be parameterized by $\tau$.
Dine {\it et al.} consider the special case in which this
path is the path 
in configuration space which an instanton follows as a function of
Euclidean time $\tau$, but this is not essential, and it is clear that they
were thinking of more general circumstances also.
The configurations discussed in subsection~\ref{T&Z} can be seen as
special cases of those described here. The energy of the configuration has
a maximum at some $\tau$ (at which the configuration crosses
the barrier) which is defined to be $\tau =0$.  Now
write down a Lagrangian which is intended to describe the 
dynamics of $\tau$ as a function of time for $\tau$ near $\tau=0$:
\be
{\cal L}(\tau,{\dot \tau})=\frac{c_1}{2\xi}{\dot \tau}^2 +
\frac{c_2}{2\xi^3} \tau^2 + 
\frac{c_3}{\xi} \frac{b}{M^2} \sigma^2\, {\dot \tau} \ .
\label{Leq}
\ee
In this expression, $c_1$, $c_2$, and $c_3$ are dimensionless constants,
different for each of the infinitely many possible 
barrier crossing trajectories.
The factors of $\xi$ have been put in by dimensional analysis treating
$\tau$ as a quantity of dimension $-1$. (However, rescaling $\tau$ by a 
dimensionful constant does not change the final result.)
This Lagrangian should be seen
as the first few terms in an expansion in powers of $\tau$ and
$\dot \tau$.  Because by assumption $\tau =0$ is a maximum
of the energy as a function of $\tau$, no odd powers of $\tau$
can appear.  
In the absence of CP violation, there can be
no odd powers of $\dot\tau$, since they make the dynamics 
for crossing the barrier from left to right different than from right
to left.  The ${\rm Tr} F\tilde F$ in the operator ${\cal O}$
includes a term which is proportional to the time derivative
of the Chern-Simons number, and this means that ${\cal O}$ must
contribute a term in ${\cal L}$ which is linear in $\dot \tau$.
It is obviously quite an over-simplification to treat barrier
crossing as a problem with one degree of freedom.  As described earlier,
the complete dynamics can be very complicated, even
for relatively simple initial conditions.  However, let's forge ahead 
with~(\ref{Leq}).

The momentum conjugate to $\tau$ is given by

\be
p = \frac{c_1}{\xi}{\dot \tau} + \frac{c_3}{\xi} \frac{b}{M^2} \sigma^2
\label{mom}
\ee
and the Hamiltonian density is therefore

\be
{\cal H}= \frac{\xi}{2c_1}\left(p - \frac{c_3}{\xi} 
\frac{b}{M^2} \sigma^2 \right)^2
-\frac{c_2}{2\xi^3} \tau^2 \ .
\label{Heq}
\ee
Before considering the thin wall case described in detail here,
it is worth pausing to consider the 
thick wall limit in which $\langle \sigma^2 \rangle$
is changing slowly and other quantities evolve adiabatically
in this slowly changing background. 
A reasonable assumption is 
that the
variables $(\tau,p)$ are Boltzmann distributed with respect to the
Hamiltonian~(\ref{Heq}) at each instant,
treating $\sigma^2$ as approximately constant. 
This implies that the distribution of $p$ is centered at

\be
p_0 = \frac{c_3}{\xi}\frac{b}{M^2} \sigma^2 \ .
\ee
However, from~(\ref{mom}), this means that the velocity
${\dot \tau}$ is Boltzmann distributed with center ${\dot \tau}= 0$. Thus,
the presence of the CP violating operator~(\ref{newop}) does not bias
the velocity of trajectories in configuration space in the thick wall
limit. This conclusion disagrees with that of \citeasnoun{DHSSi} and
\citeasnoun{DHSSii}.
There is nevertheless an effect.  Integrating the third term
in (\ref{Leq}) by parts, one obtains a term linear in $\tau$
proportional to the time derivative of $\sigma^2$.  This 
changes the shape of the potential energy surface in configuration
space during the passage of the wall, and yields an asymmetry.
In this limit, in which the wall is thick and departure from
equilibrium is small, the problem is much more easily 
treated in the language of spontaneous 
baryogenesis, as I will discuss later --
the operator ${\cal O}$ acts like a chemical potential for baryon
number. 

Now return to the thin wall case.  Immediately after the wall strikes,
the fields are not yet in equilibrium.  The idea here
is to use an impulse approximation to estimate the kick 
which $\dot\tau$ receives as the wall passes, and from this to estimate
the baryon asymmetry that results.  The equation of motion
for $\tau$ obtained from (\ref{Leq}) is

\begin{equation}
\ddot\tau = \frac{c_2}{c_1\xi^2}\tau - \frac{c_3}{c_1} 
\frac{b}{M^2}\frac{d}{dt}\sigma^2 \ .
\label{eofm}
\end{equation}
During the passage of a thin wall, the first term on the right
hand side can be neglected relative to the second.  In the
impulse approximation the passage of the wall kicks $\dot\tau$ by an amount

\be
\Delta{\dot \tau} = -\frac{c_3}{c_1}\frac{b}{M^2}\Delta \sigma^2 \ ,
\label{deltataudot}
\ee
where $\Delta \sigma^2$ is the amount by which $\sigma^2$
changes at the phase transition.
The kick $\Delta \dot\tau$ has a definite sign.
Thus, in the thin-wall limit, the distribution of the 
velocities in configuration space of 
barrier crossing
trajectories is biased and a baryon asymmetry results. How might one 
calculate the magnitude of this asymmetry?

If $\Delta \dot\tau$ is large compared to $\dot\tau_0$, the velocity
the configuration would have had as it crossed $\tau =0$ in the
absence of the action of the wall, then $\Delta \dot\tau$ 
will kick the configuration
over the barrier in the direction it favors, and will produce, say,
baryons rather than anti-baryons.  If $\Delta \dot\tau$ is small
compared to $\dot\tau_0$, it will have no qualitative effect.  The fraction
of the distribution of configurations 
with $\dot\tau_0 < \Delta\dot\tau$ is proportional
to $\Delta\dot\tau$.
Note that in this calculation it was not necessary for $\tau$
to be precisely at $\tau=0$ when the wall hits.  It was only necessary
for $\tau$ to be close enough to $\tau=0$ that the Lagrangian
(\ref{Leq}) is a good approximation.   It is difficult to quantify
what fraction $f$ of configurations satisfy this criterion
of being ``close enough to $\tau=0$'', although
it is worth noting that $f$ does not depend on the time it takes
configurations to traverse the barrier.  Nevertheless, the net number 
density of baryons produced may be estimated as

\begin{equation}
n_B \sim \Delta \dot\tau \, f \, \xi^{-3}\ ,
\end{equation}
where the constants $c_i$, $c_2$, $c_3$ have been absorbed into $f$.

At the time of the electroweak phase transition, the entropy
density of the universe is $s\sim 45 T^3$, and the baryon to entropy
ratio is therefore\footnote{This result agrees with that of 
\citeasnoun{DHSSi} although the discussion of \citeasnoun{LRT 97} 
and theirs are somewhat different.}  

\begin{equation}
\frac{n_B}{s} \sim f\, \frac{\alpha_W^3}{45}\,\frac{b}{M^2}\Delta\sigma^2\ .
\label{gettingthere}
\end{equation}
The size of the effect clearly depends on $\Delta \sigma^2$. It was 
suggested by \citeasnoun{DHSSi} that $\Delta \sigma^2$ corresponds to 
increasing $\sigma^2$ up to that value at which baryon number violating
processes become exponentially suppressed in thermal equilibrium. 
Whereas in the thick wall case, baryon number violating processes
stop when $\sigma^2$ reaches this value, this is not the 
case in the thin wall scenario.  In this setting, thermal equilibrium
is not maintained even approximately, and it can be seen from the above 
discussion that what matters is the net change in $\sigma^2$ as the wall 
passes.
Once one picks an extension of the standard model which makes
the transition strongly first order, one can compute
$\Delta \sigma^2$.  It is simplest just to take $\Delta \sigma^2 = v^2/2$,
which is approximately what is obtained in the minimal
standard model with a $35\,$GeV Higgs mass \cite{KLRS1 96}.
Putting it all together yields

\begin{equation}
\frac{n_B}{s} \sim f \,(1\times 10^{-9})\, b\, \frac{(5 {\rm TeV})^2}{M^2}\ .
\label{estimate}
\end{equation}
If, for example, $b\sim \alpha_W$ and $M\sim 1~{\rm TeV}$, the 
bound~(\ref{bound}) can be satisfied and (\ref{estimate}) suggests that a
cosmologically relevant BAU may be generated.
If CP violation is introduced
via the operator ${\cal O}$ with a coefficient $b/M^2$ satisfying
(\ref{bound}), and if the bubble walls are thin, 
then the contribution to the baryon asymmetry
of the universe from local electroweak baryogenesis
can be at an interesting level so long as the quantity $f$ is
not smaller than about a tenth.  

There are many contributions to $f$, since the treatment leading to the 
estimate (\ref{estimate}) is greatly over-simplified.  
First, there are the constants $c_i$ ($i=1,2,3$), which of course 
differ for the different configurations in different
cells of volume $\xi^3$,  and must somehow be averaged over.
Second, using the impulse
approximation is not really justified.  In reality, the wall
does not have zero thickness.  More important, even if the wall
{\it is} thin, the time during which it can affect a configuration
of size $\xi$ is at least $\xi$.  Third, 
the treatment in terms of the Lagrangian~(\ref{Leq}) only has
a chance of capturing the physics near $\tau=0$, and it is not
at all clear what fraction of configurations satisfy this.
Configurations which happen to be farther away from the
crest of the ridge between vacua when the wall hits
do receive a kick from the wall.  However, even if this kick is
large, it may not be in a suitable direction in configuration
space to be effective.  Configurations far from $\tau=0$
can contribute to $n_B$, but their contribution is hard to compute,
because there is no way to reduce the problem to one of one degree
of freedom far from $\tau=0$.
Fourth, even near the crest of the ridge for a given trajectory
the problem does not really reduce to one degree of freedom.
For the configurations of interest, $\sigma$ is a function of
space and time and the operator ${\cal O}$ and the bubble
wall conspire to affect its dynamics.  In the method described here,
the effect is described by treating $\sigma$ as constant
in space and time on either side of the wall and only changing
at the wall.
Fifth, we must face up to the specific difficulties discussed in the 
treatment of the Turok and Zadrozny mechanism.
After the wall has passed, the fields are not yet in thermal
equilibrium and their dynamics is complicated.  This may in
fact yield a further contribution to $n_B$.  It may also,
however, negate some of the contribution estimated 
in (\ref{estimate}) because some configurations kicked across
the barrier in one direction by the passage of the wall may at a later time
wander back across the barrier whence they came.
As mentioned earlier, an estimate of the magnitude
of these sorts of effects is difficult even for a restricted
class of configurations.  To sum up, $f$ is almost certainly less than $1$. 
Hence, it would be best to use (\ref{estimate}) as an upper bound on $n_B/s$, 
rather than as an estimate.  

If~(\ref{estimate}) is only used as an upper bound it 
is still interesting. 
Combined with the experimental
bound (\ref{bound}) on the coefficient of 
${\cal O}$, the result (\ref{estimate}) shows that if
the experimental sensitivity to the electric dipole moment of the
electron or the neutron can be improved by about an order of magnitude,
and if these experiments continue to yield results consistent
with zero, then the baryon asymmetry
of the universe produced by local electroweak baryogenesis 
is smaller than that observed, even if future numerical
simulations were to demonstrate that $f$ is as large as $1$. Such a 
result would rule out the operator~(\ref{newop}) as the source of CP
violation for electroweak baryogenesis. 

\subsection{Making Progress}
If the bubble walls are thick,
conditions remain close to thermal equilibrium during the passage
of the wall, and the nonequilibrium physics can be captured by
assigning nonzero chemical potentials to
various quantum numbers including baryon number.  As I have mentioned,
analytic estimates
for the BAU produced in this setting exist in the 
literature \cite{{DHSSi},{TZiii},{CKN3i},{DHSSii},{CKN3ii},{CKN3iii}}
and the need for a numerical treatment is not pressing.
If the bubble walls are thin, however, or if (as is no doubt the case)
they are comparable in thickness to other length scales in the problem,
the situation is unsettled and it seems that a large scale numerical 
treatment is necessary.

\citeasnoun{M&Tii 97} 
have recently taken a big step in this direction.
They have performed $3+1$ dimensional classical 
simulations in which a bubble wall
moves through a box converting the high temperature phase
to the low temperature phase.  To date, they have focused more
on computing quantities like the wall thickness, the wall velocity,
the surface tension, and the drag on the wall and have
only begun their treatment of local electroweak baryogenesis.
To this point, they have introduced CP violation only by
``mocking up'' the effects of (\ref{newop}) by first computing
the average wall profile $\langle \sigma \rangle (z)$ for an ensemble
of walls, and then doing a simulation in which 
one measures the distance of a given point to the nearest bubble
wall and adds 
a chemical potential for Chern-Simons number at that point
proportional to the spatial derivative of the average wall profile
at that distance.  This chemical potential is only nonzero on the
wall, as it would be if it were proportional to $d(\sigma^2)/dt$
for a moving wall.  
Nevertheless, by imposing the chemical potential
as an external driving force instead of simply introducing (\ref{newop})
in the Lagrangian and letting the dynamics do their thing
self-consistently, one
risks missing a lot of the difficulties (and potential effects)
discussed earlier.
The simulations of Moore and Turok suggest that 
a large scale numerical assault on the problem of local
electroweak baryogenesis is now possible.

\section{Nonlocal Baryogenesis}
\label{nonlocal}
If CP violation leads to an asymmetry in a quantum number other than
baryon number and this asymmetry is subsequently transformed into a baryon
excess by sphaleron effects in the symmetric phase, then this process is
referred to as {\it nonlocal} baryogenesis. 
Nonlocal baryogenesis typically involves the interaction of the 
bubble wall with the various fermionic species in the unbroken phase.
The main picture is that as a result of CP violation in the bubble wall, 
particles with opposite chirality interact differently with the wall,
resulting in a net injected chiral flux. This flux
thermalizes and diffuses into the unbroken phase where it is
converted to baryons.
In this section, for definiteness when describing these 
effects, I shall assume that the CP violation arises because of a two-Higgs
doublet structure. There are typically two distinct calculational regimes
that are appropriate for the treatment of nonlocal effects in electroweak
baryogenesis. Which regime is appropriate depends on the particular fermionic
species under consideration.

\begin{enumerate}
\item
The thin wall regime \cite{{CKN2i},{CKN2ii},{JPT2 94}}:
If the mean free path $l$ of the fermions being considered is much greater
than the thickness $\delta$ of the wall, i.e. if

\be
\frac{\delta}{l}<1 \ ,
\ee 
then we may neglect scattering effects and
treat the fermions as free particles in their interactions with the wall.   
\item 
The thick wall regime: If the mean free path of the fermions is of the
same order or less than the wall thickness, then scattering effects become
important and the non-interacting picture is no longer applicable. 
Here there are two effects,
classical force baryogenesis \cite{JPT3 94} and non-local 
spontaneous baryogenesis \cite{{JPT3 94},{CKN 94}}. 
In the former scenario, as a result of CP  
violation, an axial field emerges on the wall leading to a classical
force which perturbs particle densities, thus biasing baryon number.
In the latter, hypercharge violating
processes in the presence of an axial field on the wall are
responsible for perturbing particle densities in a CP violating
manner. When the effects of particle transport are taken into account,
both cases give rise to nonlocal baryogenesis. 
\end{enumerate}
Both these mechanisms lead to an increase in the net effective volume 
contributing to baryogenesis over that of local baryogenesis since it is no 
longer necessary to rely on anomalous interactions taking place in the 
narrow region of the face of the wall where the changing Higgs fields 
provide CP violation.

The chiral asymmetry which is converted to an asymmetry in baryon
number is carried by both quarks and leptons. However, the
Yukawa couplings of the top quark and the $\tau$-lepton are larger
than those of the other quarks and leptons respectively. Therefore, it is
reasonable to expect that the main contribution to the injected asymmetry 
comes from these particles and to neglect the effects of the 
other particles (for an alternative scenario see \citeasnoun{DRW 97}).   

When considering nonlocal baryogenesis it is convenient to write the 
equation for the rate of production of baryons in the form \cite{JPT2 94}

\begin{equation}
\frac{d n_B}{dt} = -\frac{n_f\Gamma(T)}{2T}\sum_i\mu_i \ ,
\label{nonlocal B}
\end{equation}
where the rate per unit volume for electroweak sphaleron transitions 
is given by~(\ref{unbrokenrate}).
Here, $n_f$ is again the 
number of families and $\mu_i$ is the chemical potential for left 
handed particles of species $i$. The crucial question in applying this 
equation is an accurate evaluation of the chemical potentials that bias 
baryon number production.

\subsection{Thin Bubble Walls}
Let us first consider the case where the Higgs fields change only in a 
narrow region at the face of the bubble wall. We refer to this as 
the {\it thin wall} case.
In this regime, effects due to local baryogenesis are heavily suppressed 
because CP violating processes take place only in a very small volume in 
which the rate for baryon violating processes is non-zero. However, in
contrast, we 
shall see that nonlocal baryogenesis produces an appreciable 
baryon asymmetry due to particle transport effects
\cite{{CKN2i},{CKN2ii},{JPT2 94}}.

\begin{figure}
\centerline{\leavevmode\epsfysize=10cm \epsfbox{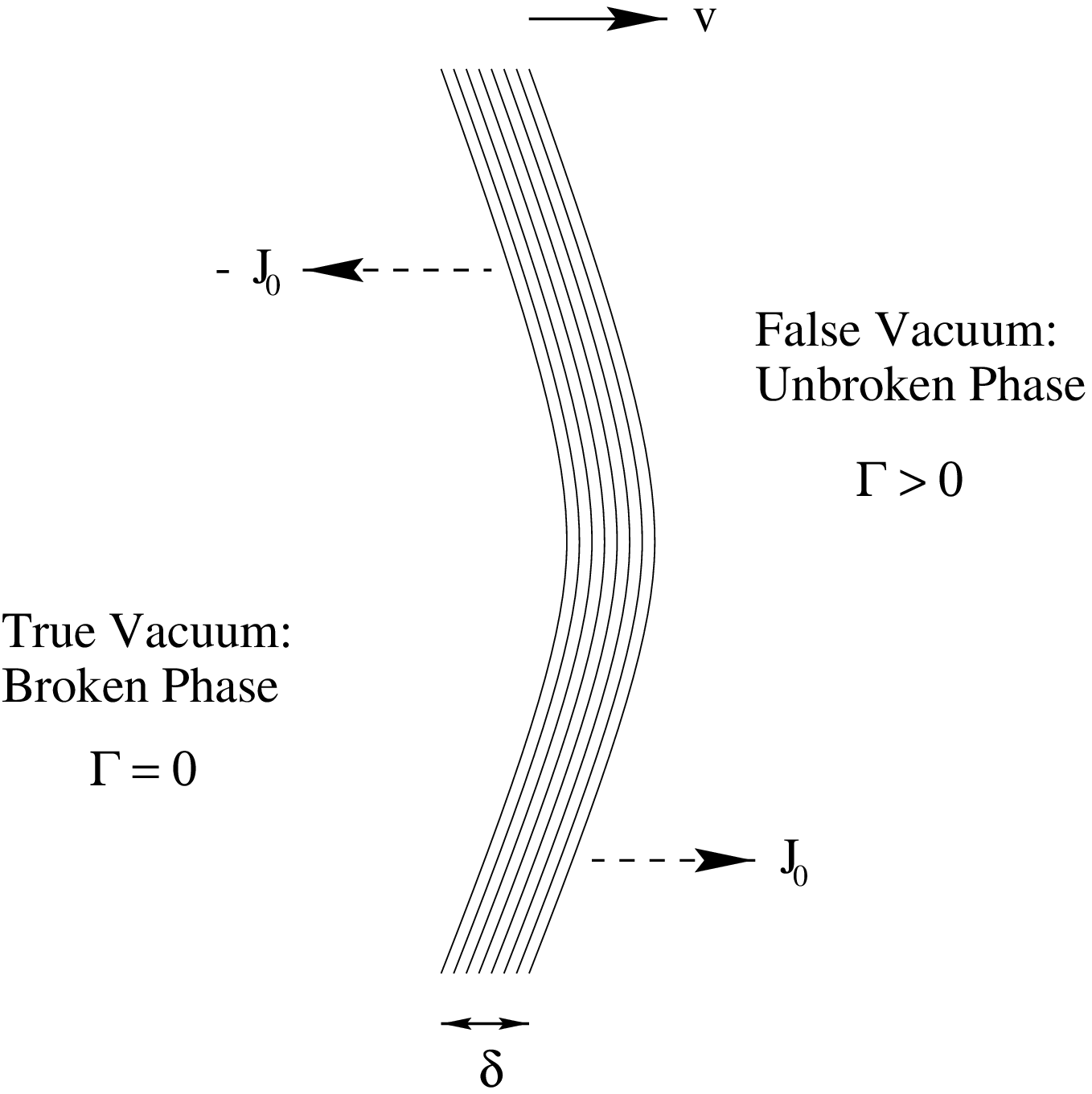}}
\vspace{0.3in}
\caption{}
\label{nonloc}
\end{figure}

In the rest frame of the bubble wall, particles see a sharp potential barrier 
and undergo CP violating interactions with the wall due to the gradient in the 
CP odd Higgs phase. 
As a consequence of CP violation, there will be asymmetric reflection
and transmission of particles, thus generating an injected current
into the unbroken phase in front of the bubble wall. As a consequence of this
injected current, asymmetries in certain quantum numbers will diffuse
both behind and in front of the wall due to particle
interactions and decays \cite{{CKN2i},{CKN2ii},{JPT2 94}}. In particular, the
asymmetric reflection and transmission of left and right handed
particles will lead to a net injected chiral flux from the wall (see figure
\ref{nonloc}).
However, there is a qualitative difference between the diffusion
occurring in the interior and exterior of the bubble.        
 
Exterior to the bubble the electroweak symmetry is restored and
weak sphaleron transitions are unsuppressed. This means that the
chiral asymmetry carried into this region by transport of the injected
particles may be converted to an asymmetry in baryon number by
sphaleron effects. In contrast, particles injected into the phase of
broken symmetry interior to the bubble may diffuse only by baryon number 
conserving decays since the electroweak sphaleron rate is exponentially 
suppressed in this region. Hence, I shall concentrate only on those particles
injected into the unbroken phase.

The net baryon to entropy ratio which results via nonlocal
baryogenesis in the case of thin walls has been calculated in several
different analyses \cite{{CKN2i},{CKN2ii}} and \cite{JPT2 94}. 
In the following I shall give a brief outline of the logic of the calculation, 
following \citeasnoun{JPT2 94}. 
The baryon
density produced is given by~(\ref{nonlocal B}) in
terms of the chemical potentials $\mu_i$ for left handed particles. 
These chemical potentials are a consequence of the asymmetric reflection and
transmission off the walls and the resulting chiral particle asymmetry.       
Baryon number
violation is driven by the chemical potentials for left handed leptons
or quarks. To be concrete, I shall focus on leptons \cite{JPT2 94} (for  
quarks see e.g. \citeasnoun{CKN2i}). If there is local thermal
equilibrium in front of the bubble walls - as I am assuming - then the
chemical potentials $\mu_i$ of particle species $i$ are related to
their number densities $n_i$ by
   
\begin{equation}  
n_i = {{T^2} \over {12}} k_i \mu_i \ ,
\end{equation}
where $k_i$ is a statistical factor which equals $1$ for fermions and $2$ 
for bosons. In deriving this expression, it is important \cite{JPT1 94}
to correctly impose the 
constraints on quantities which are conserved in the region in front of and 
on the wall.

Using the above considerations, the chemical potential $\mu_L$ for left 
handed leptons can be related to the left handed lepton number densities 
$L_L$. These are in turn determined by particle transport. The source term 
in the diffusion equation is the flux $J_0$ resulting from the 
asymmetric reflection and transmission of left and right handed leptons 
off the bubble wall.

For simplicity assume a planar wall. 
If $\vert p_z \vert$ is the momentum of the lepton perpendicular to 
the wall (in the wall frame), the analytic approximation used by
\citeasnoun{JPT2 94} allows the asymmetric reflection coefficients for 
lepton scattering to be calculated in the range

\be
m_l < \vert p_z \vert < m_H \sim {1 \over \delta} \ ,
\label{range}
\ee
where $m_l$ and $m_H$ are the lepton and Higgs masses, respectively, 
and results in

\begin{equation}
{\cal R}_{L \rightarrow R} - {\cal R}_{R \rightarrow L} \simeq 2 
\Delta \theta_{CP} {{m_l^2} \over {m_H \vert p_z \vert}} \ .
\label{reflection}
\end{equation}
The corresponding flux of left handed leptons is

\begin{equation}
J_0 \simeq {{v m_l^2 m_H \Delta \theta_{CP}} 
\over {4 \pi^2}} \ .
\end{equation}
Note that in order for 
the momentum interval in~(\ref{range}) to be non-vanishing, the condition 
$m_l \delta < 1$ needs to be satisfied.

The injected current from a bubble wall will lead to a ``diffusion tail" of
particles in front of the moving wall. In the approximation in which  the
persistence length of the injected current is much larger than
the wall thickness we may to a good approximation model it as a delta 
function source and search for a steady state solution. In 
addition, assume that the 
decay time of leptons is much longer than the time it takes for a
wall to pass so that we may neglect decays. Then the diffusion equation
for a single particle species becomes  

\begin{equation}
D_L L_L^{\prime \prime} + v L_L^{\prime} = 
\xi_L J_0 \delta (z) \ ,
\label{diffusion}
\end{equation}
where $D_L$ is the diffusion constant for leptons  and a
prime denotes the spatial derivative in the direction $z$   
perpendicular to the wall. This equation contains a parameter $\xi^L$ 
that is called the {\it persistence length} of the current in front of 
the bubble wall. This describes, and contains all uncertainties about, 
how the current thermalizes in the unbroken
phase. Equation~(\ref{diffusion}) can be immediately integrated
once, with the integration constant specified by the boundary condition

\be
\lim_{|z|\rightarrow\infty} L_L(z) = 0 \ .
\ee
This leads easily to the solution 

\begin{equation}
L_L(z) = \left\{ \begin{array}{ll}
          J_0 {{\xi_L} \over {D_L}} e^{- \lambda_D z}  & \ \ \ \ z>0 \\
          0 & \ \ \ \ z<0
         \end{array} \right. \ ,
\end{equation}
with the diffusion root 

\be
\lambda_D = \frac{v}{D_L} \ . 
\ee
Note that in this approximation the injected current does not generate
any perturbation behind the wall. This is true
provided $\xi^L \gg \delta$ is satisfied. If this  
inequality is not true,  the problem becomes significantly more complex
\cite{JPT2 94}. 

In the massless approximation the chemical potential $\mu_L$ can be
related to $L_L$ by 

\begin{equation}
\mu_L = { 6 \over {T^2}} L_L
\end{equation}
(for details see \citeasnoun{JPT2 94}).
Inserting the sphaleron rate and the above results for the chemical potential 
$\mu$ into~(\ref{nonlocal B}), the final baryon to entropy ratio becomes

\begin{equation}
\frac{n_b}{s} = \frac{1}{4 \pi^2} \kappa 
\alpha_W^4 (g^*)^{-1}
\Delta \theta_{CP} \left(\frac{m_l}{T}\right)^2 
\frac{m_H}{\lambda_D} \frac{\xi^L}{D_L} \ .
\end{equation} 
The diffusion constant is proportional to $\alpha_W^{-2}$ 
(see Ref. \cite{JPT2 94}):

\begin{equation}
{1 \over {D_L}} \simeq 8 \alpha_W^2 T \ .
\end{equation}
Hence, provided that sphalerons do not equilibrate in the diffusion tail,

\begin{equation}
\frac{n_b}{s} \sim 0.2\, \alpha_W^2 (g^*)^{-1} \kappa
\Delta \theta_{CP} \frac{1}{v} \left(\frac{m_l}{T}\right)^2 
\frac{m_H}{T} \frac{\xi^L}{D_L} \ .
\end{equation}
Since we may estimate that

\be
\frac{\xi^L}{D_L} \sim \frac{1}{T \delta} \ , 
\ee
the baryon to 
entropy ratio obtained by nonlocal baryogenesis is proportional to 
$\alpha_W^2$  and not $\alpha_W^4$ as in the result for local 
baryogenesis.   

Now consider the effects of top quarks scattering off the
advancing wall \cite{{CKN2i},{CKN2ii}}.  Several effects tend to decrease
the contribution of the top quarks relative to that of tau leptons.
Firstly, for typical wall thicknesses the thin wall approximation does not
hold for top quarks. This is because top quarks are much more strongly
interacting than leptons and so have a much shorter mean free path. 
An important effect is that the diffusion tail is cut
off in front of the wall by {\it strong sphalerons} \cite{{M&Z},{GS 94}}. 
There is an anomaly in the quark axial vector current
in QCD. This leads to chirality non-conserving processes at high 
temperatures. These processes are relevant for nonlocal baryogenesis 
since it is the chirality
of the injected current that is important in that scenario. In an analogous
expression to that for weak sphalerons, we may write the rate per unit volume 
of chirality violating processes due to strong sphalerons in the unbroken 
phase as

\be
\Gamma_s = \kappa_s(\alpha_s T)^4 \ ,
\label{strongsphaleron}
\ee
where $\kappa_s$ is a dimensionless constant \cite{GM 97}. Note that the
uncertainties in $\kappa_s$ are precisely the same as those in $\kappa$
defined in (\ref{unbrokenrate}). As such, $\kappa_s$ could easily be
proportional to $\alpha_s$, in analogy with
(\ref{newunbroken}), perhaps with a logarithmic correction.
These chirality-changing processes damp the effect of
the injected chiral flux and effectively cut off the diffusion tail in front
of the advancing bubble wall. Second, the diffusion
length for top quarks is intrinsically smaller than that for tau leptons, 
thus reducing the volume in which
baryogenesis takes place.  Although there are also enhancement factors, e.g.
the ratio of the squares of the masses $m_t^2/m_{\tau}^2$, it seems
that leptons provide the dominant contribution to nonlocal 
baryogenesis.

\subsection{Thick Bubble Walls}
If the mean free path of the fermions 
being considered is smaller than the width of the wall, we refer to the
thick wall, or adiabatic, limit, and the analysis is
more complicated. In the
case of thin bubble walls, the plasma within the walls undergoes a sharp
departure from equilibrium. When the walls are thick, however, most 
interactions within the wall will be almost in thermal equilibrium. The
equilibrium is not exact because some interactions, in particular
baryon number violation, take place on a time scale much slower than the
rate of passage of the bubble wall. These slowly-varying quantities are
best treated by the method of chemical potentials. 
For nonlocal baryogenesis, it is still useful to follow
the diffusion equation approach. However, whereas in the thin wall case it was
sufficient to model the source as a $\delta$-function, here we must
consider sources which extend over the wall. There are a number of possible
sources.

To be definite, consider the example where CP violation is due to a
CP odd phase $\theta$ in the two-Higgs doublet model. 

\subsubsection{Spontaneous Baryogenesis}
Let us begin by describing how the dynamics of $\theta$ might bias baryon 
number production. 
In order to explicitly see how $\theta$ couples to the 
fermionic sector of the theory (to produce baryons) we may remove the 
$\theta$-dependence of
the Yukawa couplings arising from the Higgs terms. We do this by
performing an anomaly-free hypercharge rotation on the fermions
\cite{CKN4i}, inducing a term in the Lagrangian density of the form

\be
{\cal L}_{CP} \propto \partial_{\mu}\theta \left[
\frac{1}{6} {\bar U}_L \gamma^{\mu}U_L +
\frac{1}{6} {\bar D}_L \gamma^{\mu}D_L +
\frac{2}{3} {\bar U}_R \gamma^{\mu}U_R -
\frac{1}{3} {\bar D}_R \gamma^{\mu}D_R -
\frac{1}{2} {\bar l}_L \gamma^{\mu}l_L -
{\bar E}_R \gamma^{\mu}E_R \right] \ .
\ee
Here $U_R$ and $D_R$ are the right handed up and down quarks 
respectively, $l_L$ are the left handed leptons and $E_R$ are the right
handed charged leptons. The quantity in the square brackets is proportional
to the fermionic part of the hypercharge current, and
therefore changes in $\theta$ provide a preferential direction for 
the production of quarks and leptons; in essence a chemical potential for
baryon number $\mu_B$ similar to that described in (\ref{1loop}).

Of course, strictly speaking, this is not a chemical potential since it 
arises dynamically rather than to impose a constraint. For this reason,
the quantity $\mu_B$ is sometimes referred to as a ``charge potential''.
The effect of the charge potential is to split the otherwise degenerate
energy levels of the baryons with respect to the antibaryons. This results
in a free energy difference which we may feed into~(\ref{balance}) to
obtain. 

\be
\frac{dn_B}{dt} = -9\frac{\Gamma(T)}{T} \mu_B \ .
\ee
The relevant baryon number produced by spontaneous baryogenesis is then
calculated by integrating this equation. 

Initially, spontaneous baryogenesis was considered as an
example of local baryogenesis. However, it has become clear that diffusion
effects can lead to an appreciable enhancement of the baryon asymmetry 
produced by this mechanism \cite{CKN 94}, and this effect has been
investigated by a number of authors \cite{{MJ 94},{CPR 95}}. In addition,
it has been shown that the final result for the baryon to entropy ratio must 
be suppressed by a factor of $m^2/T^2$ where $m$ is the relevant fermion
mass \cite{CPR 95}. 
As we saw in the case of thin walls, the major part of the work is to use
the diffusion equation to calculate the number densities of particle species,
and then to relate these densities to the chemical potentials. In the
case of thick walls the situation is complicated since one must consider 
extended sources for the appropriate diffusion equations. Nevertheless, the
calculations can be performed and I will sketch one such approach as part of 
the next subsection.

\subsubsection{Classical Force Baryogenesis}
An alternative way of generating a source for the diffusion equation was
suggested by \citeasnoun{JPT3 94}. The essential idea is that there exists 
a purely classical chiral force that acts on particles when there is a CP 
violating field on the bubble wall. I will briefly sketch this idea here.
The Lagrangian for a fermion $\psi$ in the background of a bubble wall in the
presence of the CP-odd phase $\theta$ is

\be
{\tilde {\cal L}} = i\bar{\psi}\gamma^{\mu}\left(\partial_{\mu}
+\frac{i}{2}\frac{v_2^2}{v_1^2 + v_2^2}\gamma^5\partial_{\mu}\theta\right)
\psi -m\bar{\psi}\psi \ ,
\ee
where $v_1$ and $v_2$ are the VEVs of the Higgs fields and $m$ is the fermion
mass. As usual we assume a planar wall propagating at speed $v$ in the 
$z$-direction. Further, if interactions of the Higgs 
and gauge fields with the wall have reached a stationary state, then these
fields are functions only of $z-vt$. Then, in the wall rest frame plane-wave
fermion solutions $\psi \propto e^{-ip.x}$ have the dispersion relation

\be
E=\left[p_{\perp}^2+\left(\sqrt{p_z^2+m^2}\pm\frac{1}{2}
\frac{v_2^2}{v_1^2 + v_2^2}\partial_z\theta\right)^2\right]^{1/2} \ ,
\label{dispersion}
\ee
where $\pm$ corresponds to $s_z=\pm 1/2$, with $s_z$ the $z$-component of
the spin. 

In the WKB approximation, the Hamilton equations derived from this 
relation allow calculation of the acceleration of a WKB wave-packet (i.e.
the chiral force). This force acts as a potential well that induces an
excess of chiral charge on the wall. In the diffusion tail in front of the 
wall, there exists a corresponding deficit of chiral charge. This acts
as a chemical potential which sources the production of baryon number.

This calculation is performed by solving the associated Boltzmann equation,

\be
\frac{\partial f}{\partial t} + \dot{z}\frac{\partial f}{\partial z}
+\dot{p}_z \frac{\partial f}{\partial p_z} = -C(f) \ ,
\ee
in which the collision integral $C(f)$ accounts for the fact that in the 
thick wall regime we may no longer use the non-interacting picture, and 
$\dot{z}$ and $\dot{p}_z$ are obtained from the Hamilton equations.

\citeasnoun{JPT3 94} follow a fluid approach to solving this equation. By
concentrating on particles with $|p_z|\sim T \gg m$, they may treat the 
particle and antiparticle excitations (\ref{dispersion}) as separate fluids
in the WKB approximation. This permits an analytic solution to the diffusion
equation which yields a relation between the chemical potential on the wall
for the fermion species considered and the chiral force as

\be
\mu_{\psi} = \frac{2\ln(2)}{3\zeta(3)} \frac{v}{2}
\frac{v_2^2}{v_1^2 + v_2^2}\partial_z\theta \left(\frac{m}{T}\right)^2 \ ,
\ee
where $\zeta$ is again the Riemann function. 
Integrating this in front of the wall yields a chemical potential that may
then be fed into (\ref{nonlocal B}) to yield an approximate answer for the BAU.

\subsection{Summary and Making Progress}
The advantage of nonlocal baryogenesis over local baryogenesis is that the
effective volume in which baryon number generation can occur is enhanced in
the former case. this is because the effects of transport in the plasma 
external to the expanding bubble allow baryon violating transitions in the 
unbroken phase to transform a chiral asymmetry produced on the wall into a 
baryon asymmetry. This means that in the case of nonlocal baryogenesis we 
do not need to rely on baryon number violating processes occurring in the
region where the Higgs fields are changing.

The diffusion equation approach to the problem of nonlocal baryogenesis has 
been very successful. In the thin wall case, it is a valid approximation
to assume that the source for the diffusion equation is essentially a
$\delta$-function on the wall. This is because one may ignore the effects
of particle scattering in this picture. However, in the case of thick
walls, significant particle scattering occurs and as a result it is necessary
to consider sources for the diffusion equations that extend over the wall.

New approaches to this scenario continue to appear as we try to understand
the detailed predictions of specific extensions of the standard model to
compare with upcoming accelerator tests \cite{{ERV 97},{AR1 97},{AR2 97},{CJK 97}}.

\section{A Realistic Model of EWBG: the MSSM}
\label{MSSM}
As we have seen, if we are to produce enough baryons at the electroweak phase
transition, we must go beyond the minimal Glashow-Salam-Weinberg theory. 
This is true both to obtain strong enough CP violation and to ensure
a sufficient departure from thermal equilibrium. It has been common
to invoke general two-Higgs doublet models to satisfy these conditions
\cite{{BKS 90},{ABV 92},{RP 92},{DFJM 94},{Losadaii 96}}
and the behavior of the phase transition has most recently been investigated
in that context by \citeasnoun{C&L 97}. Here 
I shall focus on a more restrictive model with good particle physics 
motivations - the minimal supersymmetric standard model (MSSM).

Supersymmetry requires that each fermion must have a bosonic superpartner and 
vice-versa. The essential feature of the MSSM, therefore, is that it contains 
twice the particle content of the MSM plus an extra Higgs doublet 
(and superpartner). Naturally, introducing a host of new free parameters
into the theory allows us to relax the constraints which were derived for
the MSM. However, many of the new parameters are constrained by supersymmetry
and by existing accelerator measurements. 

\subsection{The Electroweak Phase Transition in the MSSM}
A strongly first order phase transition can result if the theory contains
light scalars, which are strongly coupled to the Higgs field.
The MSSM contains two Higgs doublets $H_1$ and $H_2$, one linear combination
of which is a CP even Higgs boson $h$.  The mass $m_h$ of this particle
is constrained 
to be less than of order $125\,$GeV \cite{{CQW 96},{HHH 97}}. Further,
the superpartners of the top quark, the {\it stops} $\tilde{t}$, couple to 
the Higgs with strength of order the top quark Yukawa coupling. These light 
scalars are precisely the particles necessary to enable the phase transition 
to be more strongly first order than in the MSM 
\cite{{G 92},{BEQZ 94},{D& 96}}. 
In fact, for their effects to be useful, it is necessary that the lightest
stop be no heavier than the top quark itself (see, for example, 
\citeasnoun{E 96}).

The phase transition in supersymmetric electroweak theories has been
investigated very recently 
\cite{{MLi 96},{Losadai 96},{F&L 96},{Losadaii 96},{C&K 96},{BJLS 97},{C&K 97},{CQW 97}}. In addition to the
parameters mentioned above, the other important quantity is

\be
\tan\beta \equiv \frac{\langle H_2 \rangle}{\langle H_1 \rangle} \ ,
\label{tanbeta}
\ee
the ratio of the vacuum expectation values of the Higgs fields.
The methods available to estimate the region of parameter space in which
EWBG is possible in the minimal standard model can also be applied
here.  However, 
since there is more structure in the MSSM, the calculation is more involved.
Nevertheless, the calculation of the allowed region of the 
$m_h$-$m_{\tilde t}$ space has been performed using the finite temperature
effective potential computed up to all finite temperature two-loop
corrections \cite{CQW 97}. The two-loop contributions are crucial to the
accuracy of this calculation and yield the value

\be
\tan \beta \simeq 2 
\ee
and the constraints

\bea
75 {\rm GeV}\leq & m_h & \leq 105 {\rm GeV} \\
100 {\rm GeV}\leq & m_{\tilde t} & \leq m_t \ .
\label{MSSMconstraints}
\eea
These results have been confirmed by lattice numerical simulations 
by \citeasnoun{L&Ri 98}.
In the ranges (\ref{MSSMconstraints}) the phase transition in the MSSM is 
strong enough that washout of the baryon asymmetry produced is avoided.

In quoting these bounds, there is an issue I have neglected. It is possible to 
choose values of the parameters in the MSSM such that the absolute energy 
minimum in field space is a vacuum which breaks the color symmetry. Although
this would normally be forbidden, it is possible that the vacuum in which we
live is metastable with an extremely long lifetime, and hence is consistent
with observations. If we allow such parameter values, then the above 
constraints may be relaxed a little. Here I have chosen to adopt the most
conservative assumption, namely that the color preserving vacuum state is
the global minimum at all temperatures.

\subsection{Extra CP Violation in the MSSM}
In the MSSM there are extra sources of CP violation beyond that contained
in the CKM matrix of the standard model. 
First, when supersymmetry breaking occurs, as we know it must, 
the interactions of the Higgs fields
$H_1$ and $H_2$ with charginos and neutralinos at the one-loop level,
leads to a CP violating contribution to the scalar potential of the form

\be
V_{CP} = \lambda_7(H_1 H_2)^2 +\lambda_8|H_1|^2 H_1 H_2 +
\lambda_9|H_2|^2 H_1 H_2 + (h.c.) \ .
\label{MSSMCP}
\ee
The interaction of the Higgs fields with the stops is a potential further 
effect which contributes to the CP violating part of the potential.
However, in the range~(\ref{MSSMconstraints}) of stop masses required to 
maintain a strong enough phase transition, this effect is suppressed. The
nature of supersymmetry breaking can lead to the parameters being complex

\bea
\lambda_7 & = & |\lambda_7| e^{2i\alpha} \ , \\
\lambda_8 & = & |\lambda_8| e^{i\alpha} \ , \\
\lambda_9 & = & |\lambda_9| e^{i\alpha} \ , 
\label{CPphase}
\eea
The phase $\alpha$ breaks CP and can bias baryon production in a way 
similar to that described for the two-Higgs model.

Second, in SUSY extensions of the MSM, there exists a mass mixing matrix
for the charginos. This mass matrix has a similar structure to the quark
mixing matrix in the MSM. In particular, the chargino mixing matrix contains
a phase $\phi_B$ that parameterizes CP violation in this sector. Similarly,
there exists a mixing matrix for the neutralinos, which also contains this 
same phase. Finally, there is a further mass mixing matrix for the top 
squarks. This contains a phase which is a linear combination of $\phi_B$ and
a second phase $\phi_A$. These extra phases can provide enough CP violation
to be useful for baryogenesis \cite{HNii}. However, in some regimes it is 
possible to constrain their sizes through dipole moment calculations such as I 
described earlier.

The electroweak phase transition and CP violation are not the only 
features of electroweak
baryogenesis that change when one considers supersymmetric models. It 
is necessary to reanalyze the reflection of particles from bubble walls
within the context of these theories, and to reexamine the ranges of
validity of the various approximations used to obtain estimates for the
BAU. Serious attempts to do just this have only been performed in the
last two years 
\cite{{DFM 96},{AR1 97},{AR2 97},{CJK 97},{CQRVM 97},{MW 97},{toni}}.

The regions of parameter space I have mentioned above are still experimentally 
allowed. Thus, the MSSM remains a viable candidate theory to explain the BAU.
It is quite possible that the full parameter range will be 
covered in the next decade. At that time we should know whether MSSM
baryogenesis explains the BAU or whether we must turn to more complicated 
models of the electroweak scale.

\section{Topological Defects and the Departure from Thermal Equilibrium}
\label{defects}
\begin{figure}
\centerline{\epsfbox{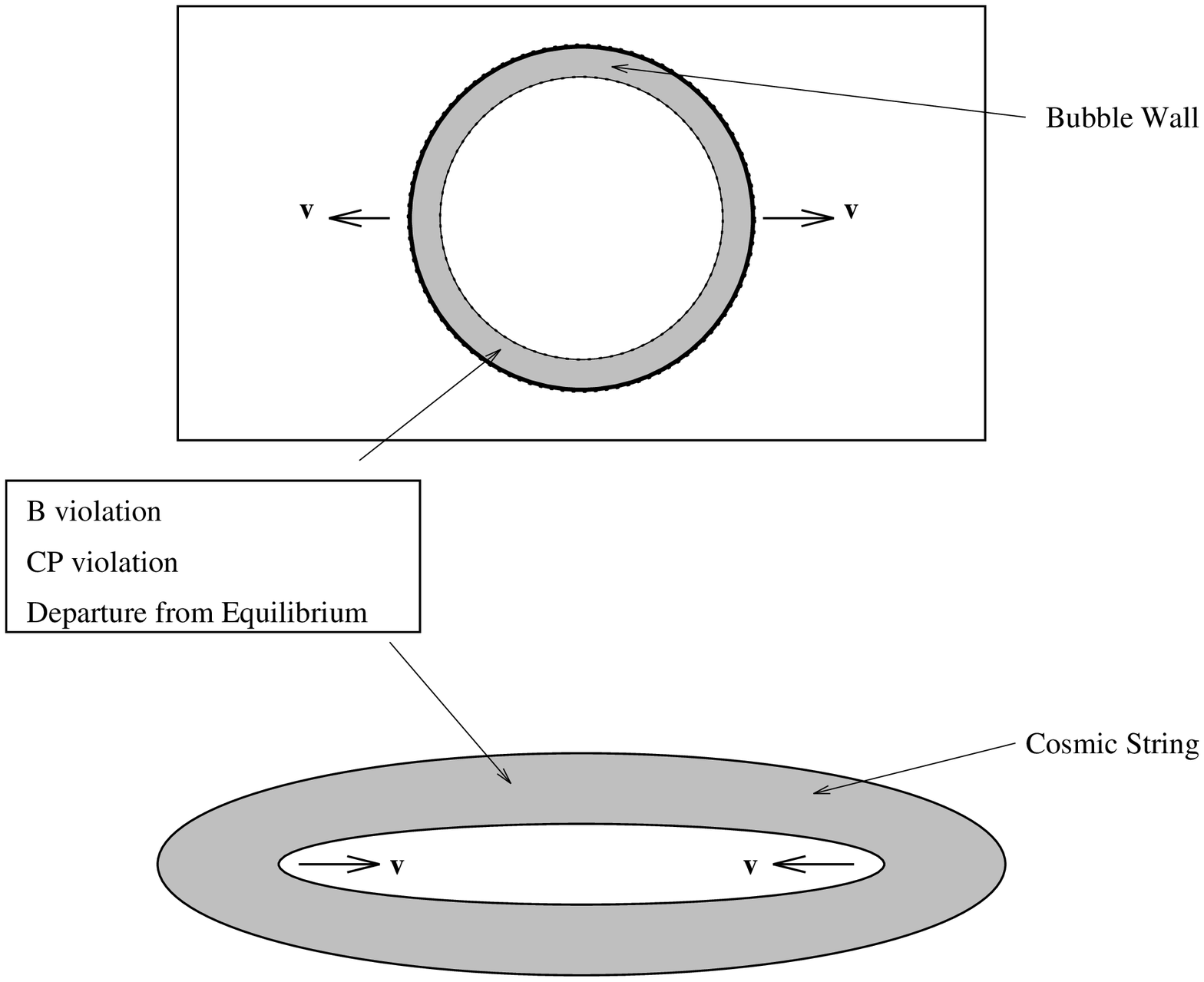}}
\vspace{0.3in}
\caption{}
\label{defect}
\end{figure}

If the dynamics of the electroweak phase transition is such that the
traditional scenarios for EWBG that I have described are inefficient, then
it is interesting to explore alternative implementations of 
baryogenesis at the electroweak scale. In this section, I describe a
particular realization of the third Sakharov condition suggested
by myself and collaborators 
\cite{{DMEWBGii},{DMEWBGv},{DMEWBGiii},{DMEWBGiv}}.
This implementation uses the out of equilibrium
evolution of a network of topological defects to realize the third 
Sakharov criterion, instead of the evolution of bubble walls (see figure
\ref{defect}).
Different
scenarios making use of topological defects formed at that temperature have
been suggested by other authors \cite{{TV 94},{MB 95}}.

Topological defects are regions of trapped energy density which can
remain after a cosmological phase transition if the topology of the 
vacuum of the theory is nontrivial. Typically, cosmological phase
transitions occur when a gauge symmetry of a particle physics theory
is spontaneously broken. In that case, the cores of the topological 
defects formed are regions in which the symmetry of the unbroken theory
is restored. There exist many excellent reviews of the physics of 
topological defects and I refer the reader to one of those, rather
than provide a lengthy aside here. For the purposes of this section, it
is sufficient to specialize to the case of cosmic strings; line-like,
solitonic solutions to spontaneously broken field theories, for which
there exist non-contractible loops in the vacuum manifold. If these objects
arise from the breakdown of a gauge symmetry, then they can have interesting
microphysics, as I will briefly describe.

\subsection{Electroweak Symmetry restoration and the Baryogenesis Volume}
Let us review the scenario for electroweak baryogenesis proposed in
\cite{DMEWBGii}. I shall begin by briefly explaining the physical principles 
behind electroweak symmetry restoration around ordinary 
(non-superconducting) defects. This is an essential ingredient of
defect-mediated electroweak baryogenesis.

For definiteness, consider a cosmic string, formed at a scale $\eta>\eta_{EW}$,
that couples to the Glashow-Salam-Weinberg model. Further, assume that the 
gauge fields 
corresponding to this higher symmetry scale acquire an extra mass at the 
electroweak scale. The simplest example is to introduce an extra gauged $U(1)$
symmetry, that breaks to the standard model

\be
SU(2)_L \times U(1)_Y \times U(1) \longrightarrow SU(2)_L\times U(1)_Y \ .
\ee
This is the model originally considered by \citeasnoun{PD 93}.
Let the string's scalar and gauge fields be $S$ (which breaks the symmetry) 
and $R_{\mu}$ respectively. The
coupling between the string and the electroweak sector is through $R_{\mu}$ 
and the electroweak Higgs field $\Phi$, in the covariant derivative

\be
D_{\mu}\Phi=\left(\partial_{\mu}- \frac{1}{2}ig{\bf \tau}.{\bf W}_{\mu} -
\frac{1}{2}ig'B_{\mu}-\frac{1}{2}ig''R_{\mu}\right)\Phi \ .
\ee
Since $\eta>\eta_{EW}$, we may consistently treat $R_{\mu}$ as a background
with Nielsen-Olesen characteristics. We now minimize the 
energy of this configuration. Writing the Higgs field in the unitary
gauge

\be
\Phi=(0,\vp(r))^T \ , 
\ee
the minimal energy configuration is achieved for

\begin{equation}
\vp(r)=\frac{\eta}{\sqrt{2}}\left\{\begin{array}{ll}
\left(\frac{r}{R_s}\right)^{a/2}
\hspace{1cm}r<R_s \\ 
1 
\hspace{2cm}r>R_s \end{array}
\right.\ ,
\end{equation}
where $a$ is a constant. With this ansatz the scale of electroweak 
symmetry restoration, $R_s$ is determined to be

\begin{equation}
R_s \sim \lambda^{-1/4} {\tilde G}^{-1/2}\eta_{\rm EW}^{-1}
\label{restore}
\end{equation}
where $\eta_{\rm EW}$ is the electroweak scale and ${\tilde G}^2=g^2+g'^2$.

Thus, the electroweak symmetry is restored out to the inverse electroweak 
scale around such a higher scale ordinary defect. If certain consistency 
conditions are satisfied (eg. sphalerons fit inside the defect so that their 
rate is not significantly suppressed \cite{WP 95})
then within this region, the rate of baryon number violation is
still given by~(\ref{unbrokenrate}) after the electroweak phase transition.

Once again, I shall assume that CP violation is due to 
a CP-odd
relative phase, $\theta$, between two electroweak Higgs doublets,
and that this phase changes by $\Delta\theta_{CP}$ during the transition from 
false to true vacuum, and by $ - \Delta\theta_{CP}$ in the reverse transition.
Thus, if $\Delta \theta_{\rm CP} >0$ for a
given process then baryon number is driven positive (an excess of baryons over
antibaryons is generated) and vice-versa.

As a defect moves, certain regions of the background space enter the core of
the defect - i.e. make the transition from true to false vacuum - while others
leave the core and make the transition from false to true vacuum. There
are certain types of motion of defects and evolutions of defect networks
that can provide
an asymmetry such that an overall baryon excess is created in the universe.

An important quantity which enters the calculation is the
suppression factor

\begin{equation}
\Lambda \sim \left(\frac{V_{\rm BG}}{V}\right) \ ,
\label{volsupp}
\end{equation}
where $V_{\rm BG}$ is the volume in which baryogenesis occurs and $V$ is the
total volume. $\Lambda$ is the factor by which defect-mediated baryogenesis is
weaker than baryogenesis with bubble walls. In the original work of 
\citeasnoun{DMEWBGii}, the processes responsible for 
the generation of the baryon asymmetry were purely local.
For a collapsing topological
defect, purely local baryogenesis restricts the baryogenesis volume to be the
initial volume of the defect because the effects of ${\dot \theta}>0$ on one
side of the string are cancelled by the effects of ${\dot \theta}<0$ on the
other. In fact, in order for $\Lambda$ not to be a prohibitively small
suppression, it is necessary that the
scale at which the defects are formed be extremely close to the 
electroweak scale.
For definiteness, we shall restrict ourselves to local baryogenesis
here, but note that
nonlocal effects lead to an appreciable enhancement, and allow the defects to
be formed at scales further above the electroweak scale.

\subsection{Local Baryogenesis and Diffusion in Defects}
In this section I shall obtain an estimate for the baryon asymmetry
produced by a topological defect as a consequence of local mechanisms.

As a topological defect passes each point in space a number density of
antibaryons is produced by local baryogenesis at the leading face of the
defect,
and then an equal number density of baryons is produced as the trailing edge
passes. Naively we would expect that these effects would cancel each other, so
that any time-symmetric motion of the defect, such as translation, 
would yield no net baryon asymmetry. Because of this reason, the analysis of 
\citeasnoun{DMEWBGii} 
was restricted the time-asymmetric motion of cosmic string loop
collapse. In that case, the cancellation effects led to the suppression
of the strength of the mechanism by the factor 

\be
\Lambda =\lambda^3 \left(\frac{\eta_{EW}}{\eta}\right)^3 \ .
\ee

However, this neglects an important effect and thus
underestimates the strength of the mechanism. The antibaryons produced at the
leading edge of the defect at a fixed point in space spend a time interval
$\tau$ inside the defect during which they may decay before the trailing edge
passes by and produces baryons at the same point. The core passage time $\tau$
is given by

\begin{equation}
\tau = \frac{L}{v_{\rm D}},
\end{equation}
where $L$ is the width of the defect and $v_{\rm D}$ is its velocity.

Thus, if $n_b^0$ is the number density of baryons (or antibaryons) produced at
either edge, we may estimate the net baryon asymmetry $B$ produced after the
defect has passed a given point once to be

\begin{equation}
B = n_b^0(1-e^{-{\bar{\Gamma}} \tau})
\label{almostnb}
\end{equation}
where ${\bar{\Gamma}}$ is the rate at which antibaryons decay and may be
related to the electroweak sphaleron rate by \cite{JPT2 94}

\begin{equation}
{\bar{\Gamma}} = 6n_f \frac{\Gamma}{T^3} = 6n_f \kappa \alpha_{\rm W}^4 T.
\label{bardec}
\end{equation}

The resulting average baryon number density $n_b$ can be estimated from
(\ref{almostnb}),
taking into account (\ref{bardec}) and the volume suppression 
(see (\ref{volsupp})):

\begin{equation}
n_b \simeq 3 n_f {{\Gamma} \over {T}} 
\mu {{\delta} \over {v_D}} (1 - e^{-{\bar \Gamma} \tau}) \Lambda,
\end{equation}
where $\delta$ is the thickness of the defect wall.
The derivative of $\theta_{CP}$ and $\delta / v_D$ combine to give $\Delta
\theta_{CP}$,
and hence the resulting net baryon to entropy ratio becomes

\begin{equation}
{{n_b} \over s} \simeq 4 \kappa \alpha_W^4 {g_*}^{-1} ({m \over T})^2 \Delta
\theta_{CP} {{V_{BG}} \over V}(1 - e^{- {\bar \Gamma} \tau}),
\end{equation}
where $g_*$ is the number of spin degrees of freedom which enters into the
equation for the entropy density. 

Note that, although there is a volume suppression factor, in some cases it is 
${\cal O}(v_{\rm D})$ because the defect network can sweep out that fraction 
of the total volume in one Hubble expansion time. That is, we are no
longer restricted to the initial volume of the defect, as we shall see shortly
in an example.
However, Even if $V_{\rm BG}/V \sim 1$ there is still a suppression of this 
mechanism over the usual bubble wall scenarios by the factor

\begin{displaymath}
(1-e^{-{\bar{\Gamma}} L/v_{\rm D}})
\end{displaymath}
This clearly distinguishes two cases. In the first case in which the defects
are ``thin", defined as $L<v_{\rm D}/{\bar{\Gamma}}$, there is a suppression
factor of approximately ${\bar{\Gamma}} L/v_{\rm D}$. However, if the 
defects are ``thick", $L>v_{\rm D}/{\bar{\Gamma}}$, then there is 
negligible suppression due to this effect.

Let us now examine how these conditions are related to the microphysical
parameters of the models. First consider non-superconducting defects. The
electroweak symmetry is restored out to a distance given by
equation~(\ref{restore}).
The defects are
considered ``thin" if the Higgs self coupling $\lambda$ satisfies

\begin{equation}
\lambda > \left(\frac{{\bar \Gamma}}{v_{\rm D}\eta_{\rm EW}}\right)^4
\frac{1}{{\tilde G}^2}
\end{equation}
and ``thick" otherwise. This quantity may be estimated by evaluating ${\bar
\Gamma}$ at the electroweak temperature, using 
${\tilde G}\sim{\cal O}(\frac{1}{30})$ and 
$v_{\rm D} \sim 0.1 -1$. This results in the condition 

\be
\lambda>10^{-23}-10^{-27} \ ,
\ee
an inequality which includes most of the parameter space
of the theory. Thus, one concludes that for the case of ordinary defects the
suppression factor ${\bar \Gamma} L/v_{\rm D}$ almost always applies.

In some cases, it is possible for the region of symmetry restoration
around a string to be enhanced. It was shown by \citeasnoun{Witten 85}
that some strings can carry supercurrents. In the simplest examples,
these currents are due to the presence of a scalar condensate field on
the string. The supercurrent is proportional to the winding of 
this condensate field. There is a large gauge flux associated with the
supercurrent and this flux, coupling to the electroweak Higgs
field, leads to an increased region of electroweak symmetry
restoration.
If, as in \citeasnoun{DMEWBGi},
we estimate the current on the defects by assuming a random walk of the winding
of the condensate field, then
we may estimate the size of the symmetry restoration region to be 
\cite{{ANO},{PD 93}}

\begin{equation}
R_s \sim \left(\frac{1}{2\lambda}\right)^{1/2}
\frac{1}{2\pi\eta_{\rm EW}}\left(
\frac{\eta}{\eta_{\rm EW}}\right)^{3/4}
\end{equation}
where $\eta$ is the scale at which the defects are formed. A similar result has
been shown to hold in the two-Higgs doublet model relevant here \cite{MT 94}.
Thus, in this case, the defects are considered ``thick" if

\begin{equation}
\eta >\left(\frac{v_{\rm D} \eta_{\rm EW}}{{\bar \Gamma}_s}
\sqrt{2\lambda}2\pi\right)^{4/3} \eta_{\rm EW}
\end{equation}
and ``thin" otherwise. Using $\lambda \sim 1$ and estimating $\Gamma$ from (9)
we obtain $\eta > 10^8 - 4 \cdot 10^{10}\,$GeV.
Therefore, if the scale of the defects is in this range then there is no
additional suppression beyond the volume suppression. If the scale lies below
this then the factor $\Gamma L/v_{\rm D}$ applies as in the case of ordinary
defects.

The above considerations allow the computation of the asymmetry in the 
baryon number
density at every point swept out by a topological defect of a given type. In
order to make a specific prediction one must consider a particular type of
defect in a given configuration and have knowledge of the evolution of the
defect network. This then provides a reliable estimate for the volume
suppression $\Lambda$ and hence the total baryon asymmetry. 

\subsection{A Specific Geometry and Examples}
First assume that the network is in the friction dominated epoch at 
$t_{\rm EW}$. In this case it is reasonable to make the approximation that 
all string loops have the
same radius. Also, note that for this example, the strings are
thin enough (for local baryogenesis) that the additional suppression
factor ${\bar \Gamma}_s L/v_{\rm D}$ mentioned above applies in both the
ordinary and superconducting cases (for a large range of Higgs self-coupling).
For nonlocal baryogenesis with defects, the suppression factor is 
linear in $L v_D / D$. 
Further, assume that there is one string loop per correlation volume at
formation, via the Kibble mechanism. In one horizon volume
the total volume taking part in baryogenesis is

\begin{equation}
V_{\rm BG} = R_s\xi(t)^2\left(\frac{t}{\xi(t)}\right)^3v_D \ ,
\end{equation}
where I have used the largest strings with radius equal to the correlation
length $\xi(t)$ and the last factor is the number of string loops per horizon
volume. Thus, dividing by the horizon volume $t^3$ yields the volume
suppression factor

\begin{equation}
\Lambda = \frac{V_{\rm BG}}{V} = \frac{R_s}{\xi(t)} \ .
\end{equation}
Using $\xi(t_f)\simeq \lambda^{-1}\eta^{-1}$ \cite{K 76} where $t_f$ is the
formation time of the string network and \cite{{KEHii},{KEHi},{KEHiii}}

\begin{equation}
\xi(t) \sim \xi(t_f)\left(\frac{t}{t_f}\right)^{5/4}
\end{equation}
gives

\begin{eqnarray}
\Lambda & = & \lambda\left(\frac{\eta_{\rm EW}}{\eta}\right)^{3/2}v_D\ \ \ \
\mbox{Ordinary Strings} \\
     & = & \lambda\left(\frac{\eta_{\rm EW}}{\eta}\right)^{3/4}v_D\ \ \ \
\mbox{Superconducting Strings}
\end{eqnarray}
These equations take into account only the dynamics during the first Hubble
expansion time after $t_{EW}$. In later expansion times, the density of strings
is diluted, and hence the above results are a good approximation of the total
effect of strings.

Now consider briefly the case where the strings are formed at a scale much 
higher
than the electroweak scale. If the strings are ordinary one still expects
the ${\bar \Gamma}_s L/v_{\rm D}$ suppression but for superconducting
defects we shall see that this is absent since the electroweak symmetry is
restored out to such a large radius that all the antibaryons may decay before
the baryons are created.

Focus again on string loops.  By the time of the electroweak phase
transition the defect network is well described by a
{\it scaling solution}. This solution is characterized by the fact that
the distribution of string loops looks the same when viewed on all
scales.
Quantitatively, the number density of string loops with
radii in the range $[R,R+dR]$ is given by \cite{{ZVi},{ZVii},{ZViii}}

\begin{equation}
n(R,t) = \left\{ \begin{array}{ll}
         \nu R^{-5/2} t^{-3/2}    & \ \ \ \mbox{$\gamma t < R < t$} \\
         \nu \gamma^{-5/2} t^{-4} & \ \ \ \mbox{$R < \gamma t$}
         \end{array} \right. \ ,
\end{equation}
where $\gamma \ll 1 $ is a constant determined by the strength of
electromagnetic radiation from the string.  Loops with radius $R=\gamma t$
decay in one Hubble expansion time.  In the above I have assumed that
electromagnetic radiation dominates over gravitational radiation.  If this is
not the case, then $\gamma$ must be replaced by $\gamma_g \, G \mu $, $\mu$
being the mass per unit length of the string $(\mu \simeq\eta^2)$ and 
\cite{{VVii},{VViii},{VVi}}
$\gamma_g \sim 100$. In other words, $\gamma$ is bounded from below
\begin{equation}
\gamma > \gamma_g G \mu\, .
\end{equation}

The suppression factor $\Lambda$ can be estimated by integrating over all the 
string loops present at $t_{\rm EW}$

\begin{equation}
\Lambda\simeq \pi \int_0^{\gamma t_{EW}} \, d R \, R^2\, R_s \, n \left( R,t_{EW}
\right) = \frac{\pi}{3} \, \nu\, \gamma^{1/2} \left(\frac{R_s}{t_{EW}}\right)
\ .
\end{equation}
Without superconductivity the suppression factor for GUT strings ($\eta =
10^{16}\,$GeV) is so small ($\sim 10^{-32}$) that the contribution is
negligible. However, for superconducting strings the suppression is

\begin{equation}
\Lambda \sim \nu \gamma^{1/2}\left(\frac{\eta}{m_{pl}}\right) \ ,
\end{equation}
so that the final baryon to entropy ratio generated by this mechanism is given
by

\begin{equation}
\frac{n_B}{s}=\frac{n^0_B}{s}\Lambda \ ,
\end{equation}
with $\Lambda$ given by the above and $n^0_B/s$ proportional to $\alpha_W^4$ 
for
local baryogenesis and to $\alpha_W^2$ for nonlocal baryogenesis.
Clearly, this lies below the observed value, and therefore, strings formed
at high energy scales are not viable candidates to mediate electroweak
baryogenesis.

\subsection{A Particle Physics Model}
The results above indicate that particle physics models which admit cosmic 
strings at or around the TeV scale are perhaps the best candidates to 
implement the defect mediated scenario. The
particle physics literature contains many such examples. Here I shall 
give just one.

The particular supersymmetric model I shall consider \cite{susy} has been
proposed as a
solution to the $\mu$-problem of the 
(MSSM) and the cosmological solar neutrino problem.

In the MSSM there exists a mixing term of the form

\begin{equation}
{\cal L}_{\mu} = \mu {\bar H}H \ ,
\label{muterm}
\end{equation}
where $H$ is the supersymmetric Higgs field.
In order to obtain radiative SUSY breaking at the weak scale it is necessary
that $\mu\sim {\cal O}(G_F^{-1/2})$ where $G_F$ is the Fermi constant. 
However, there is no natural scale in the MSSM to ensure that this is
the case.

In the model under consideration the MSSM is supplemented by two $U(1)$
symmetries. One of the extra $U(1)$'s breaks at a high scale ($\sim
10^{15}\,$GeV) and is concerned with the implementation of the MSW
\cite{{MSWii},{MSWi}}
solution of the solar neutrino problem via the seesaw mechanism. I
won't discuss that aspect of the model any further.
The $\mu$-term in this model is given in terms of a Yukawa coupling $\lambda'$
and a scalar $S$ which is a singlet under the standard model gauge group but
charged under the low energy extra $U(1)$. Thus the term (\ref{muterm}) is
forbidden since it is not invariant under the extra $U(1)$ symmetry 
and in its place we have a term

\begin{equation}
{\cal L}_{\mu} = \lambda' S{\bar H}H \ .
\end{equation}
Therefore if the low energy $U(1)$ breaks at a scale $\eta$ of the order of
1TeV then $S$ gets a VEV of this order and the $\mu$-problem is resolved.

Thus the symmetry breaking scheme of the model is

\begin{eqnarray}
SU(3)_c \times SU(2)_L \times U(1)_Y \times U(1) \times U(1) & 
\longrightarrow
&  SU(3)_c \times SU(2)_L \times U(1)_Y \times U(1) \nonumber \\
 & \stackrel{\eta}{\longrightarrow} &  SU(3)_c \times SU(2)_L \times U(1)_Y
\nonumber \\
 & \stackrel{\eta_{EW}}{\longrightarrow} &  SU(3)_c \times U(1)_{em} \ .
\end{eqnarray}
Clearly we obtain TeV scale ordinary cosmic strings from this 
final $U(1)$ breaking and this model would exhibit defect mediated 
electroweak baryogenesis.

\subsection{Summary}
I have described an alternative scenario to traditional electroweak 
baryogenesis, in which the out of equilibrium evolution of a network of
topological defects satisfies the third Sakharov condition.
The advantage of such a scenario for electroweak baryogenesis is that it
does not depend in any way on the order of the electroweak phase transition.
A further advantage of using topological defects to seed baryogenesis is that
the volume in defects decreases only as a power of time below the phase
transition temperature. Therefore, as pointed out by \citeasnoun{DMEWBGi},
defect-mediated baryogenesis remains effective even if sphalerons are in
thermal equilibrium just below the electroweak phase transition temperature.
This alleviates the problem of washout of the asymmetry.
However, a potential drawback for specific implementations is the
requirement that defects be formed at a scale rather close to the
electroweak scale in order to avoid large volume suppression factors (for
further constraints in the specific case of ordinary cosmic strings
see \cite{defectprobs}). 

One might worry that, since the baryon production in these models
occurs inhomogeneously, nucleosynthesis might proceed inhomogeneously, leading
to a conflict with observation. This possibility has been investigated
\cite{BDR 95} and, for defects formed at low enough temperatures that the
volume suppression factor is not prohibitively small, the baryons produced
are sufficiently homogeneous at the time of nucleosynthesis.

\section{Concluding Remarks and Looking to the Future}
\label{conclusions}
Modern particle cosmology consists of the union of the hot big bang model
with quantum field theories of elementary particles. Under mild
assumptions, it is a
consequence of this structure that when the universe was extremely young
and hot, the net baryon number of the universe was zero. That is, the
number of particles carrying a given baryon number in any region was
equal on average to the number of the appropriate antiparticles carrying the 
opposite
baryon number. However, on the other hand, it is a clear observational fact 
that the universe is maximally baryon - antibaryon asymmetric. This
fact is quantified from the considerations of primordial nucleosynthesis. 
These calculations are perhaps the most impressive success of the standard
cosmology, accurately predicting the abundances of the light elements from 
the single input parameter of the baryon to entropy ratio, which is 
constrained as in equation~(\ref{nucleo}). Until recently, there were two
possible explanations for this. First, the universe as a whole could be baryon
number symmetric, but baryon-antibaryon separation could have resulted in an
apparent baryon asymmetry in the local universe. the second possibility is
that some dynamical process took place as the universe evolved, causing
baryons to be preferentially produced over antibaryons. A recent analysis
\cite{CRG 97} has ruled out the former possibility over scales up to the
size of the observable universe. It
therefore appears that the latter option, {\it baryogenesis}, must have
taken place.

In this article I have tried to describe how a number of different physical
effects, all present in the standard electroweak theory at nonzero
temperature, can come together in the context of the expanding universe
to implement baryogenesis. It is an amazing fact about the
Glashow-Salam-Weinberg model and its modest extensions that they satisfy
all three Sakharov criteria for producing a baryon excess. As a result, 
over the
last decade, {\it electroweak baryogenesis} has been a very popular scenario
for the generation of the BAU.

There are four technical issues to be investigated when considering models
of electroweak baryogenesis. these are

\begin{enumerate}
\item How is the departure from equilibrium realized? Is the electroweak phase
transition strongly first order or are topological defects necessary?

\item How is sufficient CP violation obtained?

\item What is the rate of baryon number violating processes? Is this rate high
enough in the unbroken phase and can washout of any asymmetry be avoided?

\item What are the actual dynamics of baryon number production?

\end{enumerate}

I hope I have described how each of these issues has been addressed in the
literature on electroweak baryogenesis. Detailed analyses of the phase
transition, coupled with the smallness of the CP violation due to phases
in the CKM matrix, have made it clear that an extension of the standard model
is required to make the scenarios viable. While general two-Higgs doublet
theories have been considered, perhaps the most appealing candidate from
particle physics motivations is the minimal supersymmetric standard model.
The implementations of electroweak baryogenesis in this model have recently
been investigated and the range of parameters of the model for which an
appreciable BAU can be generated have been calculated. This range should be
accessible to future particle colliders and thus the scenario of EWBG in the
MSSM should be experimentally testable. If the MSSM or two-Higgs models
are not chosen by nature, then other models of electroweak baryogenesis may
be relevant. If the topological structure of the relevant theory admits
gauge solitons, then defect mediated electroweak baryogenesis may be important,
irrespective of the order of the phase transition.

Although I have argued that there exist viable scenarios of
electroweak baryogenesis, particularly in the context of supersymmetric
models, there remain a number or open questions and directions for
future research. I shall list the ones that I feel are most 
important.

\begin{enumerate}
\item At present, the best quantitative understanding of the 
electroweak phase transition comes from lattice Monte Carlo
simulations. While the results from these are impressive, they do not
provide an intuitive understanding of the microphysics of the
phase transition. The numerical results are partially supported by
some analytical approaches but these often cannot be trusted in
the physical range of Higgs masses. An analytic understanding of
the nonperturbative dynamics of the phase transition would be an
important step forward.

\item The chemical potential and ``fluid'' approaches to nonlocal
baryogenesis provide good analytical tools for understanding the
nonlocal production of baryons. However, in the case of local 
baryogenesis, analytical methods which yield believable quantitative
results have yet to be found. Again, it is encouraging that numerical
calculations, for example of Chern-Simons number diffusion, are 
providing quantitative predictions but an appropriate analytical
model is very desirable.

\item While it will strongly support electroweak baryogenesis if 
supersymmetry is verified with parameters in the correct range, it is
important to ensure that there is sufficient CP violation. To this
end, phenomenological predictions of CP violation in SUSY models
and the corresponding experimental tests are crucial hurdles that
the scenario must pass.

\end{enumerate}

If electroweak baryogenesis is correct, then no matter what the relevant 
electroweak model is, the physics involved is extremely beautiful and
diverse. The topology of gauge field theories, the physics of phase
transitions, CP violation, plasma dynamics and thermal field theory all play
a part in generating the BAU. However, perhaps the most attractive feature of
electroweak baryogenesis scenarios is that they should be testable at the 
next generation of particle colliders. It is an exciting possibility that 
we may soon understand the origin of the matter that makes up our universe.

\acknowledgments
I first want to
thank my collaborators; Robert Brandenberger, Anne Davis,
Arthur Lue, Tomislav Prokopec and Krishna Rajagopal for the pleasure of
working with them, for helping shape my ideas about baryogenesis, and
for their permission to use parts of our joint work in this article. 
I also owe thanks to Robert Brandenberger, Jim
Cline, Marcelo Gleiser, Michael Joyce, Mikko Laine, Guy Moore, 
Misha Shaposhnikov and Tanmay Vachaspati for a critical reading of the 
first draft of this work and
for their advice. Their many suggestions and comments 
were invaluable and undoubtably made this a much better article.

Finally, I should acknowledge further advice and discussions during 
the time I have spent thinking about electroweak baryogenesis with
Andy Cohen, Eddie Farhi, Jeffrey Goldstone, Alan Guth, Ken Johnson, Lawrence
Krauss, Lisa Randall, Glenn Starkman and Neil Turok.

This work was supported by the U.S. department of Energy, the National
Science Foundation and by funds provided by Case Western Reserve
University. I would also like to acknowledge the Center for Theoretical 
Physics at MIT, where I began this work.

\newpage

FIG. 1.  Zone of Concordance for primordial nucleosynthesis. 
$\tau_N$ denotes the value of the neutron lifetime used in these calculations.
Thanks to Peter Kernan and Lawrence Krauss for providing this figure
\cite{K&K 95}.

FIG. 2. The triangle diagram contributing to the anomaly in the baryon
and lepton number currents.

FIG. 3. The multiple vacua of the electroweak theory depicted along some
direction in configuration space. The sphaleron is also shown.

FIG. 4.  Sketch of the finite temperature effective potential for various
values of temperature for a first order phase transition.

FIG. 5.  Sketch of the finite temperature effective potential for various
values of temperature for a second order phase transition.

FIG. 6.  The maximum susceptibility $\chi_{max}$ as a function of
volume for different values of the Higgs mass $m_H$. The continuous lines
are mean field fits and the dashed line corresponds to the mean
field critical exponent. Thanks to Kari Rummukainen for supplying
this figure.

FIG. 7.  Evolution of the ensemble-averaged fraction $f_0(t)$ for several 
values of the scalar field self-coupling $\lambda$. Thanks to Marcelo
Gleiser for permission to use this figure, which first appeared in
\cite{B&G 95}.

FIG. 8.  Sketch of two qualitatively different possible behaviors of the 
critical surfaces $F^+=0$ (represented by a solid line) and $F^-=0$ 
(represented by a dashed line) in the space $(\beta_1,\beta_2,\ldots)$
describing initial configurations.

FIG. 9.  Figure illustrating the relevant quantities for nonlocal baryogenesis
as a bubble wall propagates through the electroweak plasma, converting false
vacuum to true.

FIG. 10.  Figure illustrating the analogy between the bubble wall mediated
baryogenesis scenario and the defect mediated scenario.

\end{document}